\documentclass[article]{elsarticle}
\usepackage{color}
\usepackage{hyperref}
\usepackage{ulem}

\journal{Journal of Computational Physics}

\usepackage{amsmath}
\usepackage{amssymb}
\allowdisplaybreaks 
\usepackage{graphicx}
\usepackage{epsfig}
\usepackage[caption = false]{subfig}
\usepackage{tabularx}
\usepackage{booktabs} 
\usepackage{multirow}
\usepackage{upgreek}
\usepackage{mathtools}
\usepackage{nomencl}
\makenomenclature
\usepackage[acronym,nomain,nonumberlist]{glossaries}
\makeglossaries
\usepackage{xcolor}

\usepackage{etoolbox}
\renewcommand\nomgroup[1]{%
    \item[\bfseries
    \ifstrequal{#1}{A}{Roman}{%
    \ifstrequal{#1}{B}{Greek}{%
    \ifstrequal{#1}{C}{Mathematical Notation}{%
    \ifstrequal{#1}{D}{Subscripts}{%
    \ifstrequal{#1}{E}{Superscripts}{}}}}}%
]}

\newcommand{\dzdt}{\frac{\partial \zeta}{\partial t}}
\newcommand{\dudt}{\frac{\partial u}{\partial t}}

\newcommand{\dpdt}{\frac{\partial p}{\partial t}}

\newcommand{\dudx}{\frac{\partial u}{\partial x}}

\newcommand{\dpdx}{\frac{\partial p}{\partial x}}
\newcommand{\dzdy}{\frac{\partial \zeta}{\partial y}}

\newcommand{\dpdy}{\frac{\partial p}{\partial y}}
\newcommand{\dzdz}{\frac{\partial \zeta}{\partial z}}

\newcommand{\dpdz}{\frac{\partial p}{\partial z}}

\newcommand{\dd}[2]{\frac{\partial #1}{\partial #2}}

\newcommand{\phin}[2]{\phi_{#1}^{#2}}

\newcommand{\nsum}[3]{\sum_{#1=#2}^{n^{#3}}}
\newcommand{\bx}{\textbf{x}}
\newcommand{\bv}{\textbf{v}}
\newcommand{\ba}{\textbf{a}}

\newcommand{\bq}{\textbf{q}}
\newcommand{\norm}[1]{\left\lVert#1\right\rVert}
\newcommand{\lp}[1]{\left #1}
\newcommand{\rp}[1]{\right #1}

\newcommand{\bZ}{\textbf{Z}}
\newcommand{\R}{\mathbb{R}}

\newcommand{\bK}{\mathbf{K}}

\newcommand{\omegax}{\omega_{x}}
\newcommand{\bA}{\textbf{A}}
\newcommand{\tc}[2]{a_{#1}^{#2}}
\newcommand{\ip}[2]{\lp{(} #1, #2 \rp{)}}
\newcommand{\amp}{\tilde{A}}
\newcommand{\freq}{\tilde{\omega}}
\newcommand{\abs}[1]{\left \lvert #1 \right \rvert}
\newcommand{\bF}{\textbf{F}}
\newcommand{\tauc}{\pmb{\hat{\uptau}}}
\newcommand{\taucc}{\hat{\tau}}
\newacronym{CFD}{CFD}{Computational Fluid Dynamics}
\newacronym{FOM}{FOM}{Full-Order Model}
\newacronym{LCO}{LCO}{Limit Cycle Oscillation}
\newacronym{ODE}{ODE}{Ordinary Differential Equation}
\newacronym{PDE}{PDE}{Partial Differential Equation}
\newacronym{POD}{POD}{Proper Orthogonal Decomposition}
\newacronym{ROM}{ROM}{Reduced-Order Model}

\begin{document}

%
%
%
%
%
%
%
\begin{frontmatter}

\title{An Efficient Proper Orthogonal Decomposition based Reduced-Order Model}

\author[addr]{Elizabeth H. Krath\corref{cor}}
\cortext[cor]{Corresponding author}
\address[addr]{Department of Aerospace Engineering, Texas A\&M University, 701 H.R. Bright Building 3141, College Station, TX, USA 77843}
\ead{ekrath@tamu.edu}

\author[addr]{Forrest L. Carpenter}
\ead{fcarpenter4@gmail.com}

\author[addr]{Paul G. A. Cizmas}
\ead{cizmas@tamu.edu}

\author[addr2]{David A. Johnston}
\ead{david.johnston.17@us.af.mil}
\address[addr2]{Turbine Engine Division, Air Force Research Laboratory, Wright-Patterson Air Force Base, Ohio, USA 45433}

\begin{abstract}
  This paper presents a novel, more efficient proper orthogonal
  decomposition (POD) based reduced-order model (ROM) for compressible
  flows. In this POD model the governing equations, {\it i.e.}, the
  conservation of mass, momentum, and energy equations were written using
  specific volume instead of density. This substitution allowed for the
  pre-computation of the coefficients of the system of ODEs that make up the
  reduced-order model. Several methods were employed to enhance the
  stability of the ODE solver: the penalty method to enforce boundary
  conditions, artificial dissipation, and a method that modifies the number
  of modes used in the POD approximation. This new POD-based reduced-order
  model was validated for four cases at both on- and off-reference
  conditions: a quasi-one-dimensional nozzle, a two-dimensional channel, a
  three-dimensional axisymmetric nozzle, and a transonic fan.  The speedup
  obtained by using the POD-based ROM vs. the full-order model exceeded four
  orders of magnitude in all cases tested.
\end{abstract}

\begin{keyword}
Proper Orthogonal Decomposition \sep Reduced-Order Model \sep Computational Fluid Dynamics 

\end{keyword}

\end{frontmatter}

\nomenclature[A, 01]{$a$}{Time coefficient}
\nomenclature[A, 02]{$A$}{Area}
\nomenclature[A, 03]{$\amp$}{Amplitude}
\nomenclature[A, 04]{$c$}{Speed of sound}
\nomenclature[A, 05]{$D$}{Dimension of domain}
\nomenclature[A, 06]{$\textbf{F}$}{Prescribed boundary condition}
\nomenclature[A, 07]{$\mathcal{F}$}{Nonlinear functional}
\nomenclature[A, 08]{$\pmb{\mathcal{G}}$}{Grassmann manifold}
\nomenclature[A, 09]{$J$}{Jacobian}
\nomenclature[A, 10]{$\mathcal{J}$}{Nonlinear functional}
\nomenclature[A, 11]{$\textbf{K}$}{Boundary condition error}
\nomenclature[A, 12]{$m$}{Total number of modes}
\nomenclature[A, 13]{$M$}{Number of snapshots}
\nomenclature[A, 14]{$n$}{Number of modes}
\nomenclature[A, 15]{$N$}{Resolution of spatial domain}
\nomenclature[A, 16]{$N_R$}{Number of reference subspaces}
\nomenclature[A, 17]{$p$}{Pressure}
\nomenclature[A, 18]{$\textbf{q}$}{State vector}
\nomenclature[A, 19]{$r$}{Radius}
\nomenclature[A, 20]{$\textbf{R}$}{Autocorrelation matrix}
\nomenclature[A, 21]{$s$}{Entropy}
\nomenclature[A, 22]{$\textbf{S}$}{Projection of Laplacian of basis functions}
\nomenclature[A, 23]{$\pmb{\mathcal{S}}$}{POD subspace}
\nomenclature[A, 24]{$t$}{Time}
\nomenclature[A, 25]{$u$}{\textit{x}-Velocity}
\nomenclature[A, 26]{$v$}{\textit{y}-Velocity}
\nomenclature[A, 27]{$\bv$}{Velocity vector}
\nomenclature[A, 28]{$w$}{\textit{z}-Velocity}
\nomenclature[A, 29]{$x$}{Spatial coordinate}
\nomenclature[A, 30]{$\bZ$}{Zeta state vector}

\nomenclature[B, 01]{$\gamma$}{Specific heat ratio}
\nomenclature[B, 02]{$\pmb{\Gamma}$}{Tangent space to POD subspace}
\nomenclature[B, 03]{$\epsilon$}{Least-squares error}
\nomenclature[B, 04]{$\varepsilon$}{Relative  error}
\nomenclature[B, 05]{$\zeta$}{Specific volume}
\nomenclature[B, 07]{$\lambda$}{Eigenvalue}
\nomenclature[B, 08]{$\tilde{\lambda}$}{Operating points}
\nomenclature[B, 09]{$\tilde{\nu}$}{Artificial dissipation parameter}
\nomenclature[B, 10]{$\tilde{\pmb{\nu}}$}{Artificial dissipation parameter diagonal matrix}
\nomenclature[B, 11]{$\rho$}{Density}
\nomenclature[B, 12]{$\tau$}{Penalty parameter}
\nomenclature[B, 13]{$\pmb{\uptau}$}{Penalty parameter diagonal matrix}
\nomenclature[B, 14]{$\phi$}{Basis function}
\nomenclature[B, 15]{$\varphi$}{Phase shift}
\nomenclature[B, 16]{$\Omega$}{Spatial domain}
\nomenclature[B, 17]{$\freq$}{Angular velocity}
\nomenclature[B, 18]{$\omegax$}{Wheel speed of blade rotating about \textit{x}-axis}

\nomenclature[C, 01]{$\langle \cdot \rangle$}{Ensemble average}
\nomenclature[C, 02]{$(\cdot,\cdot)$}{Inner product}

\nomenclature[D, 01]{BC}{Subset belonging to points along prescribed boundary}
\nomenclature[D, 02]{hub}{Subset belonging to outlet hub boundary}
\nomenclature[D, 03]{out}{Subset belonging to outlet boundary}

\nomenclature[E, 01]{$\hat{~}$}{Approximate value}
\nomenclature[E, 02]{${*}$}{Complex conjugate}
\nomenclature[E, 03]{$\tilde{~}$}{Perturbation}
\nomenclature[E, 04]{$\bar{~}$}{Spatial-averaged value}

\printnomenclature
\glsaddall
\printglossary[type=\acronymtype,title=Acronyms]

\section{Introduction}
Developments in computer hardware and infrastructure have opened up pathways
toward high-fidelity model development of complex flow problems. Despite
these advancements, the full-order model (FOM) of these complex simulations
remains computationally expensive. One can reduce the computational time of
these FOMs by improving the numerical scheme and/or grid quality. While
these approaches may be effective, the FOM still remains limited in its use
for design as well as for real-time feedback~\cite{ParkLee1998,
  Slotnick2014}.  To amend this expense, one can instead use model reduction
procedures to create reduced-order models (ROM). These model reduction
procedures replace the larger system of partial differential equations
(PDEs) of the full-order model by a much smaller system of ordinary
differential equations (ODEs) that make up the reduced-order model. One such
model reduction procedure is the Proper Orthogonal Decomposition (POD)
method.

Proper Orthogonal Decomposition was developed by Karhunen and Loeve
~\cite{Karhunen1946, 1945Loeve} and applied to fluid dynamics by
Lumley~\cite{Lumley1967}. The POD method is a common model reduction
procedure used to develop ROMs, and has been used for many applications such
as turbomachinery~\cite{Epureanu2000, A.Cizmas2003},
cavity~\cite{Nagarajan2017}, and multi-phase~\cite{Yuan2005, Brenner2010}
flows. Other model reduction methods have also been explored such as the
Balanced POD~\cite{Rowley2005}, dynamic mode
decomposition~\cite{SCHMID2010}, and bi-orthogonal
decomposition~\cite{Aubry1991}. Lucia \textit{et al.} have presented an
extensive review of several model reduction methods~\cite{Lucia2004}
including Volterra theory, harmonic balance, and the traditional POD
method. While this is not a comprehensive list, it is clear that these model
reduction procedures have become an important part of CFD model
development. Furthermore, manipulations to the traditional POD method have
been performed involving modifying the form of the governing equations,
\textit{e.g.}, through linearizing the system of equations, changing the
inner product of the Galerkin projection procedure~\cite{Barone2009}, or
extending the POD approximation to deforming computational
domains~\cite{Freno2014a}.

One drawback of the POD method is in its inability to preserve the stability
of the FOM. This limits the robustness of the ROM for handling changes in
geometry and flow conditions. While this is the case, it does not mean that
the ROM cannot be stabilized. In fact, there have been several methods
developed that can aid in regaining the stability lost through the POD
process. For example, extensions on the traditional POD method for stability
have been done for linear-time invariant ROMs~\cite{Bond2008, Amsallem2012a}
with other research focusing on eigenvalue reassignment of linear-time
invariant systems~\cite{Kalashnikova2014}. It is possible to extend the
eigenvalue reassignment of the linear-time invariant systems to nonlinear
systems through an iterative approach~\cite{Tomas-Rodriguez2013}. Another
approach is to develop stability-preserving inner products respective to the
type of flow problem~\cite{Barone2009, Kalashnikova2010}. Furthermore, it
has been found that the lack of enforcement of boundary conditions in a ROM
can lead to constrained solutions, which may lead to instabilities in the
ROM~\cite{Rempfer2000}. Therefore, enforcing boundary conditions in a ROM is
another way to preserve stability.  The penalty method can be used for such
a purpose~\cite{Hesthaven1996, Sirisup2005, Kalashnikova2012}. This method
enforces the boundary conditions by penalizing the system of ODEs if they do
not satisfy the boundary conditions of the governing system of
PDEs. Artificial dissipation has also been successfully used to stabilize
the ROM after the fact by adding an artificial dissipation term to the
governing equations~\cite{Lucia2003, Sirisup2004, Borggaard2011}.

This paper presents a novel POD method applied to the conservation of mass,
momentum, and energy equations written as a function of specific volume
instead of density. When applying the POD method to the governing equations
of fluid dynamics written as a function of the primitive or conservative
variables, a POD approximate appears in the denominator. {This increases the
  non-linearity of the system and may} require that the coefficients of the
ODEs that make up the ROM be recalculated at every time step. To correct
this limitation, the governing equations are written as a function of
specific volume instead of density~\cite{Iollo2000, Sandia2014}. This
formulation allows for the pre-computation of the coefficients of the ODEs,
which significantly reduces the computational time of the ROM. The following
section presents these governing equations written as a function of specific
volume instead of density. Then, the POD method is described along with the
stabilizing methods used for the POD-based ROM. Validation results of the
ROM are presented for four cases: a quasi-one-dimensional nozzle, a
two-dimensional channel, a three-dimensional axisymmetric nozzle, and a
three-dimensional transonic fan rotor. Results are given for both on- and
off-reference conditions. Finally, the CPU runtime of the ROM written in
terms of specific volume is compared to its FOM.

\section{Methodology}

This section presents the methods used in the development of the POD-based
ROM that used specific volume instead of density. The first part of this
section discusses the POD method. The second part presents the governing
equations of the FOM and ROM in two ways: for a stationary reference frame
and for a reference frame rotating about the \textit{x}-axis. Finally, the
third part discusses the stability methods applied to the ROM.

\subsection{Proper Orthogonal Decomposition}
Proper Orthogonal Decomposition, otherwise referred to as Principal Component Analysis or the Karhunen-Loeve Decomposition, was developed by Kar\-hu\-nen and Loeve~\cite{Karhunen1946,Loeve1945} as a statistical method to extract optimal basis sets from an ensemble of observations. The optimal basis is found in such a way that the time-averaged approximation error between the POD approximation and the full-order model is minimized in a respective norm.

To extract these optimal basis sets, consider a set of discrete snapshots of some scalar function ${q(x,t_{i})}$, $1 \le i \le M$, where $M$ is the number of snapshots. This set of snapshots is assumed to form a linear, finite-dimensional Hilbert space $L^{2}$ on a spatial domain $\Omega$. The POD method approximates the perturbation of the scalar function, $\tilde{q}=q-\bar{q}$, as a linear combination of some time-dependent orthogonal time coefficients, $a$, and time-independent orthogonal basis functions, $\phi$,
\begin{equation}
\label{PODapprox}
\tilde{q}(x,t_{i}) = \nsum{j}{1}{q} \tc{j}{ q}(t_{i}) \phin{j}{q}(x)	
\end{equation}
where $n^{q}$ is the number of terms or modes kept to approximate $\tilde{q}(x,t_{i})$. Alternatively, the POD approximation can be applied to the scalar function $q$
\begin{equation}
\label{eq:PODapprox0}
{q}(x,t_{i}) = \nsum{j}{0}{q} \tc{j}{ q}(t_{i}) \phin{j}{q}(x)	
\end{equation}
where $\tc{0}{ q}(t_{i})=1$ and $\phin{0}{q}(x)=\bar{q}$.
The reconstruction~\eqref{PODapprox} is optimal in the sense that the time-averaged least-square error of the POD approximation
\begin{equation}
\label{PODmin}
\epsilon_{n^{q}}^{q} = \lp{\langle} \norm{\tilde{q}(x,t_{i})-\nsum{k}{1}{q} \tc{k}{q}(t_{i}) \phin{k}{q}(x)}^{2} \rp{\rangle}	
\end{equation}
is a minimum for any $n^{q} \leq M$ combinations of basis functions~\cite{cipaobsy03}. Here $\norm{\cdot}$ denotes the $L^{2}$ norm, $\norm{f}=(f,f)^{1/2}$, where $(\cdot,\cdot)$ denotes an inner product, and $\langle \cdot \rangle$ denotes an ensemble average, $\langle f \rangle = \frac{1}{T} \int_{0}^{T}f(x,t)dt$. This minimization reduces to an eigenvalue problem of the form~\citep[p. 89]{holmes96}
\begin{equation}
\int_{\Omega} \lp{\langle} \tilde{q}(x) \tilde{q}^{*}(x) \rp{\rangle} \phi(y) dy = \lambda \phi(y)
\label{eigint}
\end{equation}
which is a homogenous Fredholm integral equation of the second kind~\cite{Kantorovich1964}.  The optimal basis functions are the eigenfunctions of the integral eigenvalue equation. The kernel of~\eqref{eigint} is the autocorrelation function $R(x,y) = \langle \tilde{q}(x) \tilde{q}^{*}(y) \rangle$ where $*$ denotes a complex conjugate. The integral equation in~\eqref{eigint} is solved using a Singular Value Decomposition (SVD).

If instead one has a set of discrete vector-valued observations $\bq(\bx,t_{i}) = [q(x_{1},t_{i}), q(x_{2},t_{i}), ..., q(x_{N},t_{i})]^{T}$ where $N$ is the resolution of the spatial domain, the discrete ensemble average becomes $\langle f \rangle = \frac{1}{M} \sum_{i=1}^{M} f(x,t_{i})$ and the autocorrelation function becomes a tensor product matrix
\begin{equation}
\label{autocorr}
\textbf{R} = \frac{1}{M} \sum_{i=1}^{M} \tilde{\bq}(\bx,t_{i}) \tilde{\bq}^{T} (\textbf{y},t_{i})	
\end{equation}
where $\textbf{R}$ is a symmetric, positive semi-definite matrix which by
definition has orthogonal eigenvectors~\cite{Pinnau2008}. For this discrete
set of observations, the method of snapshots developed by
Sirovich~\cite{Sirovich1987} is used to calculate the eigenvectors of
$\textbf{R}$. This method is efficient when $M \ll N$, since it reduces a
system of $N$ equations to one of $M$ equations.  Details on the
implementation of this method are given in~\cite{cipaobsy03,cizpal03}.

The only remaining unknowns after solving for the optimal set of basis functions are the time coefficients. Because the basis functions are orthogonal, the time coefficients can be calculated as
\begin{equation}
\label{tceq}
\tc{k}{q} = \frac{\ip{\tilde{q}}{\phin{k}{q}}}{\ip{\phin{k}{q}}{\phin{k}{q}}} .
\end{equation}
While~\eqref{tceq} produces the ``exact" time coefficients that best approximate $\tilde{q}$, these time coefficients cannot be used for predicting the solution at off-reference conditions, that is, for conditions different from those predicted by the full-order model. For off-reference conditions, the time coefficients need to be computed by solving the system of ordinary differential equations obtained by substituting the approximation~\eqref{PODapprox} into the governing equations and then projecting the approximate governing equations along the basis functions. 

\subsection{Governing Equations}

The governing equations of fluid motion are the mass, momentum, and energy conservation equations. Here we use the equations for a compressible, inviscid flow, otherwise known as the Euler equations. In this section, formulations of the Euler equations are given in two reference frames: a stationary and a rotating reference frame. The rotating reference frame is beneficial for modeling turbomachinery flows. Before presenting these formulations, the impact of using density $\rho$ in the governing equations is explored.

\subsubsection{Primitive Variables vs. Zeta Variables Formulation}

The Euler equations written in differential form using primitive variables are
\begin{equation}
\label{primitive}
\begin{aligned}
\dd{\rho}{t} &+ \bv \cdot \nabla \rho + \rho \lp{(} \nabla \cdot \bv \rp{)} = 0 \\
\dd{\bv}{t} &+ \lp{(} \bv \cdot \nabla \rp{)} \bv = - \frac{1}{\rho} \nabla p \\
\dpdt &+ \gamma p \lp{(} \nabla \cdot \bv \rp{)} + \lp{(} \bv \cdot \nabla \rp{)} p = 0
\end{aligned}
\end{equation}
where $\bv=(u,v,w)^{T}$ is the velocity vector, $p$ is pressure, and $\gamma$ is the specific heat ratio. The state vector for this system is $\textbf{q} = (\rho, u, v, w, p)^{T}$. One can see that density $\rho$ appears in the denominator for the conservation of momentum equations. For example, consider the one-dimensional \textit{x}-momentum conservation equation
\begin{equation*}
\dudt + u\dudx = -\frac{1}{\rho} \dpdx	
\end{equation*}
Inserting the POD approximation~\eqref{PODapprox} yields
\begin{equation*}
\nsum{i}{0}{u} \dot{a}_{i}^{u} \phin{i}{u} + \nsum{i}{0}{u} \nsum{j}{0}{u} \tc{i}{u} \tc{j}{u} \phin{i}{u} \phin{j,x}{u} = - \frac{\nsum{i}{0}{p} \tc{i}{p} \phin{i,x}{p}}{\nsum{j}{0}{\rho} \tc{j}{\rho} \phin{j}{\rho}}	
\end{equation*}
where a POD approximate appears in the denominator of the right-hand side.
Projecting along the POD basis function $\phin{k}{u}$ yields
\begin{equation}
\label{rhoissue}
\dot{a}_{k}^{u} + \nsum{i}{0}{u} \nsum{j}{0}{u} \ip{\phin{i}{u} \phin{j,x}{u}}{\phin{k}{u}} \tc{i}{u} \tc{j}{u} = 
- \ip{\frac{\nsum{i}{0}{p} \tc{i}{p} \phin{i,x}{p}}{\nsum{j}{0}{\rho} \tc{j}{\rho} \phin{j}{\rho}}}{\phin{k}{u}}.
\end{equation}
On the left-hand side, the time-derivative term is contracted due to the orthogonality of the basis functions. 
The coefficient corresponding to the convective term depends only on space,
and so can be precomputed before solving the system of ODEs. On the
right-hand side of~\eqref{rhoissue}, however, the coefficient corresponding
to the pressure gradient term depends on both space and time due to the
appearance of $\rho$ in the denominator.
%
Having a coefficient of the ODEs dependent on time is not ideal, as it requires reconstructing the solution using~\eqref{PODapprox} and then projecting along $\phin{k}{u}$ at each time step. This dependence only adds to the computational expense of the ROM, which is counter to the goal of reducing computational expense. Ideally, the coefficients of the ODEs of the ROM should be precomputed before solving the system of ODEs. 

If instead one multiplies the momentum equations of~\eqref{primitive} by $\rho$, the density is removed from the denominator and the coefficients of the ODEs can be precomputed. 
\begin{equation}
\label{pseudoconservative}
\begin{aligned}
\rho \dd{\bv}{t} &+ \rho \lp{(} \bv \cdot \nabla \rp{)} \bv = - \nabla p \\
\end{aligned}
\end{equation}

The multiplication by density, however, increases the nonlinearity of the left-hand side of~\eqref{pseudoconservative}.  To illustrate this, let us project the time-derivative term of the \textit{x}-momentum along $\phin{k}{u}$
\begin{equation*}
\ip{\rho \dudt}{\phin{k}{u}} = \ip{\nsum{i}{0}{\rho} \nsum{j}{0}{u} \tc{i}{\rho} \phin{i}{\rho} \dot{a}_j^u \phin{j}{u}}{\phin{k}{u}} = \nsum{i}{0}{\rho} \nsum{j}{0}{u} \tc{i}{\rho} \dot{a}_j^u \ip{\phin{i}{\rho} \phin{j}{u}}{\phin{k}{u}} .
\end{equation*}
It is apparent that the term cannot be fully contracted. Therefore, at every time step, a separate system of equations will need to be solved to find $\dot{a}_k^u$ rather than solving for the derivative directly. This will occur for every velocity component. For example, expanding for $n^\rho=1$ 
\begin{equation*}
	\ip{\rho \dudt}{\phin{k}{u}} = 
	\tc{1}{\rho} \dot{a}_1^u \ip{\phin{1}{\rho} \phin{1}{u}}{\phin{k}{u}} + \tc{1}{\rho} \dot{a}_2^u \ip{\phin{1}{\rho} \phin{2}{u}}{\phin{k}{u}} + \ldots + \tc{1}{\rho} \dot{a}_{n^u}^u \ip{\phin{1}{\rho} \phin{n^u}{u}}{\phin{k}{u}}
\end{equation*}
leads to a system of size $n^u$ that must be solved to determine $\dot{a}_k^u, \ k\in[1,n^u]$.  By comparison, if the time-derivative of \textit{x}-velocity is evaluated separately, then the orthogonality of the POD modes simplifies the expression
\begin{equation*}
\ip{\dudt}{\phin{k}{u}} = \ip{\nsum{i}{0}{u} \dot{a}_i^u \phin{i}{u}}{\phin{k}{u}} = \nsum{i}{0}{u} \dot{a}_i^u \ip{\phin{i}{u}}{\phin{k}{u}} = \dot{a}_k^u .
\end{equation*}

To correct both issues, the specific volume $\zeta$ is used instead of density $\rho$ in the governing equations. The following sections present the governing equations in this format.

\subsubsection{Stationary Frame}
The Euler equations written as a function of specific volume $\zeta$ instead of density $\rho$ in a stationary frame are~\cite{Iollo2000}
\begin{equation}
\label{zeta_stat}
\begin{aligned}
\dzdt &+ \bv \cdot \nabla \zeta - \zeta \lp{(} \nabla \cdot \bv \rp{)} = 0 \\
\dd{\bv}{t} &+ \bv \cdot \nabla \bv + \zeta \nabla p = 0 \\
\dpdt &+ \bv \cdot \nabla p + \gamma p \lp{(} \nabla \cdot \bv \rp{)} = 0 .
\end{aligned}
\end{equation}
The governing equations written in vectorial form are
\begin{equation} \label{eq:vecform}
\dd{\bZ_{i}}{t} + \bA_{1_{i}}\dd{\bZ_{i}}{x} + \bA_{2_{i}}\dd{\bZ_{i}}{y} + \bA_{3_{i}}\dd{\bZ_{i}}{z} = \textbf{Q}_{i}, \quad 1 \le i \le N
\end{equation}
where $\bZ_{i}$ is the state vector
\begin{equation*}
\bZ_{i} = \begin{bmatrix}
 \zeta & u & v & w & p	
 \end{bmatrix}_{i}^{T}
\end{equation*}
and
\begin{equation*}
\bA_{1_{i}} = \begin{bmatrix}
 u & -\zeta & 0 & 0 & 0 \\
 0 & u & 0 & 0 & \zeta \\
 0 & 0 & u & 0 & 0 \\
 0 & 0 & 0 & u & 0 \\
 0 & \gamma p & 0 & 0 & u	
 \end{bmatrix}_{i}
\end{equation*}
\begin{equation*}
\bA_{2_{i}} = \begin{bmatrix}
 v & 0 & -\zeta & 0 & 0 \\
 0 & v & 0 & 0 & 0\\
 0 & 0 & v & 0 & \zeta \\
 0 & 0 & 0 & v & 0 \\
 0 & 0 & \gamma p & 0 & v	
 \end{bmatrix}_{i}
\end{equation*}
\begin{equation*}
\bA_{3_{i}} = \begin{bmatrix}
 w & 0 & 0 & -\zeta & 0 \\
 0 & w & 0 & 0 & 0\\
 0 & 0 & w & 0 & 0 \\
 0 & 0 & 0 & w & \zeta \\
 0 & 0 & 0 & \gamma p & w	
 \end{bmatrix}_{i}
\end{equation*}
and $\textbf{Q}_{i} = \textbf{0}$. 

\subsubsection{Rotating Reference Frame}
The Euler equations written for a reference frame rotating about the \textit{x}-axis in terms of specific volume instead of density are
\begin{equation}
\label{zeta_rot}
\begin{aligned}
	\dzdt &+ \lp{(} \bv \cdot \nabla \rp{)} \zeta - \zeta \lp{(} \nabla \cdot \bv \rp{)} + \omegax \lp{(} z\dzdy - y\dzdz \rp{)} = 0 \\
	\dd{\bv}{t} &+ (\bv \cdot \nabla)\bv + \zeta \nabla p + \omegax \lp{(} z \dd{\bv}{y} - y\dd{\bv}{z} \rp{)} = \tilde{\textbf{Q}} \\
	\dpdt &+ \gamma p \lp{(} \nabla \cdot \bv \rp{)} + \lp{(} \bv \cdot \nabla \rp{)} p + \omegax \lp{(} z \dpdy - y \dpdz \rp{)} = 0
\end{aligned}	
\end{equation}
where $\omegax$ is the angular velocity about the \textit{x}-axis and $\tilde{\textbf{Q}} = \omegax [0, w, -v]^{T}$.  The vectorial form~\eqref{eq:vecform} is still valid, but the matrices $\bA_{2_{i}}$, $\bA_{3_{i}}$ and the vector $\textbf{Q}_{i}$ have changed to
\begin{equation*}
	\bA_{2_{i}} = \begin{bmatrix}
 		v+z \omegax & 0 & -\zeta & 0 & 0 \\
 		0 & v+z \omegax  & 0 & 0 & 0 \\
 		0 & 0 & v+z \omegax & 0 & \zeta \\
 		0 & 0 & 0 & v + z \omegax & 0 \\
 		0 & 0 & \gamma p & 0 & v + z \omegax	
 	\end{bmatrix}_{i}
\end{equation*}
\begin{equation*}
	\bA_{3_{i}} = \begin{bmatrix}
 		w - y\omegax & 0 & 0 & -\zeta & 0 \\
 		0 & w - y\omegax & 0 & 0 & 0 \\
 		0 & 0 & w - y\omegax & 0 & 0 \\
 		0 & 0 & 0 & w - y\omegax & \zeta \\
 		0 & 0 & 0 & \gamma p & w - y\omegax	
 	\end{bmatrix}_{i}
\end{equation*}
and $\textbf{Q}_{i} = \omegax [0, 0, w, -v, 0]_{i}^{T}$.

\subsection{Flow Solver}
Two models were used to solve the governing equations: a full-order model (FOM) and a reduced-order model (ROM). The FOM solved the discretized governing equations while the ROM solved the projection of the governing equations. The FOM and the ROM are covered in the following sections.

\subsubsection{Full-Order Model}
Two full-order models were used for generating the results presented herein. The first FOM solved the one-dimensional Euler equations~\eqref{primitive} written using characteristic variables~\cite{Roe1986}
\begin{equation}
\label{characteristic}
\begin{gathered}
	\dd{s}{t} + u\dd{s}{x} = 0 
	\\
	\dd{}{t} \lp{(} u + \frac{2c}{\gamma-1} \rp{)} + \lp{(} u+c \rp{)} \dd{}{x} \lp{(} u+\frac{2c}{\gamma-1} \rp{)} + cu \frac{1}{A} \dd{A}{x} = 0 
	\\
	\dd{}{t} \lp{(} u - \frac{2c}{\gamma-1} \rp{)} + \lp{(} u-c \rp{)} \dd{}{x} \lp{(} u-\frac{2c}{\gamma-1} \rp{)} - cu \frac{1}{A} \dd{A}{x} = 0
\end{gathered}
\end{equation}
where $s$ is the entropy and $c$ is the speed of sound. MacCormack's technique was used to solve the FOM system~\cite{Anderson1995}. 

The second FOM was an in-house three-dimensional unstructured solver which solved the Euler equations written in conservative, integral form using the finite volume method with Roe-Riemann flux splitting and Harten entropy fix~\cite{Han2003}. The gradients were computed using a weighted least squares with QR decomposition. The discretization is vertex-centered with a dual mesh. The FOM is second-order accurate in both space and time. 

\subsubsection{Reduced-Order Model}
\label{sec:rom}

The snapshots generated by solving the FOM were used to assemble the {\bf R} autocorrelation matrix~\eqref{autocorr}.  The POD basis functions $\phi$ were obtained as the eigenmodes of {\bf R}.  In the governing equations~\eqref{eq:vecform} the state variables were approximated using \eqref{PODapprox} which yielded
%
\begin{equation} \label{eq:sub}
\nsum{i}{0}{Z_{k}} \dot{a}_{i}^{Z_{k}} \phin{i}{Z_{k}}(\bx) = \mathcal{J}_{k} \lp{(} \ba,\pmb{\phi} \rp{)}, \quad k\in[1,D+2]
\end{equation}
where $D$ is the dimension of the domain and $\mathcal{J}$ is a nonlinear functional consisting of the combinations of the time coefficients and basis functions of each approximated state variable. The approximated governing equations,~\eqref{zeta_stat} or~\eqref{zeta_rot}, were projected along the POD basis functions, $\phin{j}{Z_{k}}, j\in[1,n^{Z_{k}}], k\in[1,D+2]$. Since the governing equations are nonlinear, the reduced system of equations is nonlinear as well.  Projecting~\eqref{eq:sub} along the POD basis functions contracts the time-derivative term
\begin{equation*}
\dot{a}_{j}^{Z_{k}} = \lp{(} \mathcal{J}_{k} \lp{(} \ba,\pmb{\phi} \rp{)}, \phin{j}{Z_{k}}(\bx) \rp{)},  \quad j\in[1,n^{Z_{k}}], \ k\in[1,D+2]
\end{equation*}
such that the system of $N$ PDEs is reduced to a system of $m=\sum_{k=1}^{D+2} n^{Z_{k}}$ ODEs where the time coefficients are the dependent variables.  The coefficients of the ODEs that appear in this equation are only dependent on space, and so can be precomputed before solving the system of ODEs. The system of ODEs can be concisely written as
\begin{equation}
\label{ODEeq}
\dot{\ba} = \pmb{\mathcal{F}} (\ba)	
\end{equation}
where $\ba \in \R^{m}$ and $\pmb{\mathcal{F}} \in \R^{m}$. The inner product selected for this operation is the $L^{2}$ inner product, which when used with the velocity components yields the rate of kinetic energy
\begin{equation*}
\ip{\dot{\bv}}{\bv} = \frac{1}{2} \dd{}{t} \norm{\bv}^{2} .
\end{equation*}
This is desirable, because the $L^{2}$ inner product reflects the physics of the flow~\cite{Barone2009}. The discrete $L^{2}$ inner product was used to project the governing equations
\begin{equation*}
\ip{f}{g} = \int_{\Omega} fg d\Omega \approx \sum_{i=1}^{N} f(x_{i}) g(x_{i}) \Omega_{i} = \sum_{i=1}^{N} f(x_{i}) g(x_{i})	
\end{equation*}
where $\Omega \in \R^{D}$ is a reference spatial domain with unit length, width, and depth. 

The governing system of ODEs of the ROM~\eqref{ODEeq} was solved using ODEPACK~\cite{ODEPACK}. The relative and absolute tolerance parameters in ODEPACK were set to 0.1 and 0.001, respectively. The DLSODE solver was selected. This solver has both nonstiff and stiff solvers imbedded into it. The nonstiff solver uses Adams' method while the stiff solver uses backward differentiation formulas. The Jacobian was generated internally. The initial conditions were taken from time coefficients derived from a FOM solution~\eqref{tceq}. A least-squares with QR decomposition method was used to evaluate the spatial derivatives in the ROM~\cite[p. 165-168]{Blazek2005}.

With the method presented here, the ROM was not developed by projecting the discretized governing equations of the FOM as was done in~\cite{yuanetal05, Brenner2012}, but was developed by projecting the differential form of the governing equations themselves. This may lead to small differences between the FOM and the ROM~\cite{Burkardt2006}. Furthermore, the ROM developed herein is not guaranteed to preserve the stability of the FOM. The following section presents the approaches used to ensure the stability of the ROM.

\subsection{Stability}

POD-based reduced-order models do not necessarily retain the stability of the FOM. This is due in part to the projection of the FOM onto the POD subspace. Although the FOM may be unconditionally stable, the POD subspace may end up seeing unstable trajectories~\cite{Rempfer2000}. Therefore, the following methods are considered for obtaining a stable ROM: (1) the penalty method, (2) artificial dissipation, and (3) a method that modifies the number of modes of the POD approximation within the governing equations.

\subsubsection{Penalty method}

The boundary conditions imposed on the FOM constrain the basis functions such that, for the on-reference cases, the ROM will satisfy the boundary conditions.  For the off-reference cases, however, the ROM might not satisfy the boundary conditions because the basis functions might not be able to reconstruct the off-reference boundary conditions. If the basis functions do not satisfy the boundary conditions of the FOM, the boundary conditions may end up becoming compatibility constraints on the ROM~\cite{Rempfer2000}. These constraints limit the combination of basis functions and time coefficients that can be used for the reconstruction~\eqref{PODapprox}.

To correct the compatibility constraints, boundary conditions are
implemented into the ROM using the penalty me\-thod~\cite{Hesthaven1996,
  Funaro1991}. The penalty method imposes a prescribed boundary condition
onto the ODEs of the ROM~\eqref{ODEeq}. Let $N_{\textrm{BC}}$ be the number
of nodes at the selected boundary.  The difference between the ROM solution
$\bZ_{\text{BC}}\in \R^{N_{\textrm{BC}} \times (D+2)}$, a subset of $\bZ$
that includes only the $N_{\text{BC}}$ nodes of the selected boundary, and
the prescribed boundary condition, $\textbf{F}(t) \in \R^{N_{\textrm{BC}}
  \times (D+2)}$, is a measure of the boundary condition error.  Let
${\phi_{\text{BC}}}_i \in \R^{N_{\text{BC}}}$, $1 \le i \le m$, be a subset
of the POD basis function $\phi_i$, which includes only the $N_{\text{BC}}$
nodes of the boundary.  $\phi_{\text{BC}_i}$ is the $i^{\text{th}}$
component of $\phi_{\text{BC}} \in \R^{N_\text{BC} \times m}$. Let $\bK_i$
define the boundary error $\bZ_\text{BC}-\bF$ projected on
${\phi_\text{BC}}_i$, $\bK_i = \left(\bZ_\text{BC}-\bF, \phi_{\text{BC}_i}
\right)$, $1 \le i \le m$.  Let $\bK \in \R^m$ be the collection of all
$\bK_i$ and let $\pmb{\uptau} \in \R^{m \times m}$ be a diagonal matrix
consisting of the penalty parameters for each ODE of \eqref{ODEeq}.  The
penalty method adds to \eqref{ODEeq} the projection of the boundary error
$\bK$ scaled by the penalty parameter $\pmb{\uptau}$
\begin{equation}
\label{ODEpenalty}
\dot{\ba} = \pmb{\mathcal{F}}(\ba) - \pmb{\uptau} \bK .
\end{equation}
A component of the vectorial equation~\eqref{ODEpenalty} is
\begin{equation*}
\begin{gathered}
	\dot{a}_{i}^{Z_{k}} = \mathcal{F}_{i}^{Z_{k}} (\ba) - \uptau_{ii}^{Z_{k}} \ip{\bZ_{\text{BC}}^{Z_{k}} - \textbf{F}^{Z_{k}}(t)}{\phi_{i}^{Z_{k}}(\bx_{\text{BC}})}, \quad i \in [1, n^{Z_{k}}], \ k\in[1,D+2]
\end{gathered}
\end{equation*}
where the superscript $Z_k$ indicates the $k^\text{th}$ component of the state variable and
\begin{equation*}
\begin{gathered}
	\uptau_{ii}^{Z_{k}} = \uptau_{\mathcal{K} \mathcal{K}}, \quad \mathcal{K} = \sum_{j=1}^{k-1} n^{Z_{j}} + i .
\end{gathered}
\end{equation*}
The prescribed boundary condition $\textbf{F}(t) \in
\R^{(D+2)N_{\text{BC}}}$ consists of the components $\textbf{F}(t) = \lp{(}
\textbf{F}^{\zeta}, \textbf{F}^{u}, \textbf{F}^{v}, \textbf{F}^{w},
\textbf{F}^{p} \rp{)}^{T}$ where $\textbf{F}^{Z_{k}} \in \R^{N_{\text{BC}}},
k\in[1,D+2]$.

The value of the penalty parameter $\uptau_{ii}^{Z_{k}}$ is typically set to some large number. Rather than randomly guessing its value, two methods are explored here for the computation of $\uptau_{ii}^{Z_{k}}$: an energy-based stabilizing method and a method that minimizes the error $\bK$ to zero. 

The energy-based stabilizing method~\cite{Hesthaven1996, Kalashnikova2012} ensures that the system energy is decreasing. Assuming $\pmb{\uptau}=\tau\textbf{I}$ in~\eqref{ODEpenalty} yields
\begin{equation}
	\label{ODEtauenergysolve}
 	\dot{\ba} = \pmb{\mathcal{F}}(\ba) - \tau \textbf{I} \bK .
\end{equation}
One can then project~\eqref{ODEtauenergysolve} along $\ba$ to get
\begin{equation*}
\ip{\dot{\ba}}{\ba} = \ip{\pmb{\mathcal{F}}(\ba)}{\ba} - \tau \ip{\bK}{\ba} .	
\end{equation*}
Note that $\ip{\dot{\ba}}{\ba} = \frac{1}{2} \dd{}{t} \norm{\ba}^{2}$, which represents the time rate of system energy. Bounds for the penalty parameter are found by ensuring that system energy is decreasing, $\frac{1}{2} \dd{}{t} \norm{\ba}^{2} \leq 0$, such that
\begin{equation*}
\tau \geq \frac{\ip{\pmb{\mathcal{F}}(\ba)}{\ba}}{\ip{\bK}{\ba}} .	
\end{equation*}

Alternately, one can calculate the penalty parameter such that the error at
the next time step $\bK$ is reduced to zero. Let us assume $\pmb{\uptau} =
\tauc$, where $\tauc \in \R^{m \times m}$ is a diagonal matrix with
$\taucc_{i_k}=\taucc_k, \ i_k\in[1,n^{Z_{k}}], \ k \in [1,D+2]$, that is,
the value of the penalty parameters are held constant for all modes $i_k$ in
their respective governing equation, $k$. The value of $\tauc$ that sets
$\bK$ to zero is found by using an iterative procedure based on Newton's
method such that $\tauc$ is the root of $\bK(\tauc)$
\begin{equation} \label{eq:root}
\taucc_{Z_{k}}^{n+1} = \taucc_{Z_{k}}^{n} - 
\frac{K(\taucc_{Z_{k}}^{n})}{dK(\taucc_{Z_{k}}^{n}) / d\taucc}, \quad k \in [1,D+2] .
\end{equation}
If $dK(\taucc_{Z_{k}}^{n}) / d\taucc$ is approximated using a first-order difference, then \eqref{eq:root} becomes
\begin{equation}
\label{tausecant}
\taucc_{Z_{k}}^{n+1} = \taucc_{Z_{k}}^{n} - \frac{K(\taucc_{Z_{k}}^{n})}{K(\taucc_{Z_{k}}^{n}) - K(\taucc_{Z_{k}}^{n-1})} \lp{(} \taucc_{Z_{k}}^{n} - \taucc^{n-1} \rp{)}, \quad k \in [1,D+2] .
\end{equation}
Numerical experiments showed that $K(\taucc_{Z_{k}}^{n})$ varied linearly with $\taucc_{Z_{k}}^{n}$, which led to a fast convergence of the method. 
This method for determining the penalty parameter $\pmb{\uptau}$ by finding the roots of $\bK$ was used herein instead of the energy-decreasing method. 

\subsubsection{Artificial Dissipation}
Artificial dissipation has been used previously to stabilize POD-based redu\-ced-order models by dissipating spurious modes~\cite{Lucia2003, Sirisup2004}. The artificial dissipation terms are projected along their respective subspace and added to the system of ODEs of the ROM such that~\eqref{ODEeq} becomes
\begin{equation}
\label{artdiss}
\dot{\ba} = \pmb{\mathcal{F}}(\ba) + \tilde{\pmb{\nu}} \textbf{S}
\end{equation}
where $\tilde{\pmb{\nu}} \in \R^{m \times m}$ is the diagonal matrix of the artificial dissipation parameters, and $\textbf{S} \in \R^{m}$ consists of the projection of the Laplacian of the basis functions along their respective subspace.  It should be noted that artificial dissipation is not applied to the zeroth (or mean) mode.  For a three-dimensional problem, \eqref{artdiss} can be written in component form as
\begin{equation*}
\begin{gathered}
\dot{a}_{i}^{Z_{k}} = \mathcal{F}_{i}^{Z_{k}} + \tilde{\nu}_{ii}^{Z_{k}} \nsum{j}{1}{Z_{k}} \ip{\phin{j,xx}{Z_{k}} + \phin{j,yy}{Z_{k}} + \phin{j,zz}{Z_{k}}}{\phin{i}{Z_{k}}} \tc{j}{Z_{k}}, \\
\ 1 \le i \le n^{Z_k}, \ 1\le k\le 5
\end{gathered}
\end{equation*}
where the artificial dissipation parameter, $\tilde{\nu}_{ii}$, must be greater than or equal to zero. 

The value of the artificial dissipation parameter $\tilde{\nu}_{ii}$ was found using an energy-based stability analysis. Therefore, artificial dissipation was added to~\eqref{ODEpenalty} to yield
\begin{equation}
\label{artdisspenalty}
\dot{\ba} = \pmb{\mathcal{F}}(\ba) + \tilde{\pmb{\nu}} \textbf{S} - \pmb{\uptau} \bK	 .
\end{equation}
Assuming $\tilde{\pmb{\nu}} = \tilde{\nu} \textbf{I}$ and projecting along $\ba$ gives
\begin{equation*}
\ip{\dot{\ba}}{\ba} = \ip{\pmb{\mathcal{F}}(\ba)}{\ba} + \tilde{\nu} \ip{\textbf{S}}{\ba} - \ip{\pmb{\uptau} \bK}{\ba}
\end{equation*}
where $\ip{\dot{\ba}}{\ba} = \frac{1}{2} \dd{}{t} \norm{\ba}^{2}$ is the system energy, which should be decreasing for stability, \textit{i.e.}, $\frac{1}{2} \dd{}{t} \norm{\ba}^{2} \leq 0$.  Consequently, the dissipation must satisfy the inequalities
\begin{equation}
	\label{artdissparam}
	0 \leq \tilde{\nu} \leq \frac{\ip{\pmb{\uptau} \bK - \pmb{\mathcal{F}}(\ba)}{\ba}}{\ip{\textbf{S}}{\ba}} .
\end{equation}

\subsubsection{Modified Number of Modes}
As mentioned in section~\ref{sec:rom}, the size of the system of ODEs~\eqref{ODEeq} is $m=\sum_{k=1}^{D+2} n^{Z_k}$.  Consequently, the value of $m$ depends on the number of modes $n^\zeta$ used to approximate the $\zeta$ variable in the mass balance equation, the number of modes $n^u$ used to approximate velocity $u$ in the $x$-component of the momentum conservation equation, and so on for the $v$, $w$ and $p$ variables.  If the number of modes used to approximate variable $u$ in the mass balance equation is less than $n^u$, the size of the system of ODEs is not changed.  The Jacobian of the reduced-order model, however, is changed.  Consequently, the eigenvalues of the Jacobian are changed, and this change affects the stability of the reduced-order model.


For example, consider the projection of the conservation of mass equation for one-dimensional flow
\begin{equation} \label{eq:ode}
\dot{a}_{k}^{\zeta} + \nsum{i}{0}{\zeta} \nsum{j}{0}{u} \ip{\phin{i,x}{\zeta} \phin{j}{u} - \phin{i}{\zeta} \phin{j,x}{u}}{\phin{k}{\zeta}} \tc{i}{\zeta} \tc{j}{u} = 0, \ k \in [1,n^{\zeta}]
\end{equation}
Assuming $n^{\zeta} = 1$ and $n^{u}=2$ in \eqref{eq:ode} yields
\begin{equation} \label{eq:ode1}
\begin{aligned}
\dot{a}_{k}^{\zeta} &+ \ip{\phin{0,x}{\zeta} \phin{0}{u} - \phin{0}{\zeta} \phin{0,x}{u}}{\phin{k}{\zeta}} + \ip{\phin{0,x}{\zeta} \phin{1}{u} - \phin{0}{\zeta} \phin{1,x}{u}}{\phin{k}{\zeta}} \tc{1}{u} + \\
& +\ip{\phin{0,x}{\zeta} \phin{2}{u} - \phin{0}{\zeta} \phin{2,x}{u}}{\phin{k}{\zeta}} \tc{2}{u} 
 + \ip{\phin{1,x}{\zeta} \phin{0}{u} - \phin{1}{\zeta} \phin{0,x}{u}}{\phin{k}{\zeta}} \tc{1}{\zeta} + \\
& + \ip{\phin{1,x}{\zeta} \phin{1}{u} - \phin{1}{\zeta} \phin{1,x}{u}}{\phin{k}{\zeta}} \tc{1}{\zeta} \tc{1}{u} 
+ \ip{\phin{1,x}{\zeta} \phin{2}{u} - \phin{1}{\zeta} \phin{2,x}{u}}{\phin{k}{\zeta}} \tc{1}{\zeta} \tc{2}{u} = 0 .
\end{aligned}
\end{equation}
If $n^{u}$ is set equal to one within this equation only, then \eqref{eq:ode1} reduces to
\begin{equation*}
\begin{aligned}
\dot{a}_{k}^{\zeta} &+ \ip{\phin{0,x}{\zeta} \phin{0}{u} - \phin{0}{\zeta} \phin{0,x}{u}}{\phin{k}{\zeta}} + \ip{\phin{0,x}{\zeta} \phin{1}{u} - \phin{0}{\zeta} \phin{1,x}{u}}{\phin{k}{\zeta}} \tc{1}{u} +
\\
&+ \ip{\phin{1,x}{\zeta} \phin{0}{u} - \phin{1}{\zeta} \phin{0,x}{u}}{\phin{k}{\zeta}} \tc{1}{\zeta} + \ip{\phin{1,x}{\zeta} \phin{1}{u} - \phin{1}{\zeta} \phin{1,x}{u}}{\phin{k}{\zeta}} \tc{1}{\zeta} \tc{1}{u} = 0 .
\end{aligned}
\end{equation*}
 %
Note that $n^{u}$ cannot be changed within the conservation of \textit{x}-momentum equation, and $n^{u}$ can only be reduced within the other equations.  

The effect of modifying the number of modes is apparent when analyzing the Jacobian of the reduced-order model
\begin{equation*}
J_{ij}^{Z_k} = \dd{\mathcal{F}_{i}^{Z_{k}}(\ba)}{a_{j}}, \quad i,j \in [1,n^{Z_{k}}], \ k\in[1,D+2]	.
\end{equation*}
The part of the Jacobian that accounts for the effect of \textit{x}-velocity $u$ on the mass balance equation is
\begin{equation} \label{eq:jac}
\dd{\pmb{\mathcal{F}}^{\zeta}}{\ba^{u}} = 
\begin{bmatrix}	
 	\dd{\mathcal{F}_{1}^{\zeta}}{\tc{1}{u}} & \dd{\mathcal{F}_{1}^{\zeta}}{\tc{2}{u}} & ... & \dd{\mathcal{F}_{1}^{\zeta}}{\tc{n^{u}}{u}} \\
 	\dd{\mathcal{F}_{2}^{\zeta}}{\tc{1}{u}} & \dd{\mathcal{F}_{2}^{\zeta}}{\tc{2}{u}} & ... & \dd{\mathcal{F}_{2}^{\zeta}}{\tc{n^{u}}{u}} \\
 	\vdots & \vdots & \ddots & \vdots \\
 	\dd{\mathcal{F}_{n^{\zeta}}^{\zeta}}{\tc{1}{u}} & \dd{\mathcal{F}_{n^{\zeta}}^{\zeta}}{\tc{2}{u}} & ... & \dd{\mathcal{F}_{n^{\zeta}}^{\zeta}}{\tc{n^{u}}{u}} \\
\end{bmatrix}
\end{equation}
If the velocity $u$ in the mass balance equation is approximated by $n^{u}-1$ terms instead of $n^u$, then all the terms of the last column of \eqref{eq:jac} are zero. Consequently, the eigenvalues of the Jacobian are modified.  The number of modes is modified in order to remove the spurious modes that have a destabilizing effect on the solution of the system of ODEs.


\subsection{Basis Function Interpolation}
The main benefit of a ROM consists in its ability to efficiently predict
flow conditions beyond what is predicted by the FOM, that is, at
off-reference conditions. To predict off-reference conditions, the basis
functions, which are functions of space, need to be interpolated to the new
flow conditions. This can be done in two ways: by enriching the snapshot
database with solutions of different flow conditions, referred to as global
POD, or by interpolating using the tangent space to a Grassmann
manifold~\cite{Amsallem2008}.  Both methods are used in this paper.

\section{On-Reference Results}
This section presents on-reference results for the zeta-variable ROM for four cases: a quasi-one-dimensional nozzle flow, a two-dimensional channel flow, a three-dimensional axisymmetric nozzle flow, and a transonic fan flow referred to as NASA Rotor 67. These four cases were used to validate the zeta-variable ROM by showing that the ROM could reproduce the FOM solution.

\subsection{Quasi-One-Dimensional Nozzle Flow}
\label{sec:1d}

The quasi-one-dimensional nozzle case was selected as an initial validation case for the zeta-variable ROM. The diverging nozzle has an area variation $A(x)$~\cite[p. 327]{Anderson1995}
\begin{equation}
\label{1Darea}
A(x) = \begin{cases}
1 + 2.2(x-0.5)^{2} & 0.0 \leq x \leq 0.5 \\
1 + 0.2223(x-0.5)^{2} & 0.5 \leq x \leq 1.0 	
\end{cases} .
\end{equation}
The FOM~\eqref{characteristic} was used to generate the snapshots for the ROM. It is important to note that while the governing equations of the FOM used the characteristic variables, the ROM used the zeta-variables~\eqref{zeta_stat}. A grid convergence study was performed for the steady flow through the nozzle by using 51, 101, and 201 nodes.  Table~\ref{t:1Dgridconvergence} shows the results of the study where the percentages are the percent change of the steady state Mach number between grid refinements.
%
\begin{table}[ht!]
\centering
\begin{tabular}{ccc}
	\toprule
	Grid & Steady-State  & Change from  \\
	Points & Outlet Mach Number & Previous (\%) \\
	\midrule
	51 & 0.3598 & - \\
	101 & 0.3603 & 0.16 \\
	201 & 0.3605 & 0.04 \\
	\bottomrule
\end{tabular}
\caption{Grid convergence of the quasi-one-dimensional nozzle.}
\label{t:1Dgridconvergence}
\end{table}
\psfull
The steady state Mach number approached an asymptotic value as the grid was refined. Due to the small change between the three grids, results are shown for the grid using 51 nodes.

To create an unsteady flow, an oscillating outlet pressure was imposed
\begin{equation}
\label{pback}
p(t,x_{\text{out}}) = f(t) = \bar{p} \lp{(} 1 + \amp \sin \lp{(} \freq t + \varphi \rp{)} \rp{)}	
\end{equation}
where the subscript ``out" denotes values at the outlet boundary, $\bar{p}$ is the mean pressure at the outlet, $\amp$ is the amplitude of the oscillation, $\freq$ is the angular velocity of the oscillation, and $\varphi$ is the phase shift.  Three cases for validating the ROM were created by varying the amplitude $\amp$ and angular velocity $\freq$ of the outlet pressure oscillation. Table~\ref{t:1dcases_on} summarizes these cases.
\begin{table}[ht!]
\centering
\begin{tabular}{ccc}
\toprule
Case & $\freq$ [rad/sec] & $\amp$ \\
\midrule
1 & 1 & 0.02 \\
2 & 1 & 0.03 \\
3 & 2 & 0.02 \\
\bottomrule
\end{tabular}
\caption{Quasi-one-dimensional nozzle on-reference cases.}
\label{t:1dcases_on}
\end{table}
For all cases $\bar{p}=0.95$ and $\varphi = 0.141 \pi$.  The FOM simulated the flow for 8 periods, and a total of 2000 snapshots were saved.

A penalty-enforced boundary condition was imposed onto the ROM. The prescribed boundary condition was applied at the outlet $F(t) = \lp{(} 0, 0, f(t) \rp{)}^{T}$ where $f(t)$ is the oscillating outlet pressure boundary condition imposed on the FOM. The penalty parameter, $\pmb{\uptau}$, was calculated by finding the roots of the error $\bK$~\eqref{tausecant}. 

\begin{figure}[ht!]
\centering
\subfloat[ROM without Penalty-Enforced BC]{\includegraphics[width=0.48\textwidth]{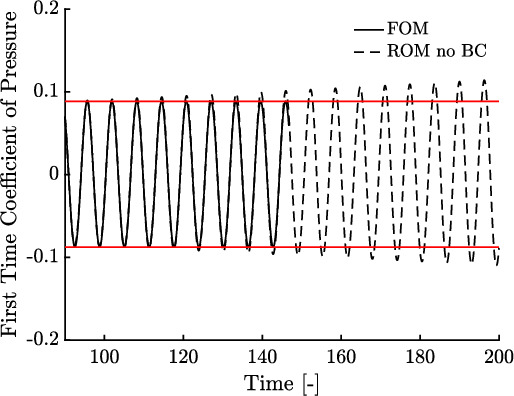}}
\hspace{0.03\textwidth}
\subfloat[ROM with Penalty-Enforced BC]{\includegraphics[width=0.48\textwidth]{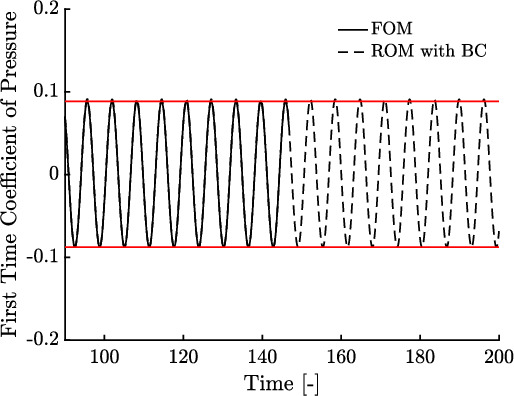}}
\caption[Effect of penalty-enforced boundary condition on ROM for a quasi-one-dimensional nozzle.]{Effect of penalty-enforced boundary condition on ROM solution for a quasi-one-dimensional nozzle case with outlet pressure oscillation with angular velocity $\freq=1$ rad/sec and amplitude $\amp=0.02$. The red lines correspond to the bounds of the time coefficients derived from the FOM solution.}
\label{fig:1Dpenalty}
\end{figure}

For brevity, only case 1 results are shown here. The other two cases showed similar results. All cases used the following number of modes $n^{\zeta}=n^{u}=n^{p}=2$. Figure~\ref{fig:1Dpenalty} shows the stabilizing effect of the penalty-enforced boundary condition on the solution of the first time coefficient of pressure. Both $\zeta$ and $u$ displayed similar results.

The time coefficients were calculated in three ways: (1) the line
corresponding to ``FOM" solved~\eqref{tceq} for the ``exact" time
coefficients, (2) the line corresponding to ``ROM with no BC"
solved~\eqref{ODEeq} for the time coefficients, and (3) the line
corresponding to ``ROM with BC" solved~\eqref{ODEpenalty} for the time
coefficients. Both ROM solutions were extrapolated past the sampling time of
the FOM. The ROM solution without the penalty-enforced boundary condition
grew over time, resulting in an unstable solution. The ROM solution with the
penalty-enforced boundary condition remained within the bounds of the time
coefficients derived from the FOM solution, resulting in a stable
solution. The penalty method, therefore, worked in stabilizing the ROM by
imposing the prescribed boundary condition.  In addition, there was also
good agreement between the ROM solution and the FOM solution once the
penalty method was imposed.

\begin{figure}[ht!]
\centering
\includegraphics[width=0.6\textwidth]{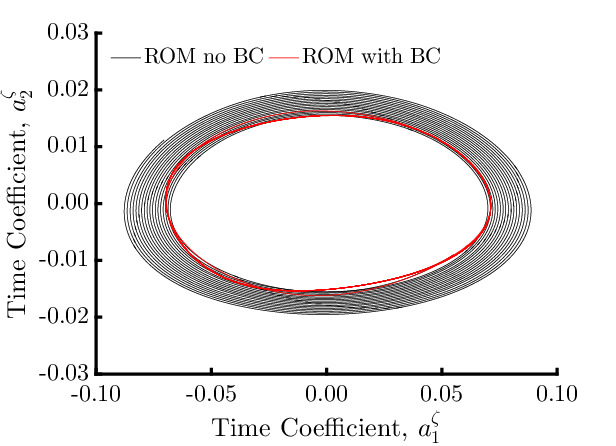}
\caption[Phase portrait of time coefficients for a quasi-one-dimensional nozzle.]{Phase portrait of the time coefficients of specific volume for a quasi-one-dimensional nozzle case with outlet pressure oscillation with angular velocity $\freq=1$ rad/sec and amplitude $\amp=0.02$.}
\label{fig:1Dphase}
\end{figure}

The phase portrait in Figure~\ref{fig:1Dphase} shows the effect of imposing the boundary conditions onto the ROM for $\zeta$. 
The ROM solution without the penalty-enforced boundary condition displays an unstable outward spiral while the ROM with the penalty-enforced boundary condition displays a stable Limit Cycle Oscillation (LCO) indicative of its oscillating outlet pressure boundary condition.

\subsection{Two-Dimensional Channel Flow}
\label{sec:2dc}

\begin{figure}[ht!]
\centering
\includegraphics[width=0.7\textwidth, trim=0.5cm 7cm 0.5cm 4cm, clip]{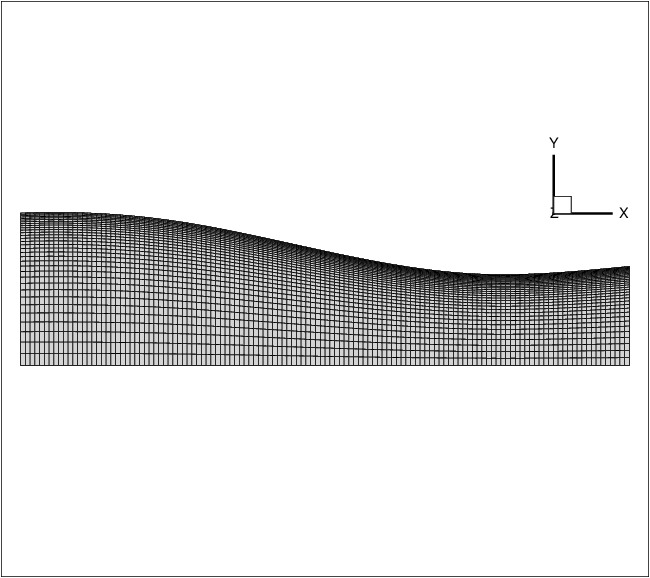}
\caption{Computational mesh of the two-dimensional channel with 4257 nodes.}
\label{fig:2dmesh}
\end{figure}

A two-dimensional channel flow was selected as the second validation case of the ROM. The area variation for the channel is given by~\eqref{1Darea} where the channel has a unit span.  The inlet stagnation pressure and temperature are 101.325~kPa and 288.15~K, respectively.  The flow enters the inlet along the $x$-axis. The mean pressure at the outlet is 98.538~kPa. The inlet was extended by a quarter of the length of the domain such that the flow enters the nozzle at a zero degree angle. A cubic spline was used to smooth out the top boundary.  Figure~\ref{fig:2dmesh} shows the computational mesh of the channel.

A grid convergence study was done assuming steady flow.
Table~\ref{t:2dgridconvergence} shows the variation of the average Mach
number at the outlet for three grids: coarse, medium and fine with 4257,
16,705, and 66,177 grid nodes,
respectively. 
\begin{table}[ht!]
\centering
\begin{tabular}{ccc}
	\toprule
	Grid & Steady-State  & Change from  \\
	~ & Outlet Mach Number & Previous (\%) \\
	\midrule
	Coarse & 0.1993 & - \\
	Medium & 0.1996 & 0.14 \\
	Fine & 0.1998 & 0.07 \\
	\bottomrule
\end{tabular}
\caption{Grid convergence of the two-dimensional channel.}
\label{t:2dgridconvergence}
\end{table}
Table~\ref{t:2dgridconvergence} also shows the percent change between grid refinements. As the grid was refined, the variation of the average Mach number decreased. Since the solution variation with grid size was small, the coarse grid results are discussed in the following paragraphs.

An oscillating outlet pressure boundary condition~\eqref{pback} was specified to generate an unsteady flow.  Three cases were created by varying the amplitude and angular velocity of the outlet pressure oscillation. Table~\ref{t:2dcases_on} summarizes these cases.
\begin{table}[ht!]
\centering
\begin{tabular}{ccc}
\toprule
Case & $\freq$ [rad/sec] & $\amp$ \\
\midrule
1 & 10 & 0.01 \\
2 & 10 & 0.02 \\
3 & 20 & 0.01 \\
\bottomrule	
\end{tabular}
\caption{On-reference cases for the two-dimensional channel.}
\label{t:2dcases_on}
\end{table}

For all cases, the pressure was $\bar{p}=98,583$ Pa and the phase shift was $\varphi=0$. The full-order model generated snapshots for approximately 8 periods, each period consisting of 300 time steps. A total of 2375 snapshots were used to generate the autocorrelation matrix $\textbf{R}$.

The penalty method was imposed onto the ROM with a prescribed boundary condition applied at the outlet
\begin{equation}
\label{penalty_zp}
\textbf{F}(t) = \lp{(} \lp{(} \gamma \textbf{f}(t) \rp{)}^{\frac{1}{\gamma}}, 0, 0, \textbf{f}(t) \rp{)}^{T}
\end{equation}
where $\textbf{f}(t) \in \R^{N_{\text{out}}}$ and $N_{\text{out}}$ is the number of nodes at the outlet boundary.  In addition, $\textbf{f}_i(t) = f(t)$ of~\eqref{pback}, $1\le i\le N_{\text{out}}$. The prescribed boundary condition for $\zeta$ was found through an isentropic relationship~\cite{Krath2018}. The penalty parameter $\pmb{\uptau}$ was solved for using~\eqref{tausecant}. 

\begin{figure}[ht!]
\centering
\includegraphics[width=0.7\textwidth]{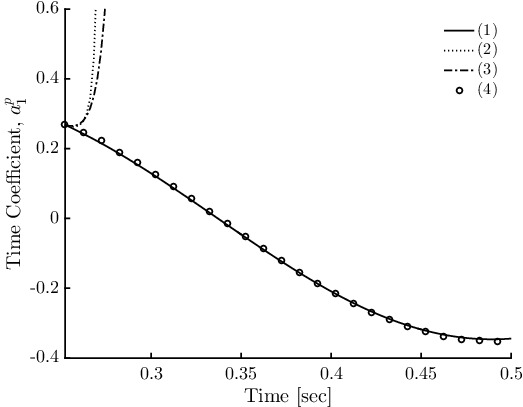}
\caption{Stabilizing effect of the penalty-enforced boundary conditions and modified number of modes on the ROM solution for a two-dimensional channel case with oscillating outlet pressure of angular velocity $\freq = 10$ rad/sec and amplitude $\amp=0.01$: (1) time coefficients derived from the FOM solution, (2) ROM solution, (3) ROM solution with penalty-enforced boundary conditions, (4) ROM solution with penalty-enforced boundary conditions and a modified number of modes.}
\label{fig:2Dstability}
\end{figure}

The ROM solution used the following number of modes, $n^{\zeta}=1, n^{u}=n^{v}=n^{p}=2$, with a modified number of modes of $n^{v}=1$ in the energy conservation equation. Modifying $n^{v}$ in the energy conservation equation altered the eigenvalues of the Jacobian of the ROM system such that the ROM was stabilized. Figure~\ref{fig:2Dstability} shows the effect of the modified number of modes on the ROM solution for case 1.

Once both the penalty method was imposed and the number of modes was modified, the ROM solution agreed well with the ``exact'' time coefficients~\eqref{tceq} derived from the FOM without solving the system of ODEs.

For brevity, results are only shown for case 1. The other cases had similar results. The relative error of the ROM solution to the FOM solution
\begin{equation}
 \epsilon = \frac{\abs{Z_{\text{FOM}} - Z_{\text{ROM}}}}{\abs{Z_{\text{FOM}}}}
 \end{equation}
was calculated at each node point to assess the accuracy of the ROM. To avoid division by zero, the relative  error for all velocity components was calculated as
\begin{equation}
\epsilon_{v} = \frac{\abs{Z_{\text{FOM}} - Z_{\text{ROM}}}}{\abs{Z_{\text{FOM}}}+1}
\end{equation}
because the velocity components could have values close to zero. 

\begin{figure}[ht!]
\centering
\subfloat[Specific Volume]{\includegraphics[width=0.48\textwidth,trim=0cm 3cm 0cm 6cm,clip]{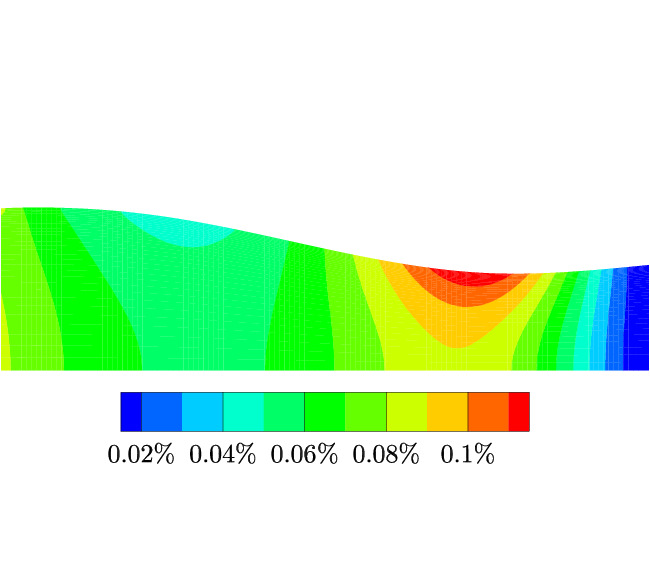}}
\hspace{0.02\textwidth}
\subfloat[\textit{x}-Velocity]{\includegraphics[width=0.48\textwidth,trim=0cm 3cm 0cm 6cm,clip]{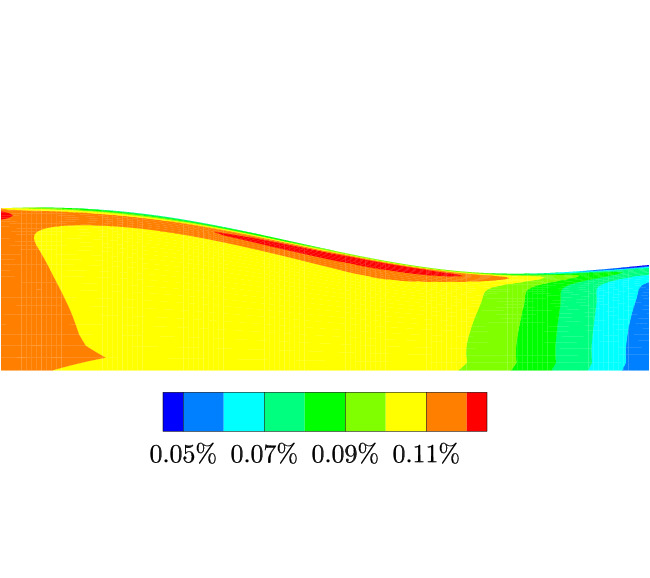}} 
\\
\subfloat[\textit{y}-Velocity]{\includegraphics[width=0.48\textwidth,trim=0cm 3cm 0cm 6cm,clip]{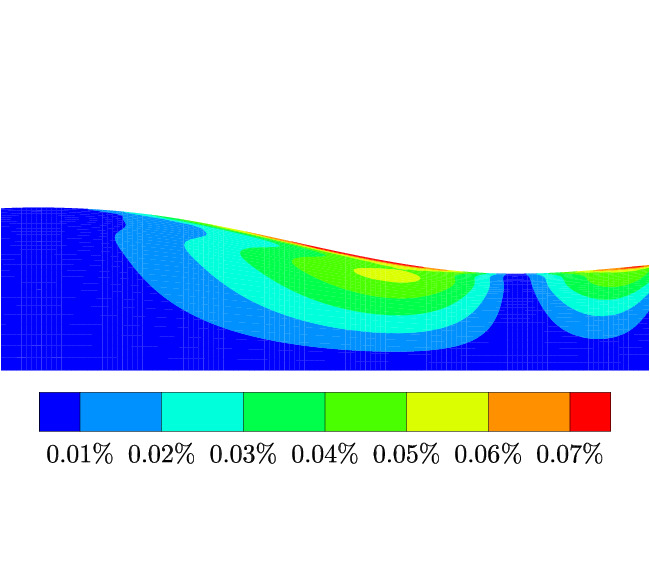}}
\hspace{0.02\textwidth}
\subfloat[Pressure]{\includegraphics[width=0.48\textwidth,trim=0cm 3cm 0cm 6cm,clip]{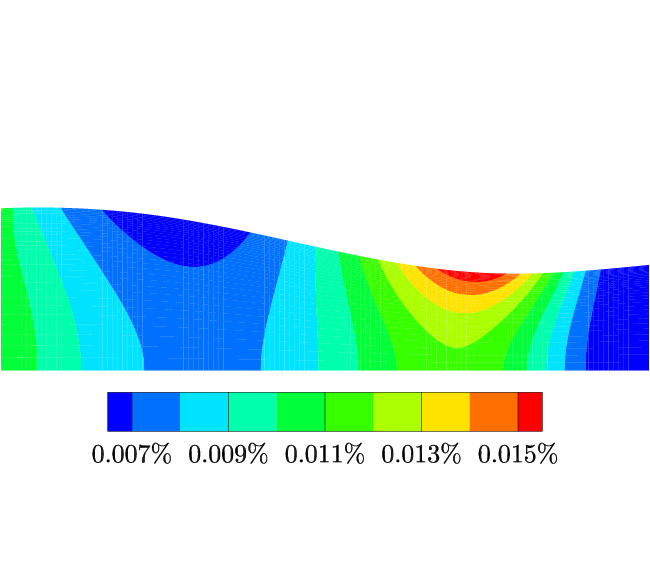}} 
\caption{Time-averaged relative  error between the ROM solution and the FOM solution for a two-dimensional channel flow with outlet pressure oscillation of angular velocity $\freq = 10$ rad/sec and amplitude $\amp = 0.01$.}
\label{fig:2dres_case1}
\end{figure}

\begin{figure}[ht!]
\centering
\includegraphics[width=0.8\textwidth]{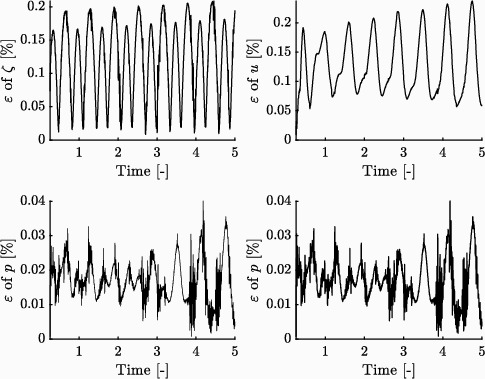}
\caption{Two-dimensional channel flow with an outlet pressure oscillation with angular velocity $\freq = 10$~rad/sec and amplitude $\amp = 0.01$: time variation of maximum relative error between the ROM and FOM solutions across all nodes.}
\label{fig:2dmax_case1}
\end{figure}

Figure~\ref{fig:2dres_case1} shows the time-averaged relative  error of the ROM solution with respect to the FOM solution for case 1. 
The ROM solution agreed well with the FOM solution as the error was less than 0.11\% for all variables. To assess the variation of the error in time, Fig.~\ref{fig:2dmax_case1} shows the maximum relative error across all nodes versus time for case 1.
The maximum relative  error across all nodes shows an oscillatory motion, but remains small for all state variables. To summarize the results of all cases, Table~\ref{t:2dmax_allcases_on} shows the maximum spatial and temporal relative error for all three validation cases. 
\begin{table}[ht!]
\centering
\begin{tabular}{ccccc}
\toprule
& \multicolumn{4}{c}{$\max_{x,t} (\epsilon)$ (\%)} \\
Case & $\zeta$ & $u$ & $v$ & $p$ \\
\midrule
1 & 0.20 & 0.19 & 0.19 & 0.04 \\ 
2 & 0.40 & 0.29 & 0.27 & 0.09 \\ 
3 & 0.23 & 0.21 & 0.13 & 0.07 \\ 
\bottomrule
\end{tabular}
\caption{Two-dimensional channel flow: maximum spatial and temporal relative error between the ROM and FOM, for all on-reference cases.}
\label{t:2dmax_allcases_on}
\end{table}
All cases match well with the FOM solution, the maximum relative error being less than 0.5\%. 

\subsection{Three-Dimensional Axisymmetric Nozzle Flow}
\label{sec:axi}

The third validation case for the ROM is a three-dimensional axisymmetric nozzle flow. The nozzle is a converging-diverging nozzle with area variation given by \cite{Liou1987}
\begin{equation*}
A(x) = \begin{cases}
	1.75 - 0.75 \cos \lp{(} \pi \lp{(} 2x/\ell - 1 \rp{)} \rp{)} & 0 \leq x \leq \ell/2 \\
	1.25 - 0.25 \cos \lp{(} \pi \lp{(} 2x/\ell - 1 \rp{)} \rp{)} & \ell/2 \leq x \leq \ell
\end{cases} .
\end{equation*}
The length of the nozzle $\ell$ was 10 inches while the throat was located at 5 inches downstream from the inlet. The inlet stagnation pressure and temperature were 104.191~kPa and 290.435~K, respectively.  The flow entered the inlet along the $x$-axis. The mean outlet pressure was 102.107~kPa.

\begin{figure}[ht!]
\centering
\includegraphics[width=0.6\textwidth]{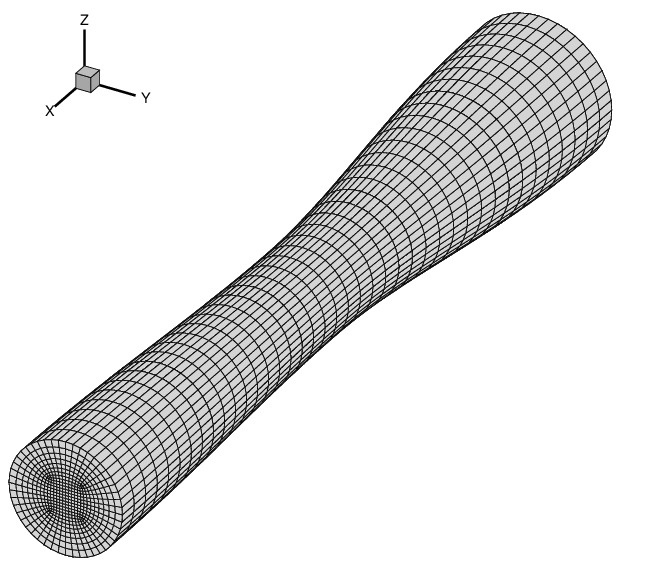}
\caption{Computational mesh of the axisymmetric nozzle with 22,673 nodes.}	
\label{fig:3dnozzlemesh}
\end{figure}

The FOM was used to simulate the steady flow in the axisymmetric nozzle for three grids: coarse, medium and fine with 3045, 22,673, and 175,041 grid points, respectively. Table~\ref{t:3dnozzlegrids} summarizes the information on the three grids.  Figure~\ref{fig:3dnozzlemesh} shows the medium grid for the axisymmetric nozzle.
\begin{table}[ht!]
\centering
\begin{tabular}{ccccc}
\toprule
\multirow{2}{*}{Type} & Nodes in & Nodes in circum- & Nodes in & Total number  
\\
& \textit{x}-direction & ferential direction & \textit{yz}-plane & of nodes $N$ 
\\
\midrule
Coarse & 21 & 24 & 145 & 3045 \\
Medium & 41 & 48 & 553 & 22,673 \\
Fine & 81 & 96 & 2161 & 175,041 \\
\bottomrule
\end{tabular}
\caption{Grids refinement of the axisymmetric nozzle.}
\label{t:3dnozzlegrids}
\end{table}

Table~\ref{t:3dnozzlegridconvergence} shows the grid convergence of the steady-state Mach number at the outlet.  As the grid was refined, the variation of Mach number decreased. The percent change between grid refinements is also shown in Table~\ref{t:3dnozzlegridconvergence}. Since the variation between the medium and fine grid was small, results for the medium grid are shown herein.
%
%
\begin{table}[ht!]
\centering
\begin{tabular}{ccc}
	\toprule
	\multirow{2}{*}{Grid} & Steady-State & Change from  \\
	 & Outlet Mach Number &  Previous (\%) \\
	\midrule
	Coarse & 0.1552 & - \\
	Medium & 0.1600 & 3.12 \\
	Fine & 0.1644 & 2.74 \\
	\bottomrule
\end{tabular}
\caption{Grid convergence of the axisymmetric nozzle.}
\label{t:3dnozzlegridconvergence}
\end{table}

The nozzle flow was made unsteady in two ways: (i) through an oscillating outlet pressure boundary condition~\eqref{pback}, and (ii) through an oscillating wall deformation. The results of both cases are given in the following sections.

\subsubsection{Oscillating Outlet Pressure}
An outlet pressure oscillation boundary condition~\eqref{pback} was imposed onto the axisymmetric nozzle.  Table~\ref{t:3dnozzlepressurecases} shows the three cases created for validating the ROM by varying the angular velocity and amplitude of the outlet pressure oscillation. 
\begin{table}[ht!]
\centering
\begin{tabular}{ccc}
\toprule 
Case & $\freq$ [rad/sec] & $\amp$ \\
\midrule
1 & 10 & 0.01 \\
2 & 10 & 0.02 \\
3 & 20 & 0.01 \\
\bottomrule
\end{tabular}
\caption{Axisymmetric nozzle on-reference cases for an outlet pressure oscillation boundary condition.}
\label{t:3dnozzlepressurecases}
\end{table}

For all cases the pressure was $\bar{p}=101,065$~Pa and the phase shift was $\varphi = 0$. The FOM generated snapshots for 3 periods, each period consisting of approximately 666 time steps. A total of 2000 snapshots were used to generate the autocorrelation matrix $\textbf{R}$. 

For all cases, three procedures were used to stabilize the ROM solution. The penalty method was imposed onto the ROM by enforcing a prescribed boundary condition at the outlet
\begin{equation*}
\textbf{F}(t) = \lp{(} \lp{(} \gamma \textbf{f}(t) \rp{)}^{\frac{1}{\gamma}}, 0, 0, 0, \textbf{f}(t) \rp{)}^{T}	
\end{equation*}
where $\textbf{f}(t) \in \R^{N_{\text{out}}}$ and   $\textbf{f}_i(t)=f(t)$ of~\eqref{pback}, $1\le i \le N_{\text{out}}$. The penalty parameter $\pmb{\uptau}$ was solved for using~\eqref{tausecant}. Artificial dissipation was used to obtain a stable solution~\eqref{artdisspenalty}. All cases used $n^{\zeta}=n^{p}=1, n^{u}=3,$ and $n^{v}=n^{w}=2$ modes with a modified number of modes of $n^{u}=2$ in the \textit{y}- and \textit{z}-momentum equations and in the conservation of energy equation.

\begin{figure}[ht!]
\centering
\subfloat[Specific Volume]{\includegraphics[width=0.47\textwidth,trim=0cm 5cm 0cm 7cm,clip]{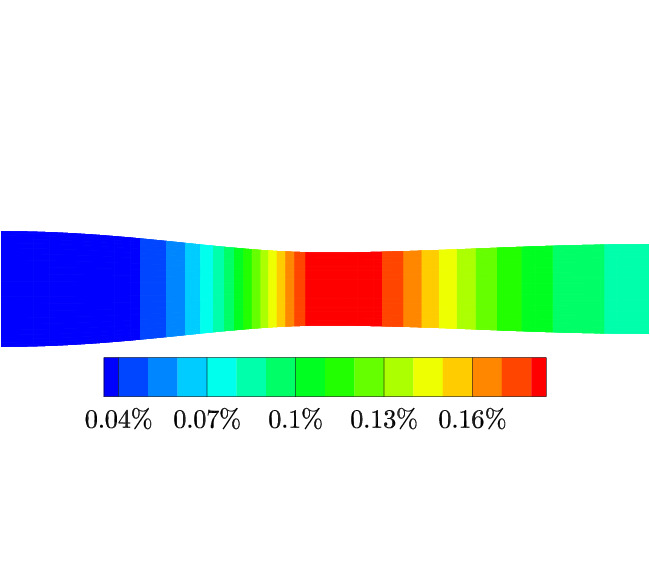}}
\hspace{0.02\textwidth}
\subfloat[\textit{x}-Velocity]{\includegraphics[width=0.47\textwidth,trim=0cm 5cm 0cm 7cm,clip]{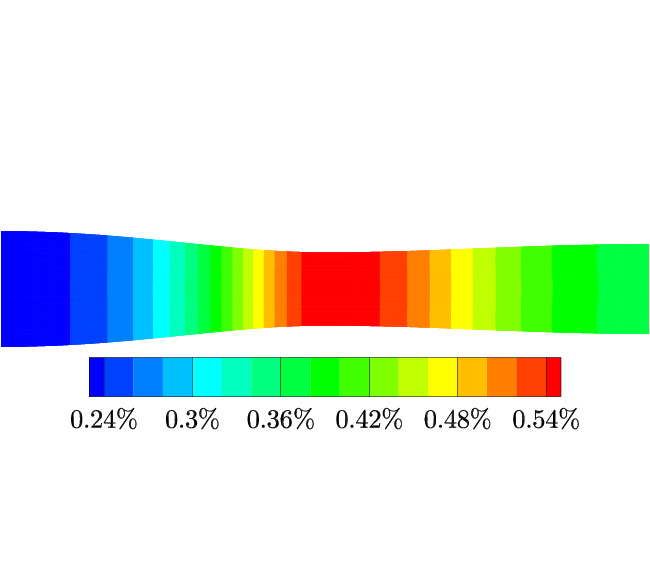}}
\\
\subfloat[\textit{y}-Velocity]{\includegraphics[width=0.47\textwidth,trim=0cm 5cm 0cm 7cm,clip]{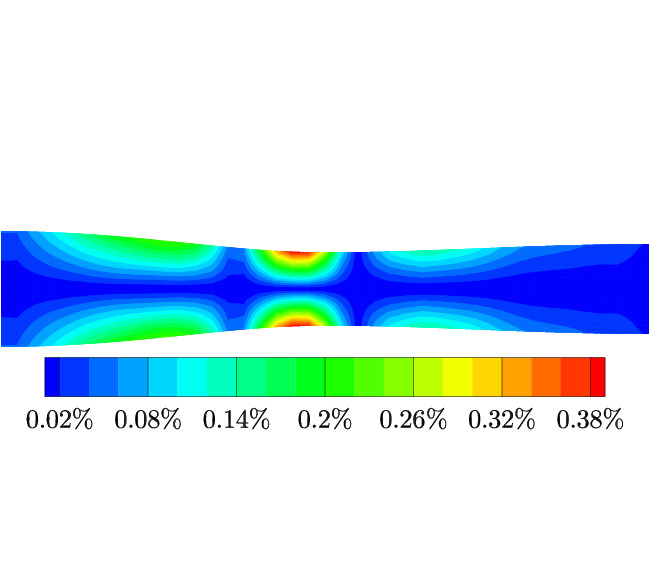}}
\hspace{0.02\textwidth}
\subfloat[Pressure]{\includegraphics[width=0.47\textwidth,trim=0cm 5cm 0cm 7cm,clip]{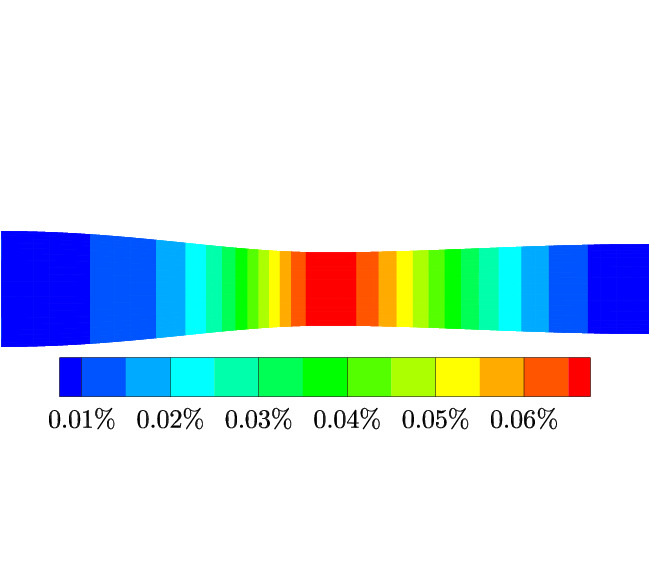}}
\caption{Time-averaged relative error between the ROM and FOM solutions  for an axisymmetric nozzle with outlet pressure oscillation of angular velocity $\freq=10$ rad/sec and amplitude $\amp=0.01$.}
\label{fig:3dnozzle_case1}
\end{figure}

Results are shown only for case 1 since all other cases showed similar results. Case 1 had values for the artificial dissipation parameter of $\tilde{\nu} = (2 , 0.45, 4, 4, 2)^{T}$ $\times 10^{-4}$. 
Figure~\ref{fig:3dnozzle_case1} shows the time-averaged relative  error between the ROM and the FOM solutions. The comparison for \textit{z}-velocity is not shown since it is identical to \textit{y}-velocity for an axisymmetric nozzle. The ROM solution for \textit{x}-velocity used only one mode for its reconstruction~\eqref{PODapprox}, while three modes were needed while solving for stability.
The ROM and FOM solutions are in good agreement, with a time-averaged relative error below 1\% for all state variables. For conciseness, the maximum spatial and temporal relative error is shown in Table~\ref{t:3dnozzle_maxerr_pressure_on} for all on-reference cases. 
\begin{table}[ht!]
\centering
\begin{tabular}{cccccc}
\toprule
& \multicolumn{5}{c}{$\max_{x,t} (\epsilon)$ (\%)} \\
Case & $\zeta$ & $u$ & $v$ & $w$ & $p$ \\
\midrule
1 & 0.29 & 0.70 & 0.48 & 0.48 & 0.13 \\ 
2 & 0.49 & 0.56 & 0.64 & 0.25 & 0.62 \\ 
3 & 0.34 & 0.69 & 0.68 & 0.29 & 0.38 \\ 
\bottomrule
\end{tabular}
\caption{Axisymmetric nozzle with outlet pressure oscillation: maximum spatial and temporal relative error between the ROM and FOM solutions for all on-reference cases.}
\label{t:3dnozzle_maxerr_pressure_on}
\end{table}
The ROM solution agreed well with the FOM in all cases, the maximum relative error being less than 1\%.  All three stabilizing methods were necessary to obtain a converged solution.

\subsubsection{Deforming Boundary}
\label{sec:axid}

\begin{figure}[ht!]
	\centering
	\subfloat[Minimum, $t=-\frac{1}{4 {\freq}}$]{\includegraphics[width=0.3\textwidth]{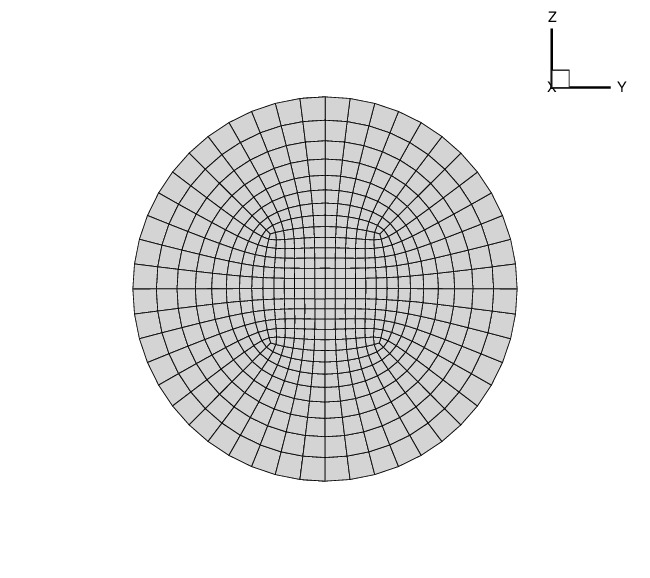}}
	\subfloat[Undeformed, $t=0$]{\includegraphics[width=0.3\textwidth]{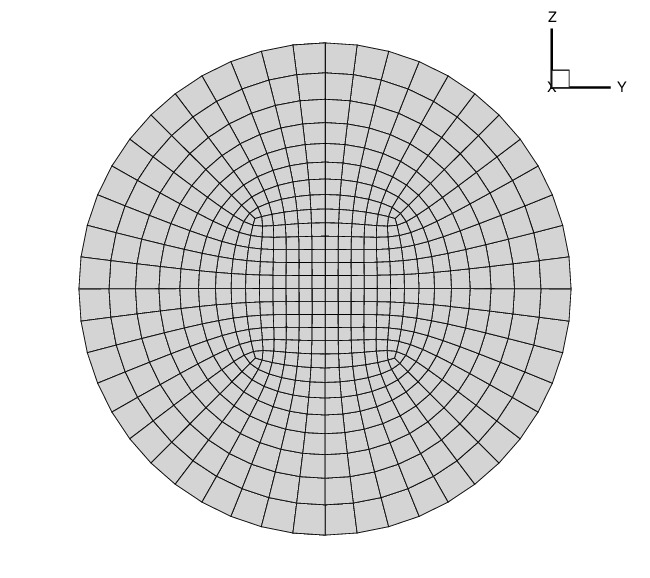}}
	\subfloat[Maximum, $t=\frac{1}{4 {\freq}}$]{\includegraphics[width=0.3\textwidth]{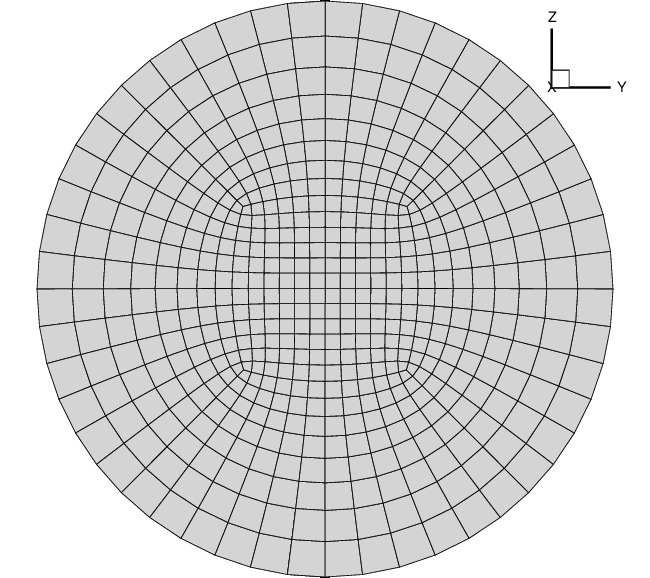}}
	\caption{Nozzle inlet deformation for amplitude $\amp=0.2$.}
	\label{fig:3dnozzle_deform}
\end{figure}

An oscillating wall boundary deformation was imposed on the axisymmetric nozzle by varying in time the radius of the nozzle at every \textit{x} location according to
\begin{equation}
\label{rforced}
r_{\text{forced}}(t) = \bar{r} \lp{(} 1 + \amp \sin \lp{(} 2 \pi \freq t \rp{)} \rp{)}	
\end{equation}
where $r_\text{forced}$ is the radius of the nozzle that is deformed, $\bar{r}$ is the nozzle undeformed radius, $\amp$ is the amplitude of the deformation, and $\freq$ is the angular velocity of the deformation.  An example of the deformation for an amplitude $\amp=0.2$ is shown in Figure~\ref{fig:3dnozzle_deform}.

To validate the zeta-variable POD-based ROM, three cases were investigated by varying the amplitude and angular velocity of the nozzle deformation. Table~\ref{t:3dnozzledeformcases} summarizes these cases.
\begin{table}[ht!]
\centering
\begin{tabular}{cccc}
	\toprule
	Case & $\freq$ [rad/sec] & $\amp$ [-] & Type \\
	\midrule
	1 & 10 & 0.2 & On-reference \\
	2 & 10 & 0.4 & On-reference \\
	3 & 20 & 0.2 & On-reference \\
	\bottomrule
\end{tabular}
\caption{Axisymmetric nozzle on-reference cases for an oscillating nozzle deformation.}
\label{t:3dnozzledeformcases}
\end{table}	
A back pressure of 102.1~kPa was held constant at downstream infinity during the deformation. The FOM solutions were generated for five periods, each period consisting of 100 snapshots. The zeta-variable ROM used the following number of nodes: $n^{\zeta}=n^{u}=n^{v}=n^{w}=n^{p}=1$.

To stabilize the zeta-variable ROM and preserve accuracy, the penalty me\-thod was applied to each state variable. To apply it to each state variable, a prescribed boundary condition needed to be defined for each state variable. To define the prescribed boundary condition, several approximate parameters were defined. The spatial and temporal average of the state vector across a prescribed boundary of the FOM is
\begin{equation}
\label{Zhat}
\hat{Z}_{k} = \lp{\langle} \bar{Z}_{k} (\bx_{\text{BC}}) \rp{\rangle}	
\end{equation}
where $\bar{\cdot}$ denotes a spatial average, $\langle \cdot \rangle$ denotes a time average, and $\bx_{\text{BC}}$ are the nodes along the prescribed boundary. Recall that the zeta-variable state vector is $\bZ = (\zeta, u, v, w, p)^{T}$, such that $Z_{1} = \zeta$, $Z_{2} = u$, \textit{etc.} The amplitude is defined using the maximum and minimum values over time across a prescribed boundary
\begin{equation}
\label{Ahat}
\hat{A}_{k} = \frac{\max \lp{(} \bar{Z}_{k} (\bx_{\text{BC}}) \rp{)} - \min \lp{(} \bar{Z}_{k} (\bx_{\text{BC}}) \rp{)}}{2 \hat{Z}_{k}}	.
\end{equation}
Using definitions~\eqref{Zhat} and~\eqref{Ahat}, an approximation of the solution at the prescribed boundary is defined
\begin{equation}
\label{approxeq}
Z_{k_{\text{approx}}} = \hat{Z}_{k} \lp{(} 1 + \hat{A}_{k} \sin \lp{(} \freq t - \hat{\varphi}_{k} \rp{)} \rp{)}, \ k \in [1,D+2]	
\end{equation}
where the value of the phase $\hat{\varphi}_{k}$ is determined such that the peak values of the FOM solution and its approximation are in phase, that is, $\freq t^{\text{peak}} - \hat{\varphi}_{k} = \freq t^{\text{peak}}_{\text{FOM}}$. Consequently,
\begin{equation}
\label{phihat}
\hat{\varphi}_{k} = \freq \lp{(} t^{\text{peak}} - t^{\text{peak}}_{\text{FOM}} \rp{)} .
\end{equation}
The penalty method was imposed onto the ROM using these approximate equations, where the prescribed boundary condition was applied at the outlet
\begin{equation*}
\textbf{F}(t) = \lp{(} \bZ_{\zeta_{\text{approx}}}, \bZ_{u_{\text{approx}}}, \bZ_{v_{\text{approx}}}, \bZ_{w_{\text{approx}}}, \bZ_{p_{\text{approx}}} \rp{)}^{T}
\end{equation*}
each component being a vector of size $N_{\text{out}}$.  The penalty parameter $\pmb{\uptau}$ was calculated for each variable by solving~\eqref{tausecant}. 

The spatial and temporal averages $\hat{Z}_{k}$, $\hat{A}_{k}$ and $\hat{\varphi}_{k}$ were calculated based on the FOM solutions and are shown in Table~\ref{t:3dnozzle_deform_approxcases}.
%
%
\begin{table}[ht!]
\centering
\begin{tabular}{c|ccc}
\toprule
~ & \multicolumn{3}{c}{Case} \\
~ & 1 & 2 & 3 \\
\midrule
 $\hat{\zeta}$ & 1.00 & 1.00 & 1.00 \\
 $\hat{A}_\zeta \times 10^{-5}$ & 1.52 & 2.93 & 1.73 \\
 $\hat{\varphi}_\zeta $ & -0.82 & -0.69 & -0.75 \\ 
\midrule
 $\hat{u}$ & 0.16 & 0.16 & 0.16 \\
 $\hat{A}_u \times 10^{-3}$ & 0.81 & 1.68 & 0.89 \\
 $\hat{\varphi}_u$ & 3.64 & 3.77 & -1.26 \\ 
\midrule
 $\hat{v}$ $\times 10^{-6}$ & -5.11 & -5.39 & -5.16 \\
 $\hat{A}_v$ & -0.34 & -0.65 & -0.40 \\
 $\hat{\varphi}_v$ & 9.36 & 28.27 & -28.27 \\
\midrule
 $\hat{w}$ $\times 10^{-6}$ & -5.94 & -6.06 & -5.95 \\
 $\hat{A}_w$ & -0.27 & -0.54 & -0.30\\
 $\hat{\varphi}_w$ & 15.77 & 15.77 & -9.42 \\
\midrule
 $\hat{p}$ & 0.72 & 0.72 & 0.72 \\
 $\hat{A}_p \times 10^{-5}$ & 0.97 & 1.74 & 0.72 \\
 $\hat{\varphi}_p$ & 9.36 & 9.42 & 28.40 \\ 
\bottomrule
\end{tabular}
\caption{Mean values of state variables, amplitudes and phase angles for a nozzle deformation of the axisymmetric nozzle.}
\label{t:3dnozzle_deform_approxcases}
\end{table}
One can note that $\hat{Z}_{k}$ remains approximately constant for all cases while the amplitude $\hat{A}_{k}$ scales approximately with the actual amplitudes $\tilde{A}$. For example, case 2 has an amplitude, $\amp=0.4$, that is twice as large as the amplitude for case 1, $\amp=0.2$. Similarly, using conservation of mass as an example, the approximate amplitude of case 2, $\hat{A}_{\zeta}=2.93 \times 10^{-5}$, is approximately twice as large as the approximate amplitude for case 1, $\hat{A}_{\zeta} = 1.52 \times 10^{-5}$.

\begin{figure}[ht!]
\centering
\subfloat[Specific Volume]{\includegraphics[width=0.47\textwidth,trim=0cm 4cm 0cm 6cm,clip]{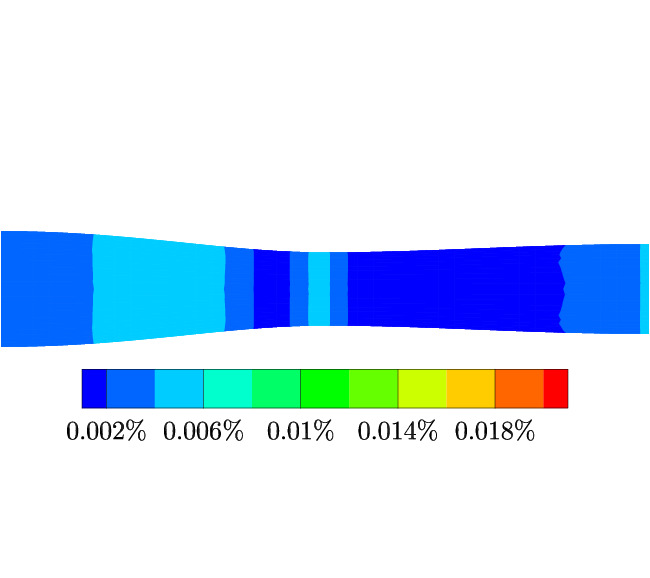}}
\hspace{0.02\textwidth}
\subfloat[Specific Volume]{\includegraphics[width=0.47\textwidth,trim=0cm 4cm 0cm 6cm,clip]{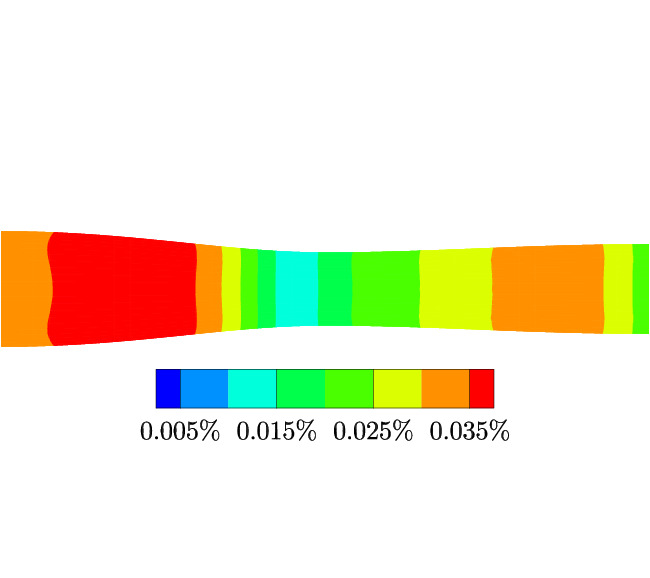}}
\\
\subfloat[Specific Volume]{\includegraphics[width=0.47\textwidth,trim=0cm 4cm 0cm 6cm,clip]{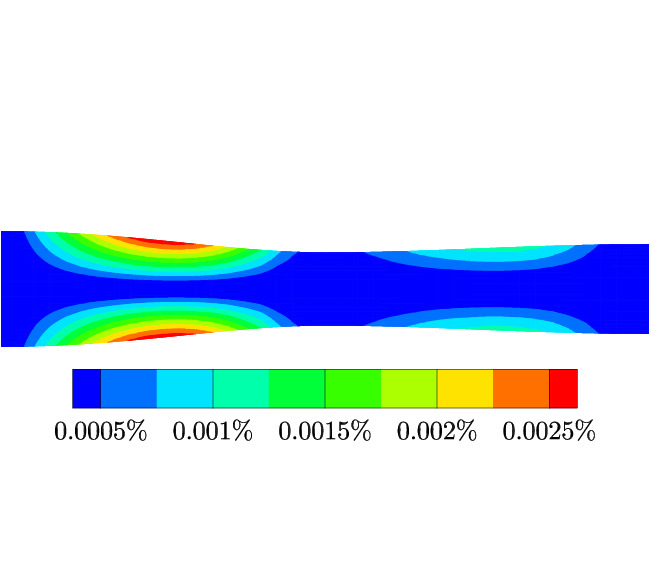}}
\hspace{0.02\textwidth}
\subfloat[Specific Volume]{\includegraphics[width=0.47\textwidth,trim=0cm 4cm 0cm 6cm,clip]{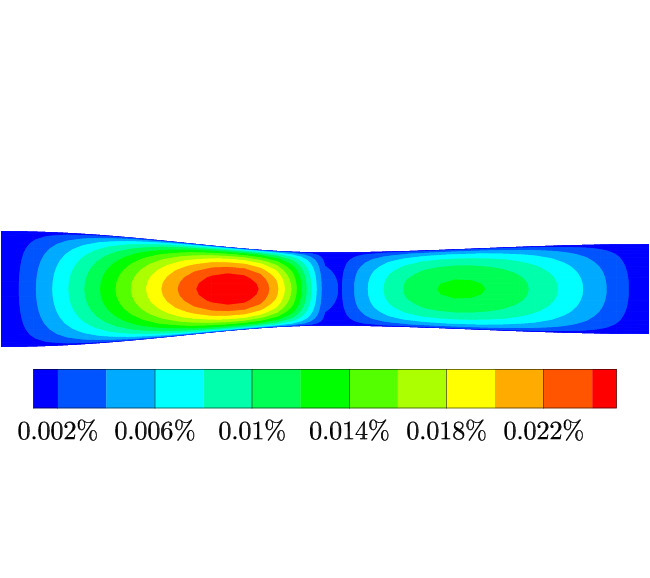}}
\caption{Time-averaged relative error between the ROM and FOM solutions for an axisymmetric nozzle deforming with angular velocity $\freq = 10$~rad/sec and amplitude $\amp = 0.2$.}
\label{fig:3dnozzle_deform_case1}
\end{figure}

Only case 1 results are shown here since all other cases showed similar results. Figure~\ref{fig:3dnozzle_deform_case1} shows the time-averaged relative error between the ROM and FOM solutions. 
The \textit{z}-velocity results are not shown since they were identical to
those of the \textit{y}-velocity. The time-averaged relative errors showed a
good agreement between the ROM and the FOM solutions.  The maximum spatial
and temporal relative error is given for each case in
Table~\ref{t:3dnozzle_maxerr_deform_on}.  The zeta-variable ROM was able to
reproduce the FOM solution with a relative error of less than 0.5\%.

\begin{table}[ht!]
\centering
\begin{tabular}{cccccc}
\toprule
& \multicolumn{5}{c}{$\max_{x,t} (\epsilon)$ (\%)} \\
Case & $\zeta$ & $u$ & $v$ & $w$ & $p$ \\
\midrule
1 & 0.04 & 0.08 & 0.01 & 0.04 & 0.02 \\ 
2 & 0.06 & 0.17 & 0.05 & 0.09 & 0.03 \\ 
3 & 0.06 & 0.42 & 0.06 & 0.36 & 0.15 \\ 
\bottomrule	
\end{tabular}
\caption{Maximum relative  error across all nodes and time of the ROM solution to the FOM solution for all on-reference cases for an axisymmetric nozzle deformation.}
\label{t:3dnozzle_maxerr_deform_on}
\end{table}

\subsection{NASA Rotor 67}
\label{sec:r67}

\begin{figure}[ht!]
\centering
\subfloat[Computational domain]{\includegraphics[width=0.99\textwidth,trim=-1cm 7.5cm -1cm 6cm, clip]{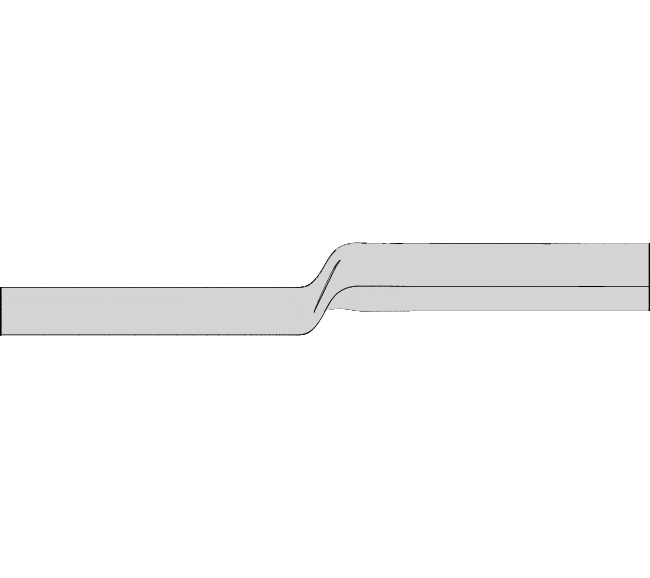}}
\\
\subfloat[Detail of hub-to-tip mesh]{\includegraphics[width=0.4\textwidth]{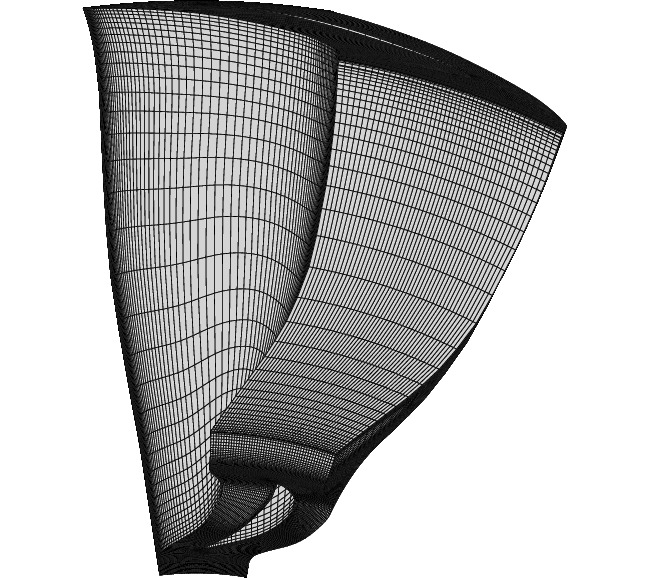}}
\hspace{0.01\textwidth}
\subfloat[Detail of mesh near airfoil]{\includegraphics[width=0.4\textwidth]{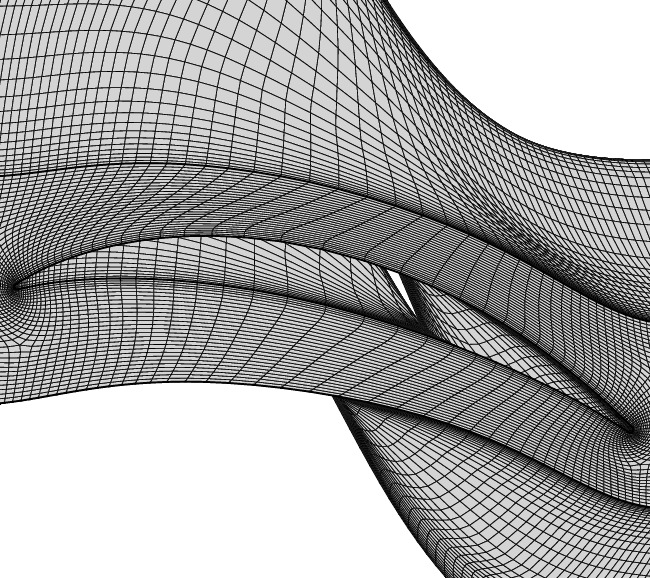}}
\caption{NASA Rotor 67: computational mesh with $N=299,844$ grid nodes.}
\label{fig:3drotor67mesh}
\end{figure}

Rotor 67 is a twenty-two-blade first-stage rotor of a two-stage fan developed and tested at NASA Lewis~\cite{Urasek1979}.  The inlet stagnation pressure and temperature are 124.453~kPa and 305.74~K, respectively.  The flow enters the inlet along the x-axis. The mean outlet pressure at the hub is 136.898~kPa.  A FOM solution for this case was generated and validated against the experimental data~\cite{Carpenter2016}. This FOM solved the Euler equations written in a rotating reference frame about the \textit{x}-axis~\eqref{zeta_rot}. Figure~\ref{fig:3drotor67mesh} shows the computational mesh of Rotor 67 that used $299,844$ grid nodes. 

The inlet and outlet boundaries were located approximately 5 chords away from the airfoil to reduce spurious boundary reflections. Periodic boundary conditions were used to reduce the computational domain to one blade passage. The pressure variation from hub to tip across the outlet was solved using the axisymmetric radial momentum equation
\begin{equation}
\label{radialmom}
\dd{p}{r} = \frac{\rho v_{\theta}^{2}}{r}	
\end{equation}
where $v_{\theta} = (vz-wy)/r$.  The angular velocity about the \textit{x}-axis was $\omegax = 16,520$~RPM. Integrating~\eqref{radialmom} from the hub to a radial location $r$, $r \leq r_\text{tip}$, yields
\begin{equation*}
    p(r) - p(r_{\text{hub}}) = \int_{r_\text{hub}}^r \frac{\rho v_{\theta}^{2}}{r} dr .
\end{equation*}
The unsteadiness was introduced in the FOM through an oscillating outlet pressure at the hub $p(r_{\text{hub}})$ that followed the form of~\eqref{pback}. 
Two amplitudes $\amp$ and two angular velocities $\freq$ were considered, as shown in Table~\ref{t:rotor67_cases_on}. 
\begin{table}[ht!]
\centering
\begin{tabular}{ccc}
\toprule
Case & $\freq$ [rad/sec] & $\amp$ \\
\midrule
1 & 38,059 & 0.05 \\
2 & 38,059 & 0.10 \\
3 & 37,582 & 0.05 \\
\bottomrule	
\end{tabular}
\caption{Rotor 67 on-reference cases for an oscillating outlet pressure boundary condition.}
\label{t:rotor67_cases_on}
\end{table}

The FOM generated snapshots for five periods. Fifty snapshots were saved for each period. All cases had a pressure $\bar{p}=136.9$ kPa and a phase shift $\varphi=0$. 

The penalty method was imposed onto the ROM for each state variable using an approximation of the boundary solution~\eqref{approxeq}. To avoid solving~\eqref{radialmom} from hub to tip at each time step in the ROM, the penalty method was applied only at the hub and not at the other radial locations.  
The prescribed boundary condition was
\begin{equation*}
\textbf{F}(t) = \lp{(} \lp{(} \gamma \textbf{f}(t) \rp{)}^{\frac{1}{\gamma}}, \bZ_{u_{\text{approx}}}, \bZ_{v_{\text{approx}}}, \bZ_{w_{\text{approx}}}, \bZ_{p_{\text{approx}}} \rp{)}^{T}
\end{equation*}
where $\textbf{f}(t) \in \R^{N_{\text{out}}}$ and $\bZ_{k_{\text{approx}}} \in \R^{N_{\text{hub}}}, k\in[2,D+2]$, with $N_{\text{hub}}$ being the number of nodes along the outlet hub boundary. The isentropic relation was used for determining the boundary condition for the specific volume as done for the channel and the axisymmetric nozzle.
The penalty parameter $\pmb{\uptau}$ was solved for using~\eqref{tausecant} for each state variable. The approximate values of~\eqref{approxeq}  were determined from the FOM solution for each case using~\eqref{Zhat} to~\eqref{phihat}, and these values are shown in Table~\ref{t:r67_approx_on}.
\begin{table}[ht!]
\centering
\begin{tabular}{c|ccc}
\toprule
~ & \multicolumn{3}{c}{Case} \\
~ & 1 & 2 & 3 \\
\midrule
 $\hat{u}$ & 0.50 & 0.50 & 0.50 \\
 $\hat{A}_u \times 10^{-2}$ & 0.95 & 1.91 & 0.97 \\ 
 $\hat{\varphi}_u$ & -13.83 & -13.83 & -19.86 \\
\midrule
 $\hat{v}$ & -0.67 & -0.67 & -0.67 \\
 $\hat{A}_v \times 10^{-4}$ & -2.22 & -4.47 & -2.27 \\ 
 $\hat{\varphi}_v$ & 4.53 & 4.65 & -1.49 \\
\midrule
 $\hat{w}$ & 0.12 & 0.12 & 0.12 \\
 $\hat{A}_w \times 10^{-2}$ & 0.73 & 1.46 & 0.74 \\ 
 $\hat{\varphi}_w$ & -19.99 & -19.99 & -19.86 \\
\midrule
 $\hat{p}$ & 1.08 & 1.08 & 1.08 \\
 $\hat{A}_p \times 10^{-2}$ & 0.52 & 1.04 & 0.52 \\ 
 $\hat{\varphi}_p$ & -23.13 & -23.13 & 2.23 \\
\bottomrule
\end{tabular}
\caption{NASA Rotor 67 with outlet pressure oscillation: mean values of state variables, amplitudes and phase angles.}
\label{t:r67_approx_on}
\end{table}
Like the axisymmetric nozzle deformation cases, the approximate values scale roughly the same as the actual values.

\begin{figure*}
	\centering
	\setcounter{subfigure}{-2}
 	\subfloat[Specific Volume]{\subfloat{\includegraphics[width=0.55\textwidth,trim=0cm 4.5cm 0cm 7.5cm, clip]{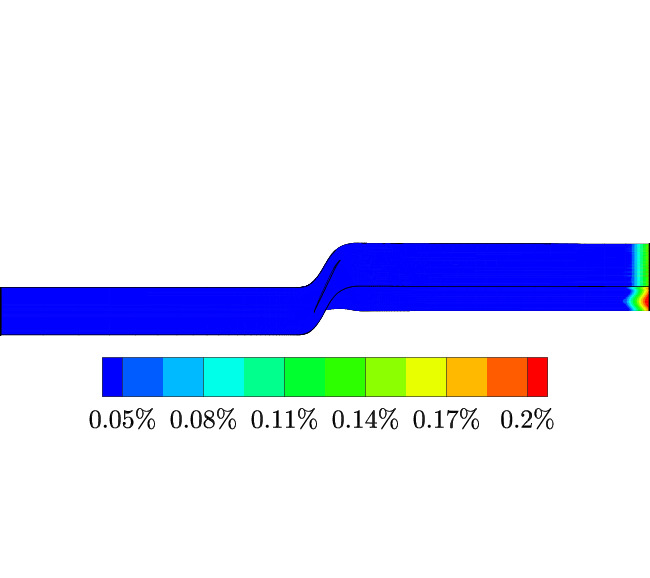}} \subfloat{\includegraphics[width=0.2\textwidth]{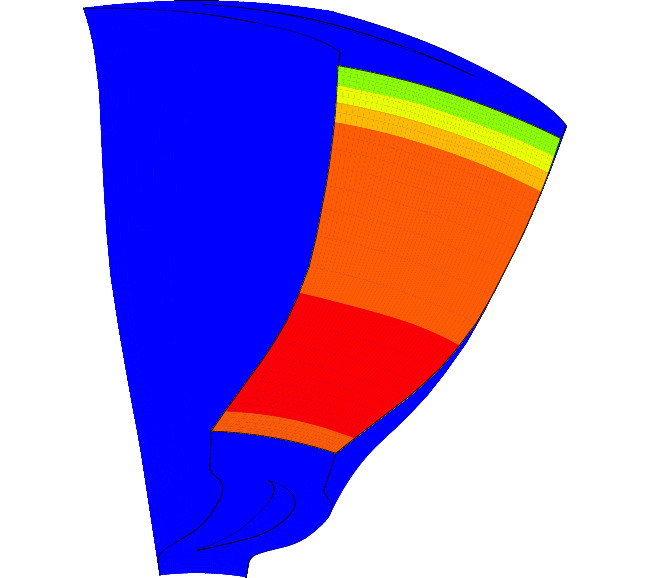}}}
 	\\
 	\setcounter{subfigure}{-1}
 	\subfloat[\textit{x}-Velocity]{\subfloat{\includegraphics[width=0.55\textwidth,trim=0cm 4.5cm 0cm 7.5cm, clip]{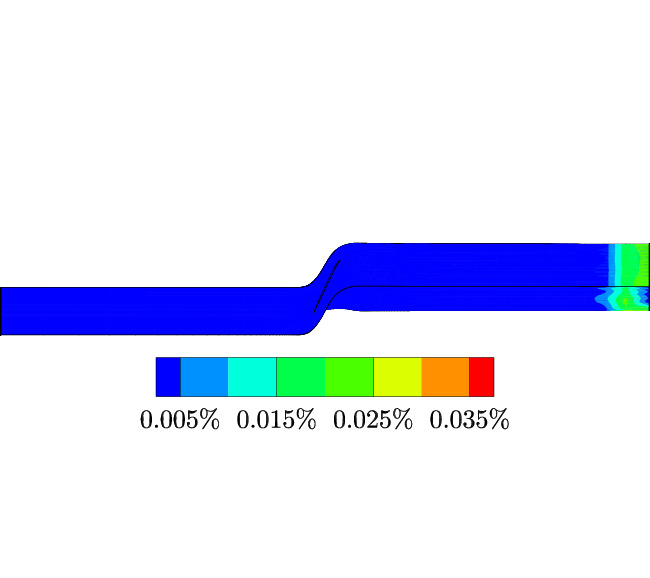}} \subfloat{\includegraphics[width=0.2\textwidth]{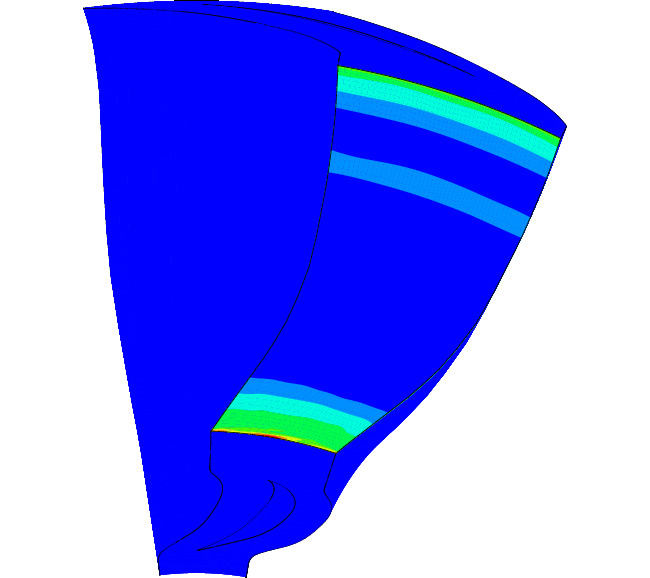}}} 
 	\\
 	\setcounter{subfigure}{0}
 	\subfloat[\textit{y}-Velocity]{\subfloat{\includegraphics[width=0.55\textwidth,trim=0cm 4.5cm 0cm 7.5cm, clip]{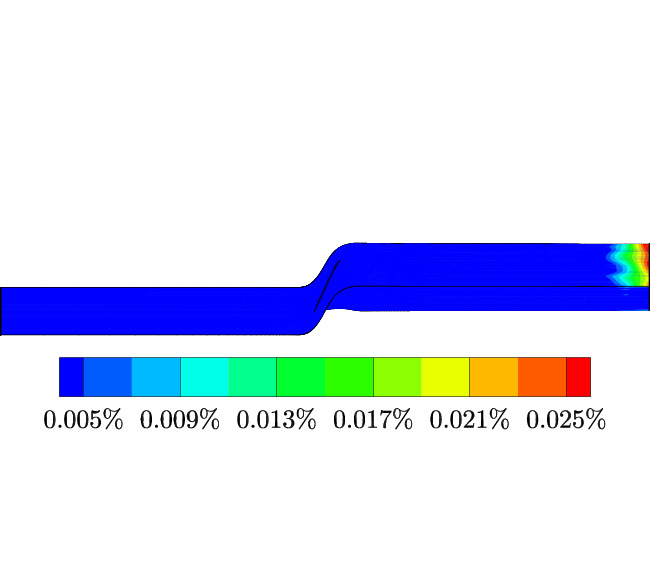}} \subfloat{\includegraphics[width=0.2\textwidth]{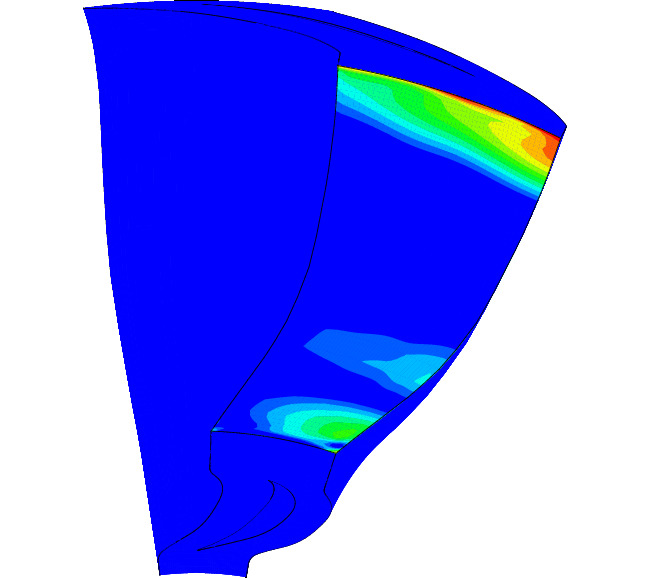}}}
 	\\
 	\setcounter{subfigure}{1}
 	\subfloat[\textit{z}-Velocity]{\subfloat{\includegraphics[width=0.55\textwidth,trim=0cm 4.5cm 0cm 7.5cm, clip]{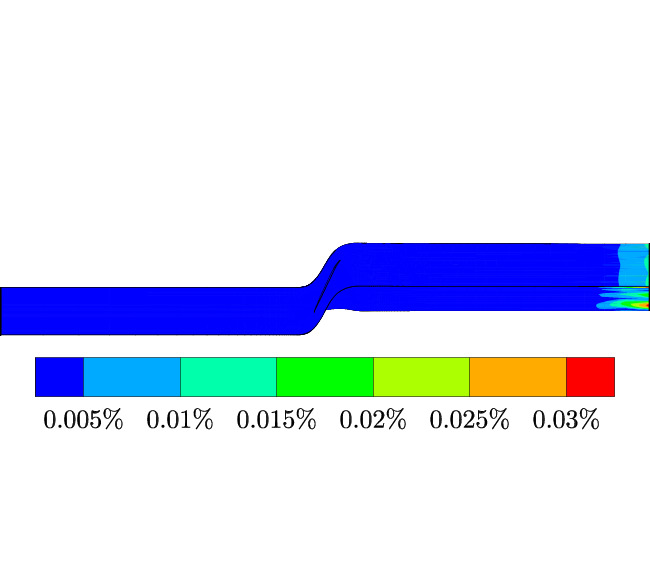}} \subfloat{\includegraphics[width=0.2\textwidth]{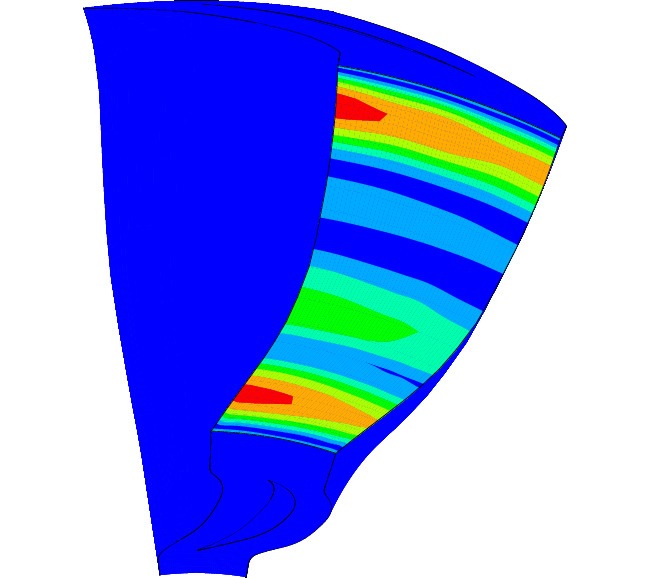}}} 
 	\\
 	\setcounter{subfigure}{2}
 	\subfloat[Pressure]{\subfloat{\includegraphics[width=0.55\textwidth,trim=0cm 4.5cm 0cm 7.5cm, clip]{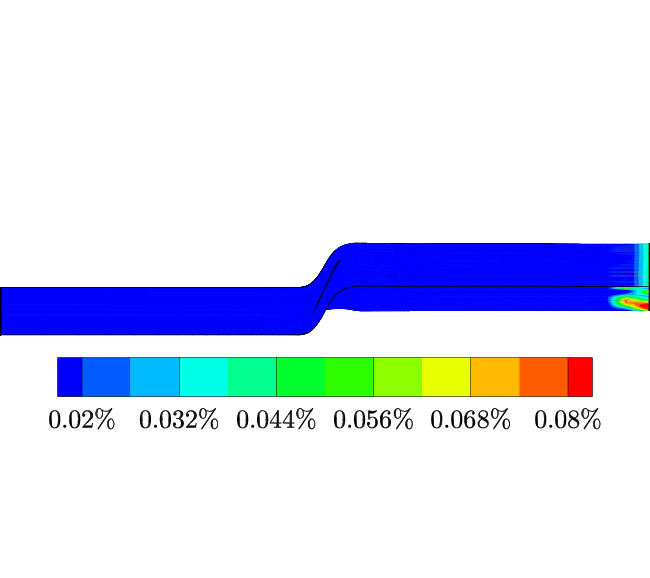}} \subfloat{\includegraphics[width=0.2\textwidth]{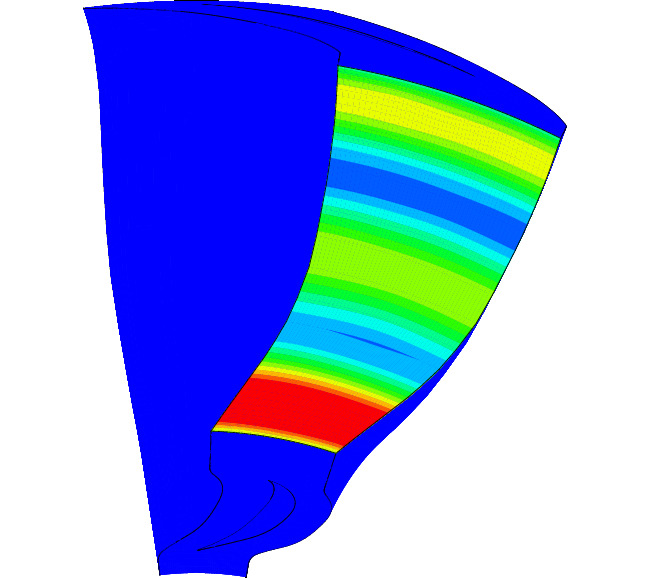}}}
	\caption{Time-Averaged Relative Error of the ROM to the FOM for Rotor 67 with an outlet pressure oscillation with angular velocity $\freq = 38,059$ rad/sec and amplitude $\amp = 0.05$.}
	\label{fig:r67_case1_res}
\end{figure*}

For conciseness and since all cases showed similar results, only case 1
results are shown here. Figure~\ref{fig:r67_case1_res} shows the
time-averaged relative error between the FOM and ROM solutions.  The
time-averaged relative error is less than 0.5\% for all state variables. The
largest errors occur at the outlet, where the penalty-imposed boundary
conditions were applied. Of note in this case is that there exists a shock
wave near the leading edge of the blade. The error values are not increased
at the shock location, therefore the zeta-variable ROM solution accurately
captures the flow discontinuity from the FOM
solution. Table~\ref{t:r67_maxerr_allcases_on} summarizes the results of all
on-reference cases for Rotor 67.
\begin{table}[ht!]
\centering
\begin{tabular}{cccccc}
\toprule
& \multicolumn{5}{c}{$\max_{x,t} (\epsilon)$ (\%)} \\
Case & $\zeta$ & $u$ & $v$ & $w$ & $p$ \\
\midrule
1 & 0.50 & 0.19 & 0.04 & 0.05 & 0.39 \\ 
2 & 0.78 & 0.39 & 0.07 & 0.11 & 0.78 \\ 
3 & 0.60 & 0.28 & 0.04 & 0.07 & 0.54 \\ 
\bottomrule	
\end{tabular}
\caption{NASA Rotor 67 with outlet pressure oscillation: maximum spatial and temporal relative errors between the ROM and FOM solutions, for all on-reference cases.}
\label{t:r67_maxerr_allcases_on}
\end{table}
Each case shows good agreement between the zeta-variable ROM solution to its respective FOM solution with a maximum relative  error over all time and space of less than 1\%.

\section{Off-Reference Results}
This section presents the off-reference results for the zeta-variable
ROM. The POD basis functions were interpolated to new flow conditions by
either enriching the snapshot database or using Grassmann
interpolation. Results are shown for three configurations: (1) a
quasi-one-dimensional nozzle flow, (2) a three-dimensional axisymmetric
nozzle flow, and (3) the NASA Rotor 67.

\subsection{Quasi-One-Dimensional Nozzle Flow}

Two cases were used to validate the zeta-variable ROM at off-reference conditions. Table~\ref{t:1d_allcases} shows these cases in addition to the on-reference cases analyzed in section~\ref{sec:1d}.
\begin{table}[ht!]
\centering
\begin{tabular}{cccc}
\toprule
Case & $\freq$ [rad/sec] & $\amp$ & Type \\
\midrule
1 & 1.0 & 0.020 & on-reference \\
2 & 1.0 & 0.030 & on-reference \\
3 & 2.0 & 0.020 & on-reference \\
4 & 1.0 & 0.025 & off-reference \\
5 & 1.5 & 0.020 & off-reference \\
\bottomrule
\end{tabular}
\caption{On- and off-reference cases for the quasi-one-dimensional nozzle with oscillating outlet pressure.}
\label{t:1d_allcases}
\end{table}
To interpolate the basis functions, the snapshots of cases 1 and 2 were combined to develop the autocorrelation matrix~\eqref{autocorr} for case 4. The snapshots of cases 1 and 3 were combined to develop the autocorrelation matrix for case 5. It is important to note that the zeta-variable ROM used the interpolated basis functions developed by enriching the snapshot database to solve for the off-reference solutions. 

The penalty method for the off-reference cases was used in a manner similar to that for the on-reference cases. Consequently, the penalty method was used to constrain the solution of the zeta-variable ROM such that it matched the desired flow conditions, \textit{i.e.}, the angular velocity and amplitude of the outlet pressure oscillation. The penalty parameter $\pmb{\uptau}$ was determined by finding the roots of the error $\bK$. 

\begin{figure}[ht!]
\centering
\subfloat[Specific Volume]{\includegraphics[width=0.334\textwidth]{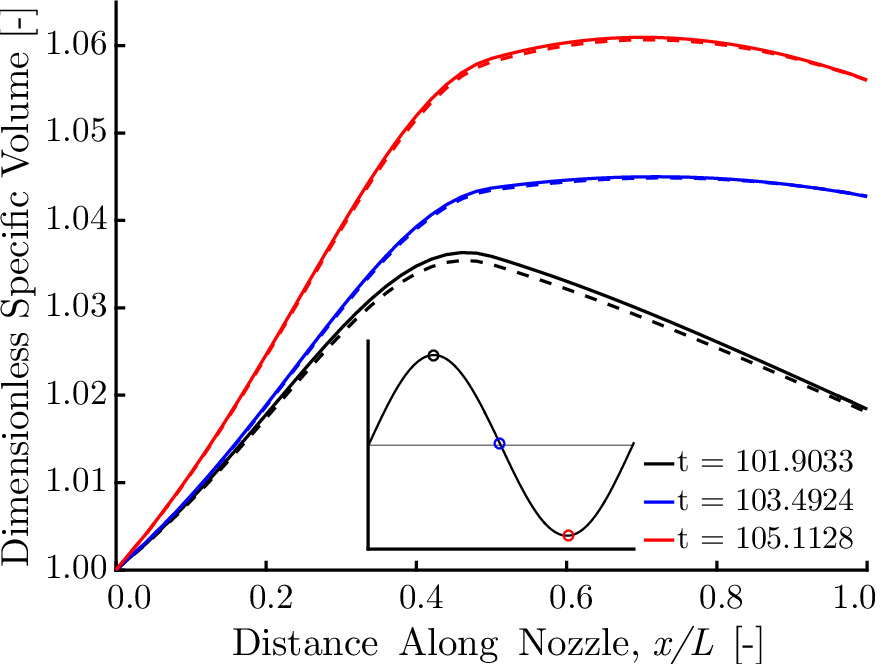}}
\subfloat[\textit{x}-Velocity]{\includegraphics[width=0.33\textwidth]{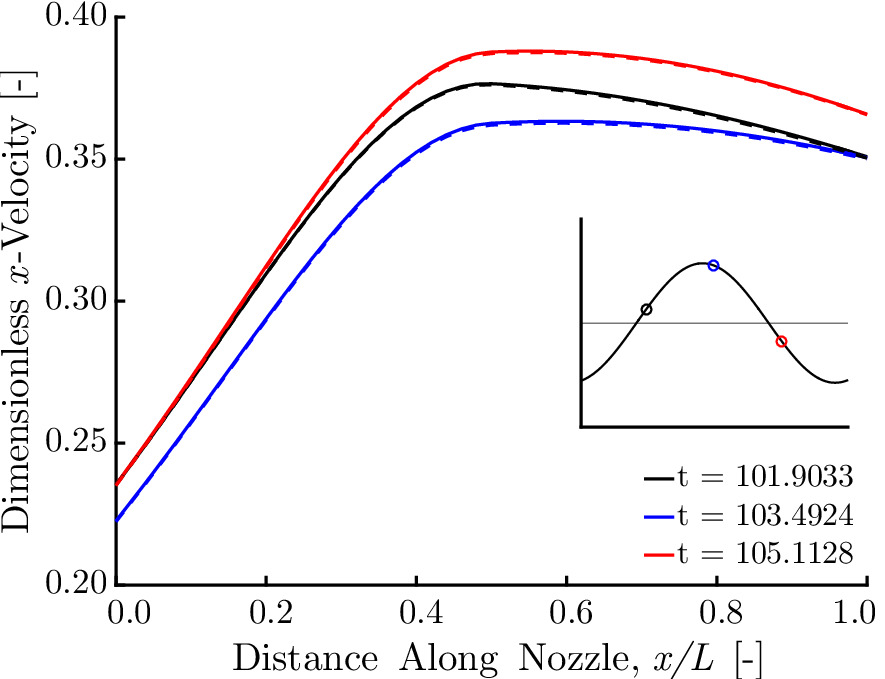}}
\subfloat[Pressure]{\includegraphics[width=0.33\textwidth]{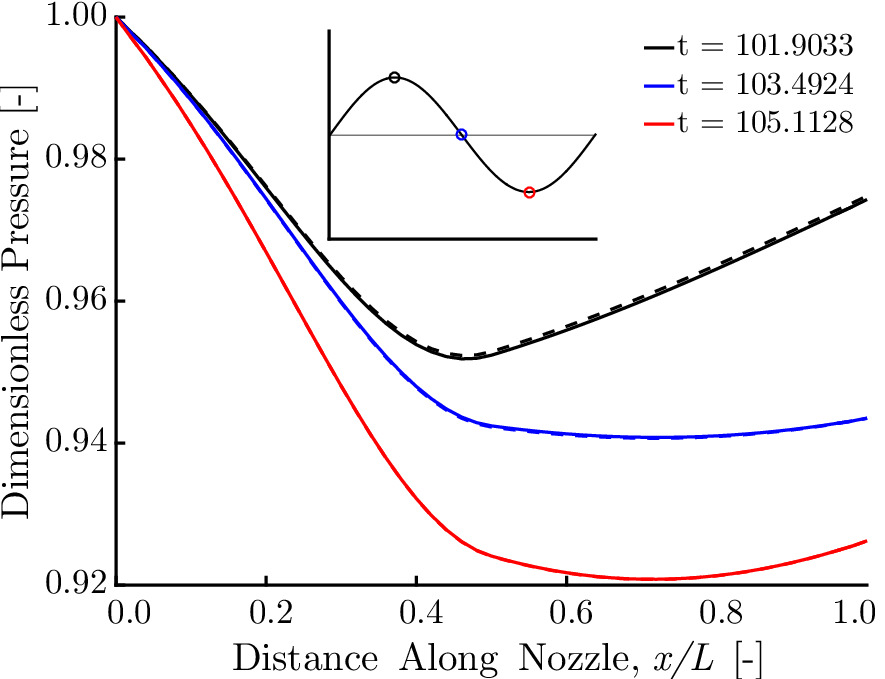}}
\caption{Off-reference ROM solution vs. FOM solution for a quasi-one-dimensional nozzle with an outlet pressure oscillation with angular velocity $\freq$ = 1~rad/sec and amplitude $\amp=0.025$. The dashed lines are the ROM solution and the solid lines are the FOM solution.}
\label{fig:1d_case4}
\end{figure}

The zeta-variable ROM was solved for both cases using the following number of modes $n^{\zeta}=n^{u}=n^{p}=2$. Figure~\ref{fig:1d_case4} shows a comparison between the reconstructed zeta-variable ROM solution and the FOM solution for case 4 at three different times corresponding to the minimum, mean, and maximum of the outlet pressure oscillation. It is important to note that this respective FOM solution was only developed for comparison, and the snapshots of the respective FOM solution were not used to develop the autocorrelation matrix.  The ROM solution shows good agreement with the respective FOM solution for all three times. 

\begin{figure}[ht!]
\centering
\subfloat[Specific Volume]{\includegraphics[width=0.334\textwidth]{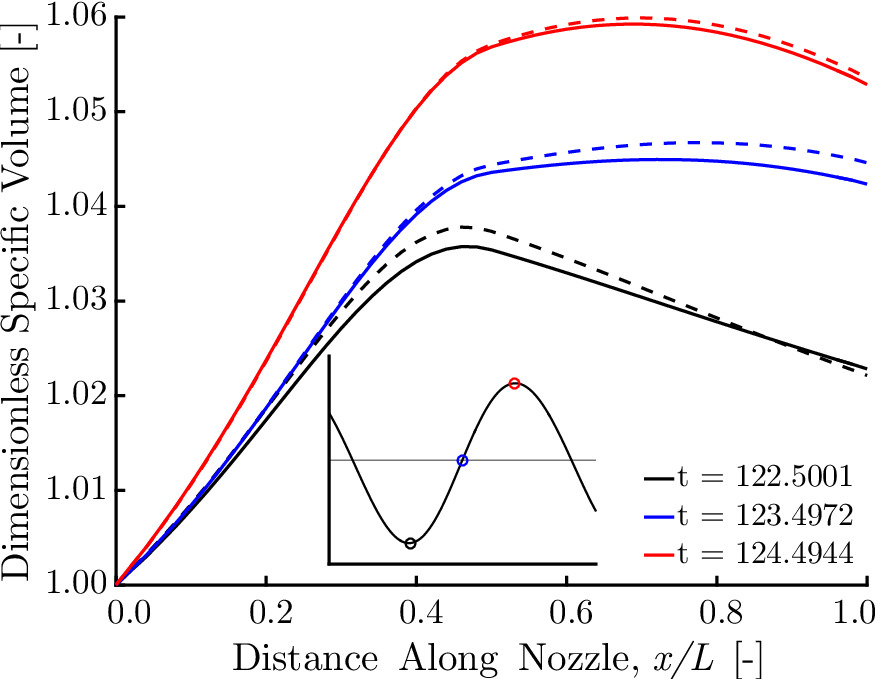}}
\subfloat[\textit{x}-Velocity]{\includegraphics[width=0.33\textwidth]{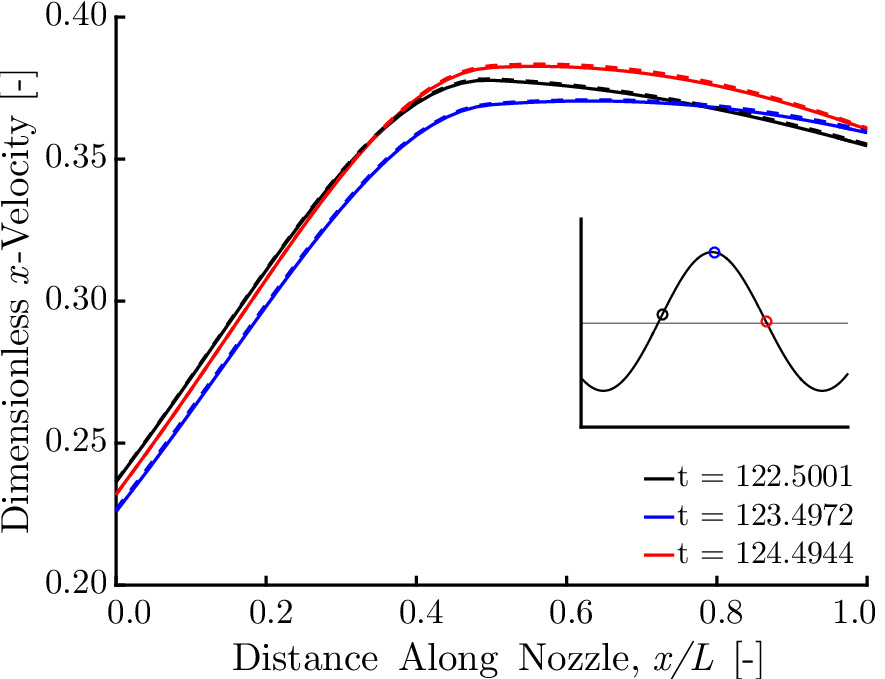}}
\subfloat[Pressure]{\includegraphics[width=0.33\textwidth]{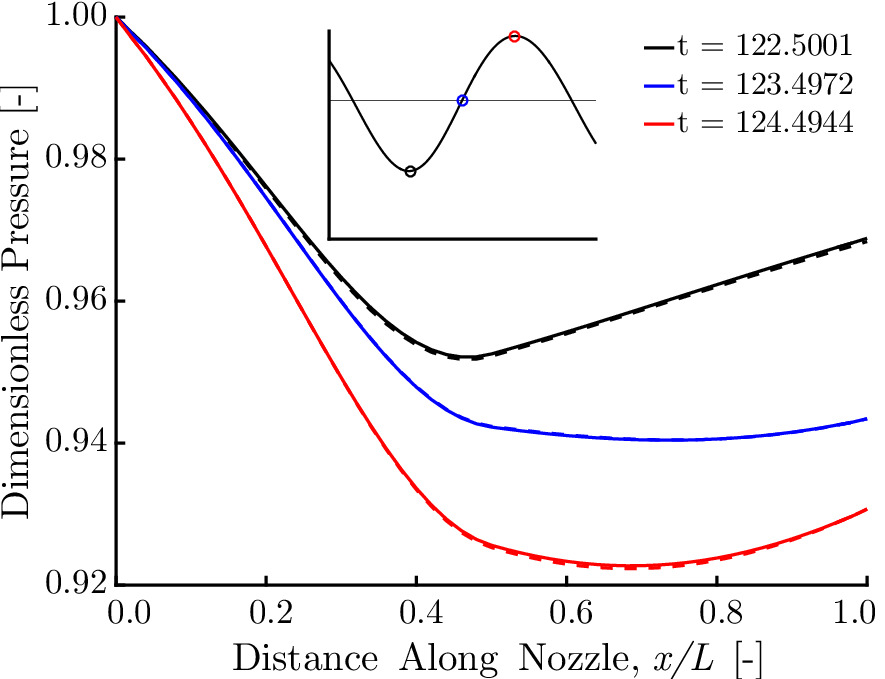}}
\caption{Off-reference ROM solution vs. FOM solution for a quasi-one-dimensional nozzle with an outlet pressure oscillation with angular velocity $\freq$ = 1.5~rad/sec and amplitude $\amp=0.02$. The dashed lines are the ROM solution and the solid lines are the FOM solution.}
\label{fig:1d_case5}
\end{figure}

Figure~\ref{fig:1d_case5} shows a comparison between the reconstructed zeta-variable ROM solution and the FOM solution for case 5.
The zeta-variable ROM again agreed well with its respective FOM solution.

\subsection{Three-Dimensional Axisymmetric Nozzle Flow}
The zeta-variable ROM was validated at off-reference conditions for a three-dimensional axisymmetric nozzle flow with two boundary conditions: an oscillating outlet pressure boundary condition~\eqref{pback} and an oscillating wall boundary condition~\eqref{rforced}. 

\subsubsection{Oscillating Outlet Pressure}

Two cases were used to validate the off-reference zeta-variable ROM. Table~\ref{t:3dnozzle_pressure_allcases} shows these cases in addition to the on-reference cases analyzed in section~\ref{sec:axi}. 
\begin{table}[ht!]
\centering
\begin{tabular}{cccc}
\toprule
Case & $\freq$ [rad/sec] & $\amp$ & Type \\
\midrule
1 & 10 & 0.010 & on-reference \\
2 & 10 & 0.020 & on-reference \\
3 & 20 & 0.010 & on-reference \\
4 & 10 & 0.015 & off-reference \\
5 & 15 & 0.010 & off-reference \\
\bottomrule
\end{tabular}
\caption{On- and off-reference cases for the axisymmetric nozzle with an oscillating outlet pressure.}
\label{t:3dnozzle_pressure_allcases}
\end{table}
To develop the basis functions for case 4, the snapshots of cases 1 and 2 were combined to produce the autocorrelation matrix.  The basis functions for case 5 were generated using the snapshots of cases 1 and 3. 

The same stabilizing methods used for the on-reference cases of the axisymmetric nozzle were applied to the off-reference cases.  The penalty parameter $\pmb{\uptau}$ was determined by using~\eqref{tausecant}. Artificial dissipation was applied to limit the growth of spurious modes. 
The artificial dissipation parameters were given the upper bound value of~\eqref{artdissparam}. 

\begin{figure}[ht!]
\centering
\subfloat[Specific Volume]{\includegraphics[width=0.47\textwidth,trim=0cm 4cm 0cm 7cm,clip]{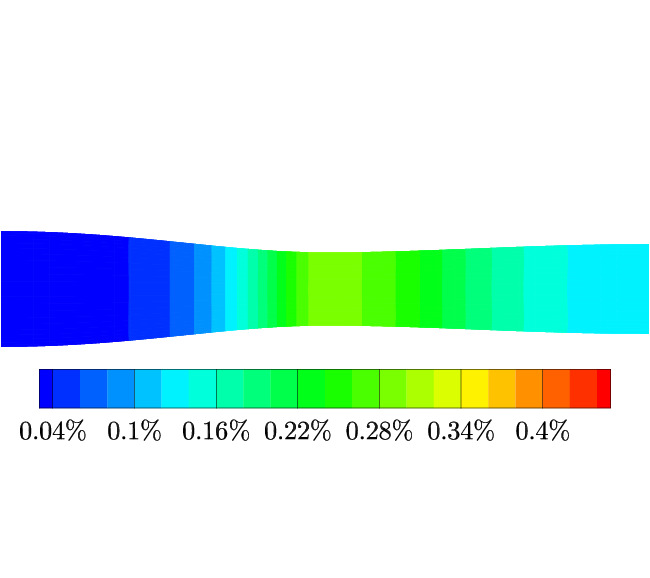}}
\hspace{0.02\textwidth}
\subfloat[\textit{x}-Velocity]{\includegraphics[width=0.47\textwidth,trim=0cm 4cm 0cm 7cm,clip]{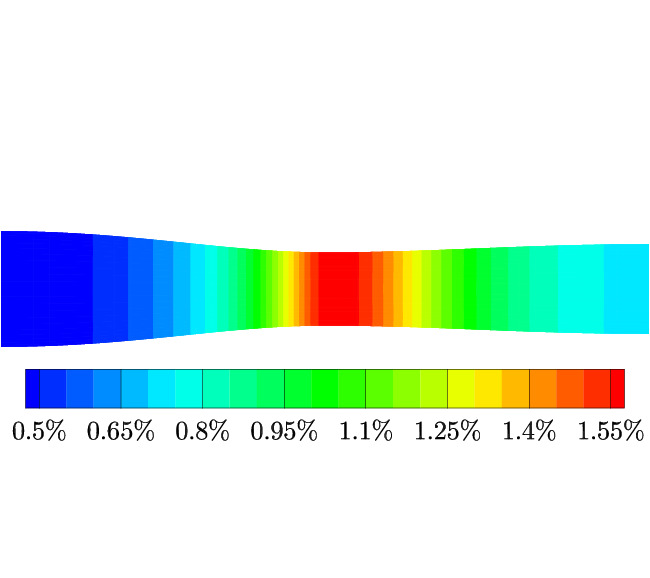}}
\\
\subfloat[\textit{y}-Velocity]{\includegraphics[width=0.47\textwidth,trim=0cm 4cm 0cm 7cm,clip]{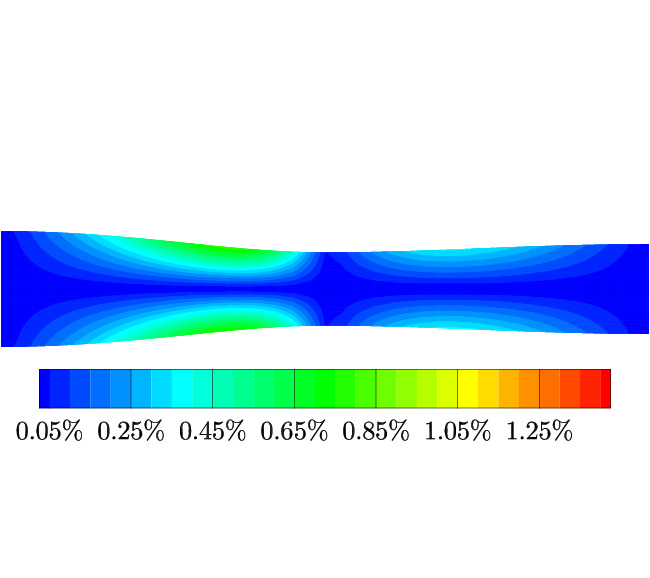}}
\hspace{0.02\textwidth}
\subfloat[Pressure]{\includegraphics[width=0.47\textwidth,trim=0cm 4cm 0cm 7cm,clip]{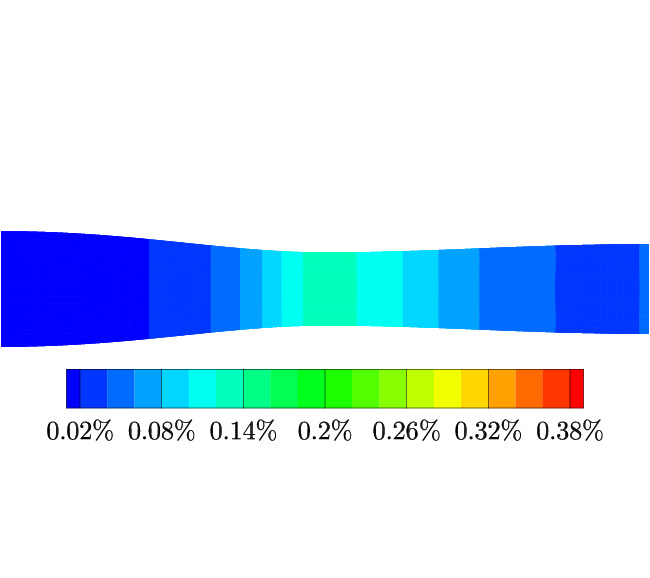}}
\caption{Time-averaged relative error between the ROM off-reference solution and the FOM solution for an axisymmetric nozzle with outlet pressure oscillation with angular velocity $\freq=10$ rad/sec and amplitude $\amp=0.015$.}
\label{fig:3dnozzle_case4}
\end{figure}

Figure~\ref{fig:3dnozzle_case4} shows the time-averaged relative error of the ROM solution with respect to the FOM solution for case 4. 
The FOM solution was only generated for the purpose of comparison, and its snapshots were not used in the development of the autocorrelation matrix. For this case, artificial dissipation was only applied to the velocity components. The ROM used the following number of modes: $n^{\zeta}=n^{p}=1$ and $n^{u}=n^{v}=n^{w}=3$. There was good agreement between the time-averaged ROM and FOM solutions. The largest error occurred in the \textit{x}-velocity and was less than 1.6\%. 

\begin{figure}[ht!]
\centering
\subfloat[Specific Volume]{\includegraphics[width=0.47\textwidth,trim=0cm 4cm 0cm 7cm,clip]{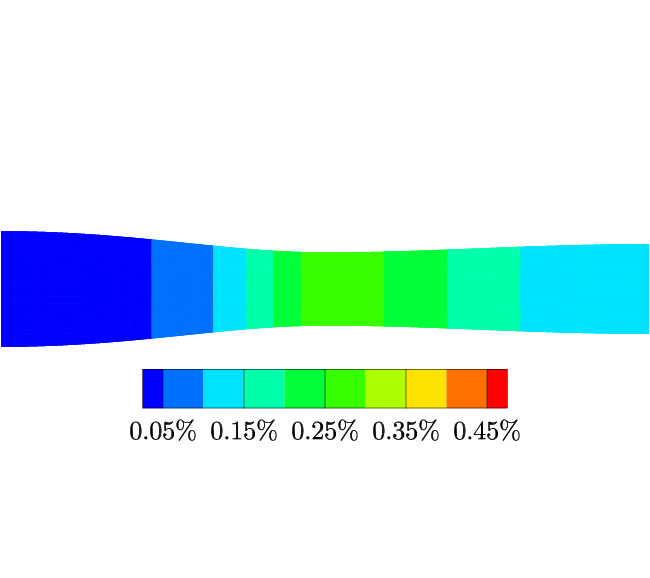}}
\hspace{0.02\textwidth}
\subfloat[\textit{x}-Velocity]{\includegraphics[width=0.47\textwidth,trim=0cm 4cm 0cm 7cm,clip]{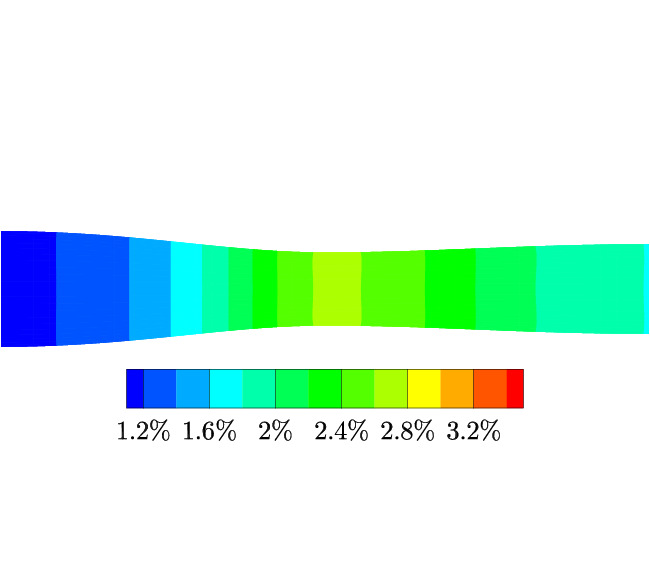}}
\\
\subfloat[\textit{y}-Velocity]{\includegraphics[width=0.47\textwidth,trim=0cm 4cm 0cm 7cm,clip]{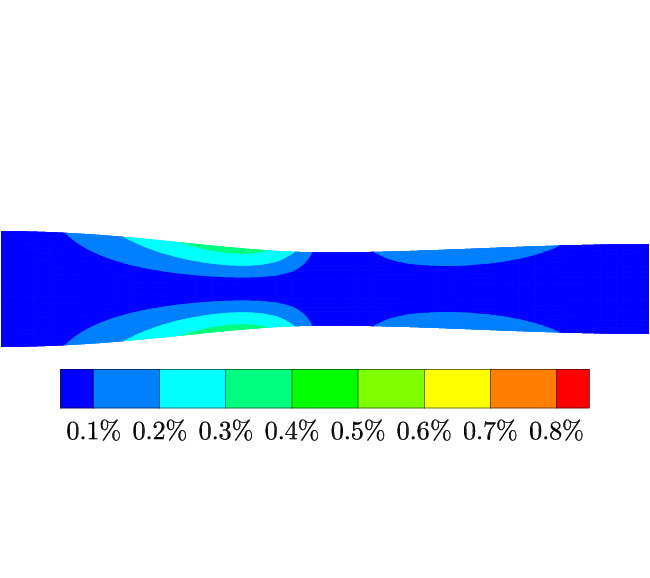}}
\hspace{0.02\textwidth}
\subfloat[Pressure]{\includegraphics[width=0.47\textwidth,trim=0cm 4cm 0cm 7cm,clip]{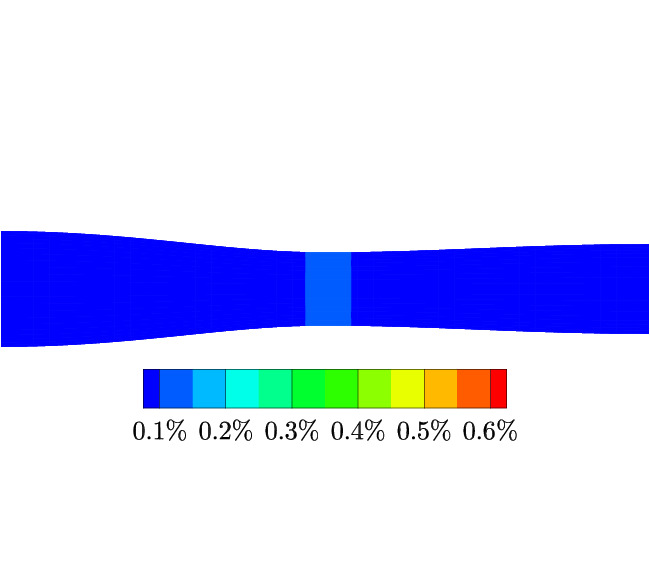}}
\caption{Time-averaged relative  error between the ROM off-reference solution and the FOM solution for an axisymmetric nozzle with outlet pressure oscillation with angular velocity $\freq=15$~rad/sec and amplitude $\amp=0.01$.}
\label{fig:3dnozzle_case5}
\end{figure}

Figure~\ref{fig:3dnozzle_case5} shows the time-averaged relative error of the ROM solution with respect to the FOM solution for case 5. The ROM used the following number of modes: $n^{\zeta}=n^{p}=1$ and $n^{u}=n^{v}=n^{w}=2$. Similarly to case 4, artificial dissipation was only applied to the velocity components. 
The ROM solution agreed well with the FOM solution, the largest error being in \textit{x}-velocity.  

Table~\ref{t:3dnozzle_maxerr_pressure_off} summarizes the maximum spatial and temporal relative error for the off-reference cases.
\begin{table}[ht!]
\centering
\begin{tabular}{cccccc}
\toprule
& \multicolumn{5}{c}{$\max_{x,t} (\epsilon)$ (\%)} \\
Case & $\zeta$ & $u$ & $v$ & $w$ & $p$ \\
\midrule
4 & 0.92 & 4.32 & 1.49 & 1.44 & 1.24 \\ 
5 & 0.64 & 4.96 & 1.00 & 0.62 & 0.91 \\ 
\bottomrule
\end{tabular}
\caption{Axisymmetric nozzle with outlet pressure oscillation: maximum spatial and temporal relative error between the ROM and FOM solutions, for all off-reference cases.}
\label{t:3dnozzle_maxerr_pressure_off}
\end{table}
The largest error occurred in \textit{x}-velocity and was less than 5\% for both cases.  All other variables had errors less than 2\%. 

\subsubsection{Deforming Boundary}

Two cases were used to validate the off-reference zeta-variable ROM with deforming boundary.  Table~\ref{t:3dnozzle_deform_allcases} shows these two cases in addition to the on-reference cases analyzed in section~\ref{sec:axid} for an oscillating boundary deformation~\eqref{rforced}.
\begin{table}[ht!]
\centering
\begin{tabular}{cccc}
\toprule
Case & $\freq$ [rad/sec] & $\amp$ & Type \\
\midrule
1 & 10 & 0.2 & on-reference \\
2 & 10 & 0.4 & on-reference \\
3 & 20 & 0.2 & on-reference \\
4 & 10 & 0.3 & off-reference \\
5 & 15 & 0.2 & off-reference \\
\bottomrule
\end{tabular}
\caption{On- and off-reference cases for the axisymmetric nozzle with an oscillating boundary deformation.}
\label{t:3dnozzle_deform_allcases}
\end{table}

Grassmann interpolation was used to find the interpolated basis. Cases 1 and 2 were used to develop the interpolated basis functions for case 4. Cases 1 and 3 were used to find the interpolated basis functions for case 5. 


The penalty method was applied at the outlet using an outlet approximation~\eqref{approxeq} for each variable as the prescribed boundary condition, similarly to what was done for the on-reference cases. Unlike the on-reference cases, the approximate values, which were found from the FOM solution, cannot be directly computed using~\eqref{Zhat}, \eqref{Ahat}, and \eqref{phihat}. Therefore, the approximate values, $\hat{q}_k$, $\hat{A}_k$, and $\hat{\varphi}_k$, are found by interpolating from the on-reference solutions.

As noted in section~\ref{sec:axid}, $\hat{Z}_{k}$ are almost independent of
the case number, while the amplitudes $\hat{A}_{k}$ scale with the actual
amplitudes
$\amp_{k}$. 
Table~\ref{t:3dnozzle_approxcases_off} shows the mean values $\hat{Z}_{k}$,
$\hat{A}_{k}$, and $\hat{\varphi}_k$ for cases 4 and 5.  The table includes
``exact'' values, that is, derived directly from respective FOM solutions
evaluated at the amplitudes and angular velocities of cases 4 and 5, and
values scaled from case 1. The values scaled from case 1 matched well the
``exact'' values.
\begin{table}[ht!]
\centering
\begin{tabular}{c|cc|cc}
\toprule
~ & \multicolumn{4}{c}{Case} \\
& \multicolumn{2}{c}{Exact (FOM)} & \multicolumn{2}{c}{Scaled} \\
& 4 & 5 & 4 & 5 \\
\midrule
 $\hat{\zeta}$ & 1.00 & 1.00 & 1.00 & 1.00 \\
  $\hat{A}_\zeta \times 10^{-5}$ & 2.26 & 1.70 & 2.28 & 1.52 \\
 $\hat{\varphi}_\zeta $ & 11.81 & 17.63 & -0.75 & -0.90 \\
\midrule
 $\hat{u}$ & 0.16 & 0.16 & 0.16 & 0.16 \\
 $\hat{A}_u \times 10^{-3}$ & 1.23 & 0.66 & 1.21 & 0.81 \\
 $\hat{\varphi}_u$ & -2.58 & 41.54 & 3.71 & 2.26 \\
\midrule
 $\hat{v}$ $\times 10^{-6}$ & -5.05 & -5.05 & -5.11 & -5.11 \\
 $\hat{A}_v$ & -0.52 & -0.38 & -0.51 & -0.34 \\
 $\hat{\varphi}_v$ & 3.08 & 40.42 & 18.82 & -3.58 \\
\midrule
 $\hat{w}$ $\times 10^{-6}$ & -5.82 & -5.87 & -5.94 & -5.94 \\
 $\hat{A}_w$ & -0.42 & -0.30 & -0.41 & -0.27 \\
 $\hat{\varphi}_w$ & 22.05 & 27.95 & 15.77 & 8.29 \\
\midrule
 $\hat{p}$ & 0.72 & 0.72 & 0.72 & 0.72 \\
 $\hat{A}_p \times 10^{-5}$ & 1.41 & 0.96 & 1.46 & 0.97 \\
 $\hat{\varphi}_p$ & 21.99 & 9.10 & 9.39 & 17.67 \\
  \bottomrule
\end{tabular}
\caption{Axisymmetric nozzle with oscillating boundary deformation: exact and scaled mean values of state variables, amplitudes and phase angles.}
\label{t:3dnozzle_approxcases_off}
\end{table}

The phase angle for case 4 was determined by linear interpolation between cases 1 and 2.  Similarly, the phase angle for case 5 was determined by linear interpolation between cases 1 and 3.
The penalty parameter $\pmb{\uptau}$ was calculated by solving~\eqref{tausecant}. 
To generate the off-reference solutions, the ROM used the following modes: $n^{\zeta}=n^{u}=n^{v}=n^{w}=n^{p}=1$ for all cases.

\begin{figure}[ht!]
\centering
\subfloat[Specific Volume]{\includegraphics[width=0.47\textwidth,trim=0cm 4cm 0cm 6cm,clip]{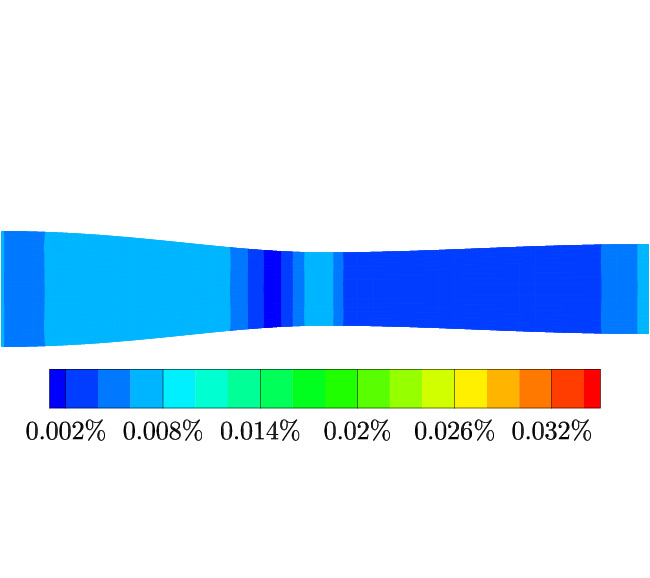}}
\hspace{0.02\textwidth}
\subfloat[\textit{x}-Velocity]{\includegraphics[width=0.47\textwidth,trim=0cm 4cm 0cm 6cm,clip]{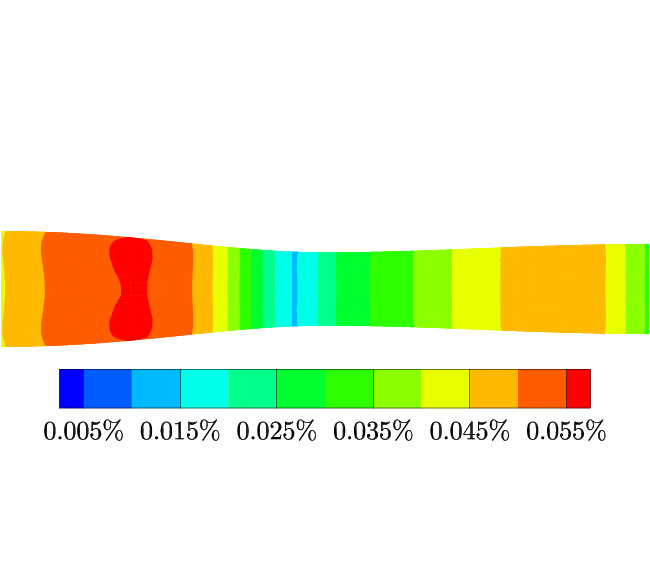}}
\\
\subfloat[\textit{y}-Velocity]{\includegraphics[width=0.47\textwidth,trim=0cm 4cm 0cm 6cm,clip]{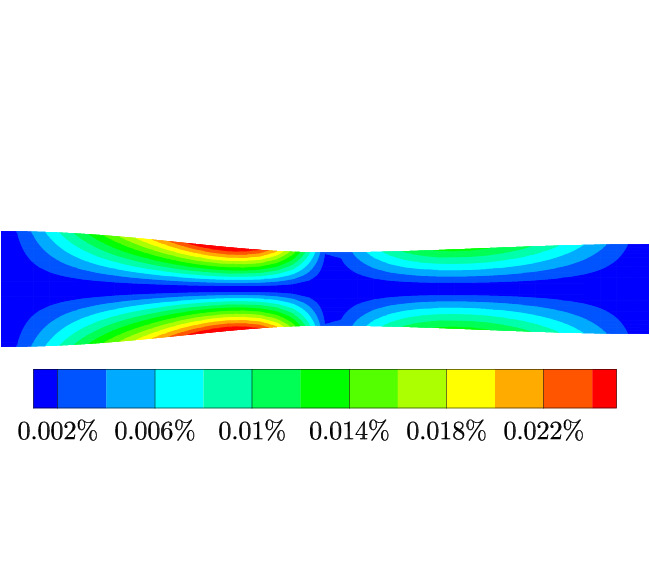}}
\hspace{0.02\textwidth}
\subfloat[Pressure]{\includegraphics[width=0.47\textwidth,trim=0cm 4cm 0cm 6cm,clip]{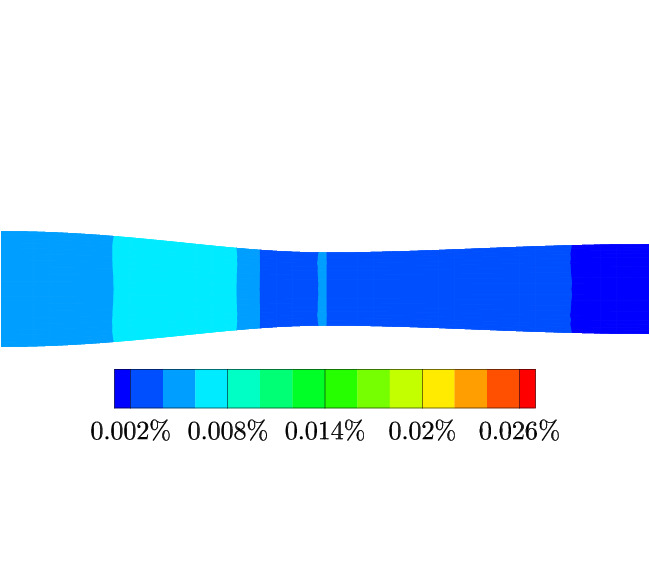}}
\caption{Time-averaged relative error between the ROM off-reference and FOM solutions for an axisymmetric nozzle with walls oscillating with angular velocity $\freq = 10$~rad/sec and amplitude $\amp = 0.3$.}
\label{fig:3dnozzle_deform_case4}
\end{figure}

\begin{figure}[ht!]
\centering
\subfloat[Specific Volume]{\includegraphics[width=0.47\textwidth,trim=0cm 4cm 0cm 6cm,clip]{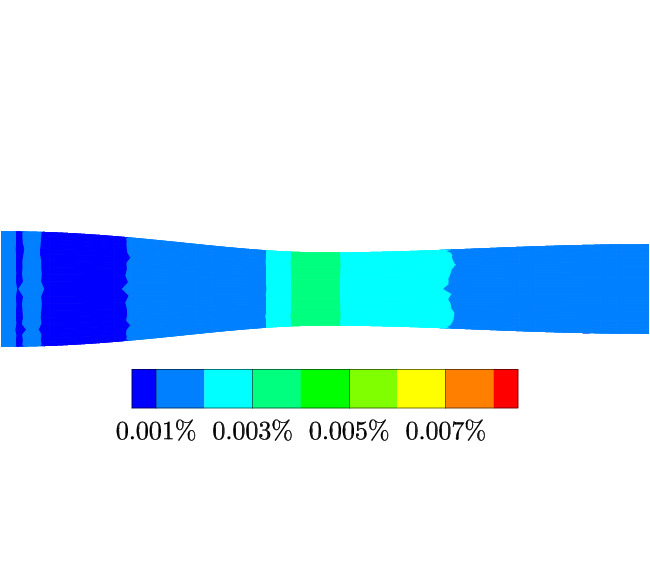}}
\hspace{0.02\textwidth}
\subfloat[\textit{x}-Velocity]{\includegraphics[width=0.47\textwidth,trim=0cm 4cm 0cm 6cm,clip]{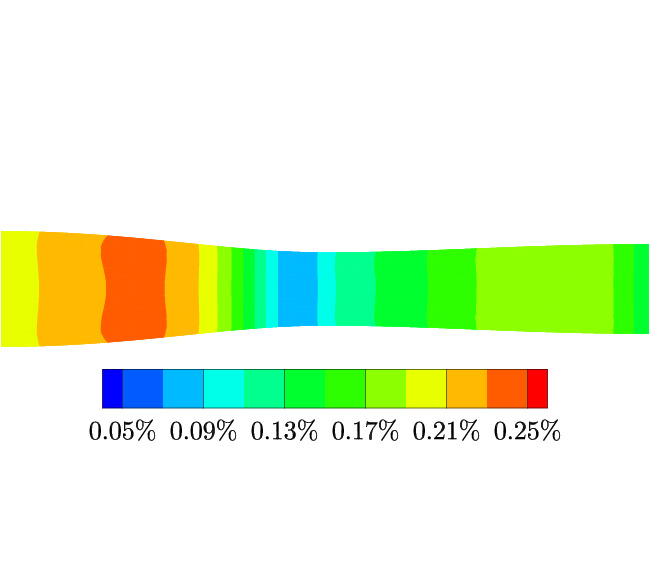}}
\\
\subfloat[\textit{y}-Velocity]{\includegraphics[width=0.47\textwidth,trim=0cm 4cm 0cm 6cm,clip]{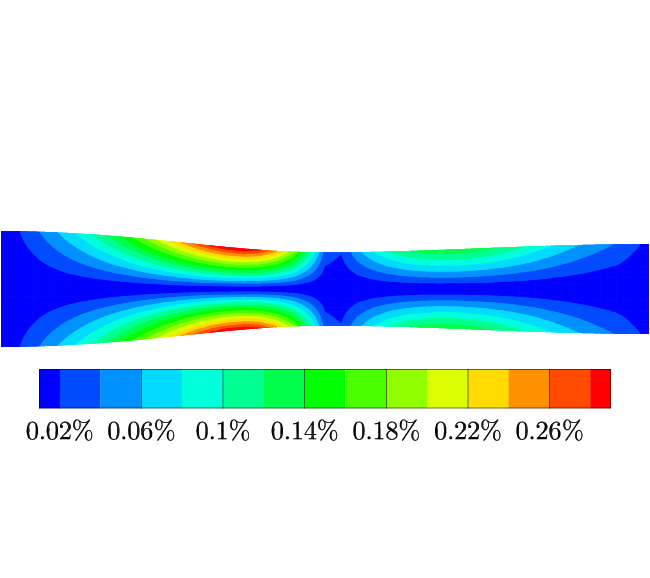}}
\hspace{0.02\textwidth}
\subfloat[Pressure]{\includegraphics[width=0.47\textwidth,trim=0cm 4cm 0cm 6cm,clip]{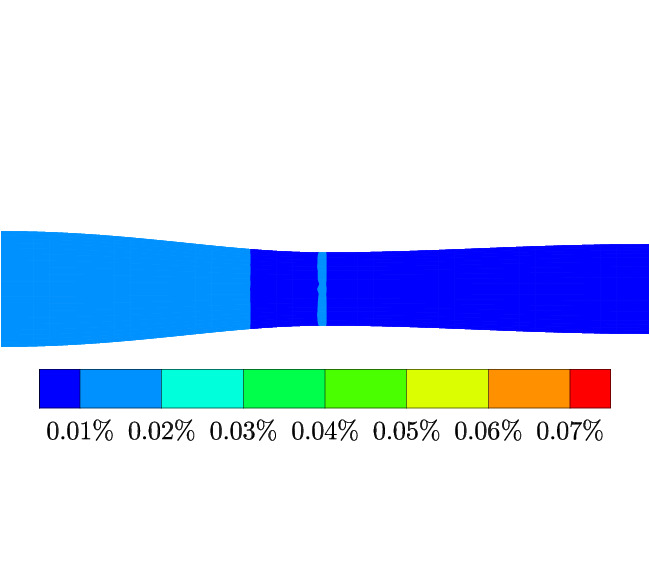}}
\caption{Time-averaged relative error between the ROM off-reference and FOM solutions for an axisymmetric nozzle with walls oscillating with angular velocity $\freq = 15$~rad/sec and amplitude $\amp = 0.2$.}
\label{fig:3dnozzle_deform_case5}
\end{figure}

Figures~\ref{fig:3dnozzle_deform_case4} and \ref{fig:3dnozzle_deform_case5} show the time-averaged relative errors of the ROM solution with respect to the FOM solution for cases 4 and 5, respectively.
In both cases, the off-reference ROM solution matched well the FOM solution, the time-averaged relative error being less than 0.1\%.

Table~\ref{t:3dnozzle_maxerr_deform_off} shows the maximum spatial and temporal relative error for both cases 4 and 5.  The maximum relative error between the off-reference ROM and the FOM solutions was less than 1\%. 
\begin{table}[ht!]
\centering
\begin{tabular}{cccccc}
\toprule
& \multicolumn{5}{c}{$\max_{x,t} (\epsilon)$ (\%)} \\
Case & $\zeta$ & $u$ & $v$ & $w$ & $p$ \\
\midrule
4 & 0.05 & 0.12 & 0.96 & 0.48 & 0.06 \\ 
5 & 0.04 & 0.24 & 0.48 & 0.31 & 0.11 \\ 
\bottomrule
\end{tabular}
\caption{Axisymmetric nozzle with oscillating walls: maximum spatial and temporal relative error between the ROM and FOM solutions.}
\label{t:3dnozzle_maxerr_deform_off}
\end{table}
%


\subsection{NASA Rotor 67}

Two cases were used to validate the ROM for Rotor 67 at off-reference conditions.  Table~\ref{t:r67_allcases} shows these two off-reference cases in addition to the three on-reference cases analyzed in section~\ref{sec:r67}.
\begin{table}[ht!]
\centering
\begin{tabular}{cccc}
\toprule
Case & $\freq$ [rad/sec] & $\amp$ & Type \\
\midrule
1 & 38,059 & 0.050 & on-reference \\
2 & 38,059 & 0.100 & on-reference \\
3 & 37,582 & 0.050 & on-reference \\
4 & 38,059 & 0.075 & off-reference \\
5 & 37,765 & 0.050 & off-reference \\
\bottomrule
\end{tabular}
\caption{NASA Rotor 67: on- and off-reference cases for oscillating outlet pressure.}
\label{t:r67_allcases}
\end{table}

The Grassmann interpolation was used to interpolate the basis functions to a new flow condition. Cases 1 and 2 were used to find the interpolated basis of case 4. Cases 1 and 3 were used to find the interpolated basis of case 5. 

The penalty method was imposed similarly to the way it was applied to the on-reference cases. The approximate values of the amplitudes were scaled based on case 1 values given in Table~\ref{t:r67_allcases}. The approximate values of the phase angles of case 4 were scaled based on the phase angles of cases 1 and 2.  The phase angles of case 5 were scaled based on the phase angles of cases 1 and 3.

Table~\ref{t:r67_approx_off} shows the mean values $\hat{Z}_{k}$, $\hat{A}_{k}$ and $\hat{\varphi}_k$ for cases 4 and 5.  The table includes  ``exact'' values, that is, derived directly from respective FOM solutions evaluated at the amplitudes and angular velocities of cases 4 and 5, and values scaled from case 1. The values scaled from case 1 matched well with the ``exact'' values.  The phase angle was interpolated using a linear interpolation.
\begin{table}[ht!]
\centering
\begin{tabular}{c|cc|cc}
\toprule
~ & \multicolumn{4}{c}{Case} \\
~ & \multicolumn{2}{c}{Exact (FOM)} & \multicolumn{2}{c}{Scaled} \\
& 4 & 5 & 4 & 5 \\
\midrule
 $\hat{u}$ & 0.50 & 0.50 & 0.50 & 0.50 \\
 $\hat{A}_u \times 10^{-2}$ & 1.43 & 0.96 & 1.43 & 0.97 \\
 $\hat{\varphi}_u$ & -13.83 & 5.36 & -13.83 & -17.48 \\
\midrule
 $\hat{v}$ & -0.67 & -0.67 & -0.67 & -0.67 \\
 $\hat{A}_v \times 10^{-4}$ & -3.34 & -2.25 & -3.33 & -2.27 \\
 $\hat{\varphi}_v$ & 4.53 & -13.97 & 4.53 & 0.80 \\
\midrule
 $\hat{w}$ $\times 10^{-6}$ & 0.12 & 0.12 & 0.12 & 0.12 \\
 $\hat{A}_w \times 10^{-2}$ & 1.09 & 0.74 & 1.09 & 0.74 \\
 $\hat{\varphi}_w$ & -19.99 & -0.87 & -19.99 & -19.82 \\
\midrule
 $\hat{p}$ & 1.08 & 1.08 & 1.08 & 1.08 \\
 $\hat{A}_p \times 10^{-3}$ & 7.77 & 5.22 & 7.77 & 5.22 \\
 $\hat{\varphi}_p$ & -23.13 & -16.59 & -23.13 & -7.41 \\
  \bottomrule
\end{tabular}
\caption{NASA Rotor 67: exact and scaled mean values of state variables, amplitudes and phase angles.}
\label{t:r67_approx_off}
\end{table}

\begin{figure}[ht!]
	\centering
	\includegraphics[draft=false,width=0.85\textwidth,trim=0cm 6cm 0cm 0cm,clip]{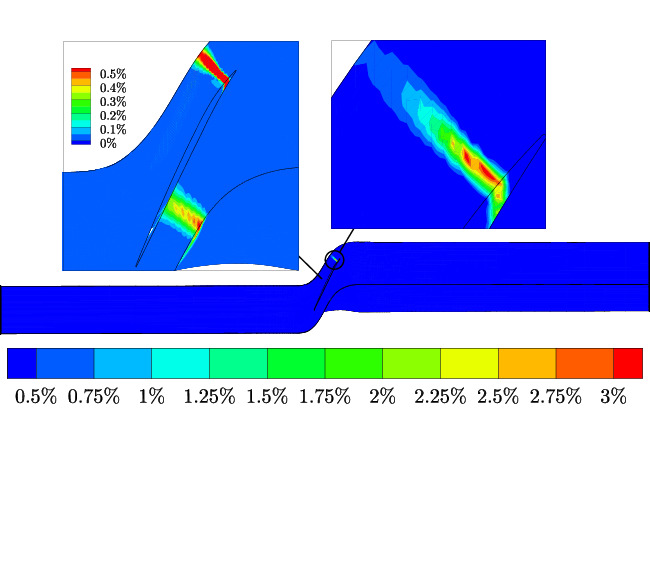}
	\caption{NASA Rotor 67: time-averaged relative error between the off-reference ROM and FOM pressures for an outlet pressure oscillating with angular velocity $\freq=38,059$ rad/sec and amplitude $\amp=0.075$. A close-up of the error around the blade is shown at blade midspan while a close-up of the error around the shock is shown at around 95\% blade span location measured from the hub.  Note that the legend for the midspan contour plots is different from the other two plots.}
	\label{fig:r67_case4_res}
\end{figure}

The off-reference ROM used the following number of modes: $n^{\zeta}=n^{u}=n^{v}=n^{w}=n^{p}=1$.  A FOM solution was generated for assessing the accuracy of the ROM solution.  This FOM solution was not used to generate basis functions. Figure~\ref{fig:r67_case4_res} shows the time-averaged relative error between the ROM and FOM pressures, for case 4. Pressure had the largest error, approximately 3\%, that occurred near the tip of the airfoil where a shock developed.
%

\begin{figure*}
  	\centering
  	\setcounter{subfigure}{-2}
  	\subfloat[Specific Volume]{\subfloat{\includegraphics[draft=false,width=0.55\textwidth,trim=0cm 4.5cm 0cm 7.5cm, clip]{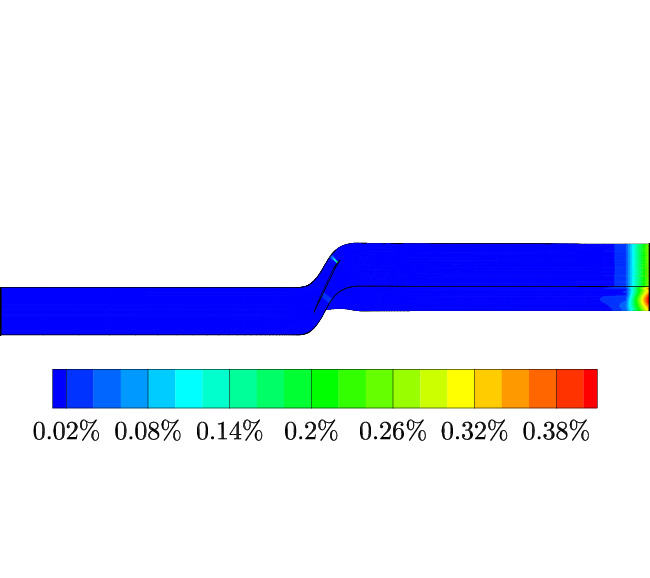}} \subfloat{\includegraphics[draft=false,width=0.2\textwidth]{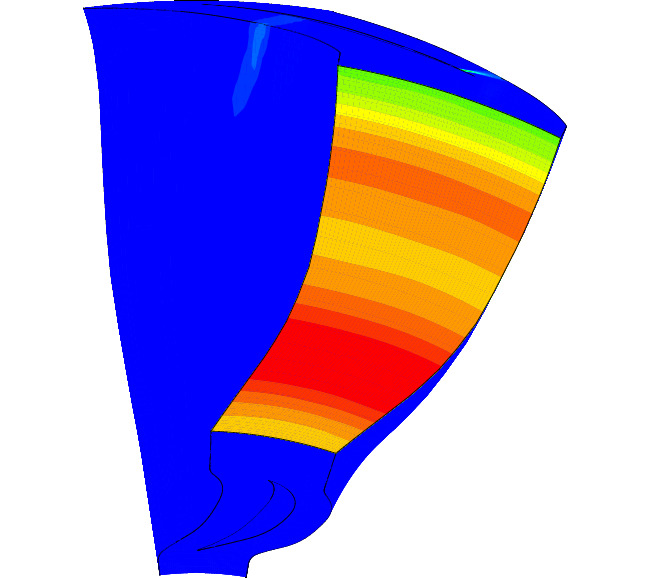}}}
  	\\
  	\setcounter{subfigure}{-1}
  	\subfloat[\textit{x}-Velocity]{\subfloat{\includegraphics[draft=false,width=0.55\textwidth,trim=0cm 4.5cm 0cm 7.5cm, clip]{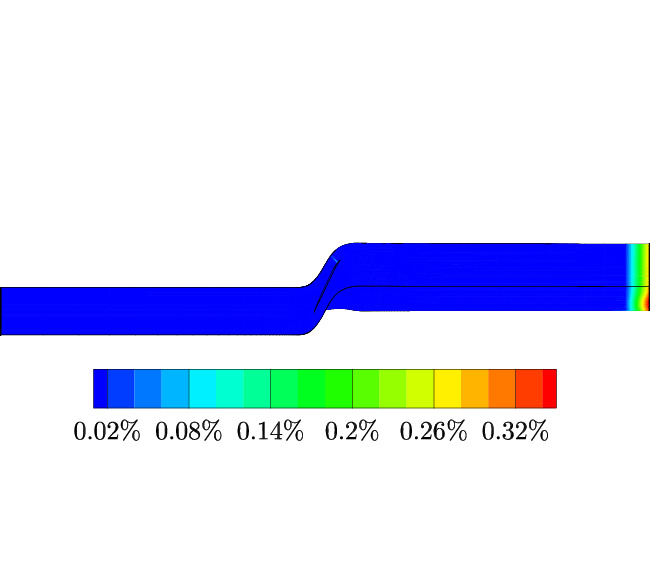}} \subfloat{\includegraphics[draft=false,width=0.2\textwidth]{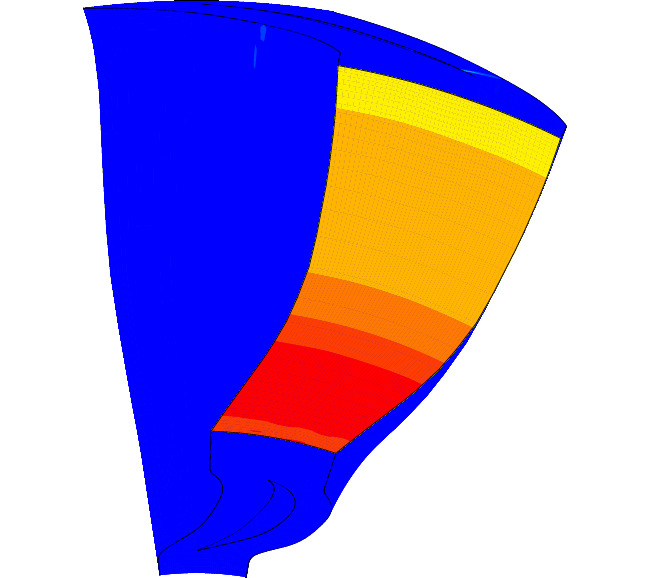}}} 
  	\\
  	\setcounter{subfigure}{0}
  	\subfloat[\textit{y}-Velocity]{\subfloat{\includegraphics[draft=false,width=0.55\textwidth,trim=0cm 4.5cm 0cm 7.5cm, clip]{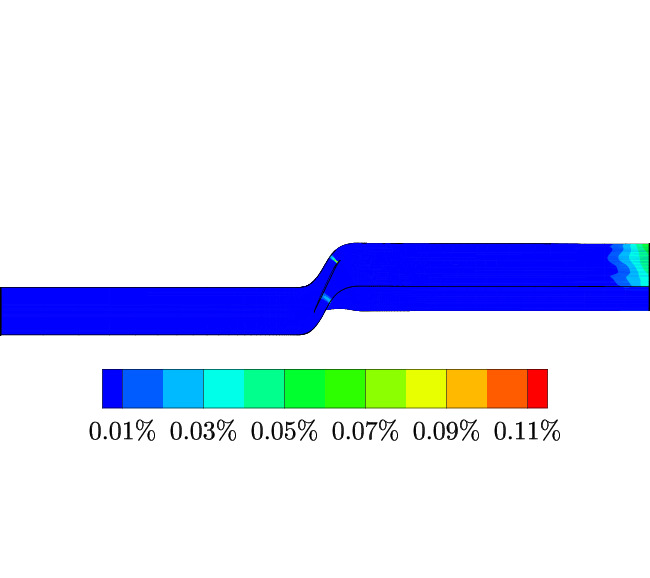}} \subfloat{\includegraphics[draft=false,width=0.2\textwidth]{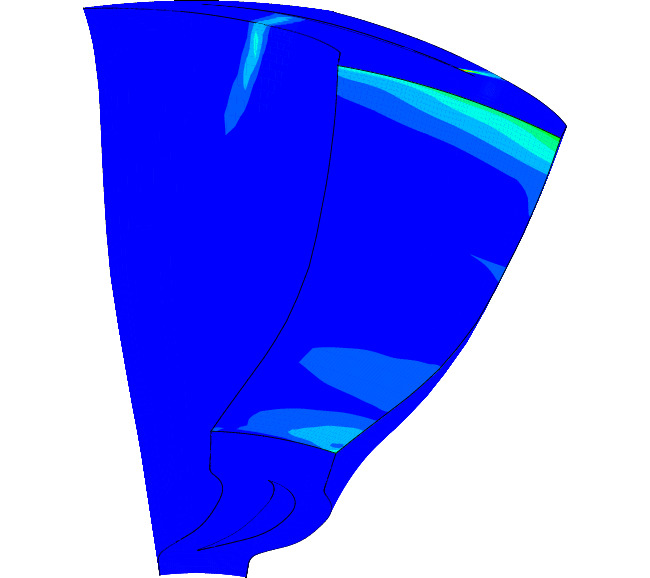}}}
  	\\
  	\setcounter{subfigure}{1}
  	\subfloat[\textit{z}-Velocity]{\subfloat{\includegraphics[draft=false,width=0.55\textwidth,trim=0cm 4.5cm 0cm 7.5cm, clip]{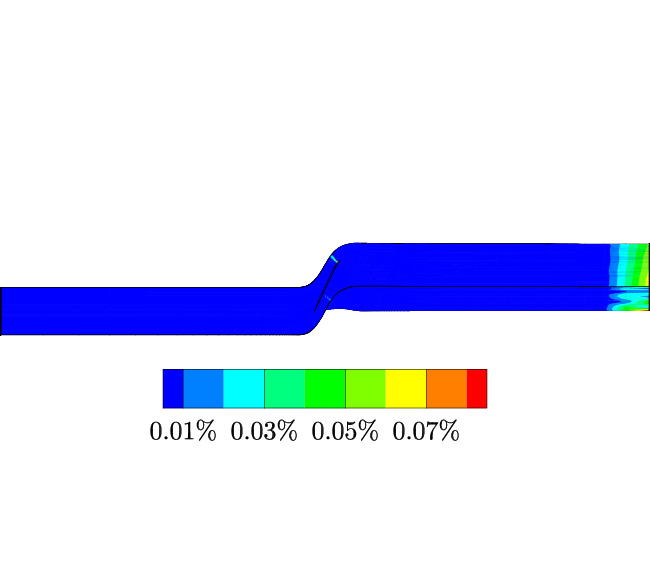}} \subfloat{\includegraphics[draft=false,width=0.2\textwidth]{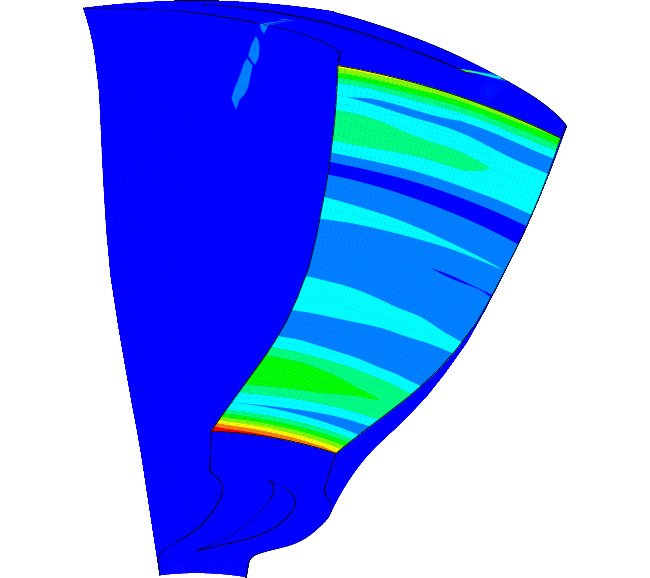}}} 
  	\\
  	\setcounter{subfigure}{2}
  	\subfloat[Pressure]{\subfloat{\includegraphics[draft=false,width=0.55\textwidth,trim=0cm 4.5cm 0cm 7.5cm, clip]{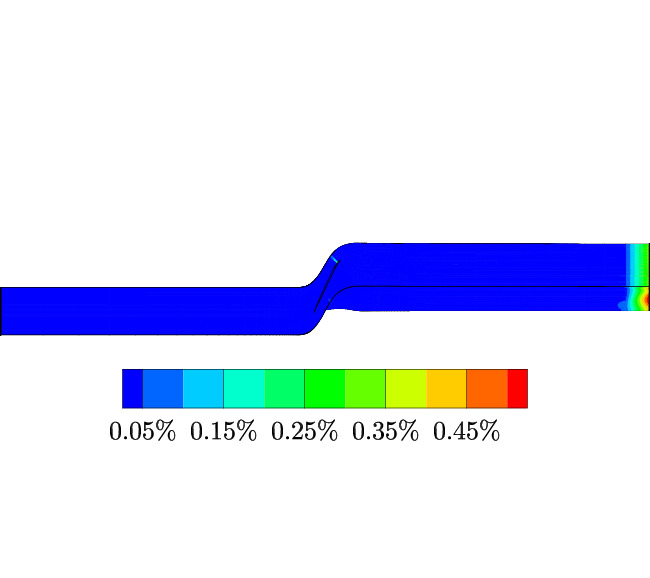}} \subfloat{\includegraphics[draft=false,width=0.2\textwidth]{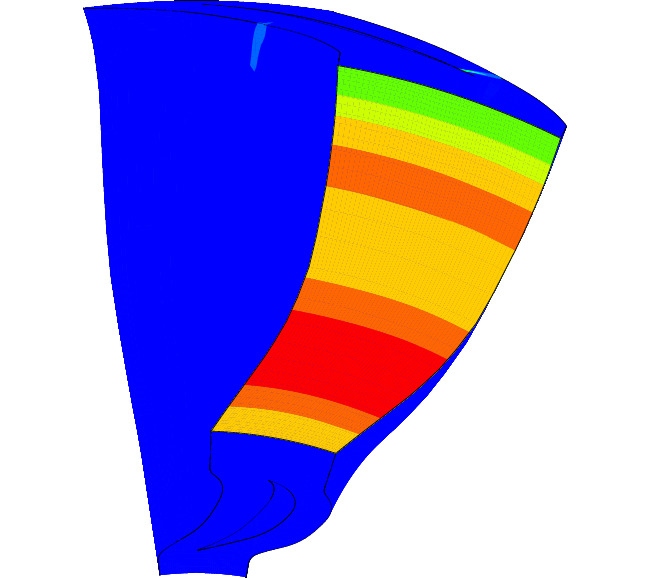}}}
\caption{NASA Rotor 67: time-averaged relative error between the off-reference ROM and FOM state variables, for an outlet pressure oscillating with angular velocity $\freq=37,765$~rad/sec and amplitude $\amp=0.05$.}
\label{fig:r67_case5_res}
\end{figure*}

Figure~\ref{fig:r67_case5_res} shows the time-averaged relative error
between the off-reference ROM and FOM state variables for case 5.  The
largest errors occurred at the outlet, as opposed to the shock location, and
their values were smaller than 0.5\%.

Table~\ref{t:r67_maxerr_off} gives the maximum spatial and temporal relative error between the off-reference ROM and FOM state variables, for cases 4 and 5.  The small error implies that the zeta-variable ROM predicted accurately off-reference conditions even when using approximate equations boundary conditions.  The zeta-variable ROM captured well the shock wave that formed on the suction side of the blade. 
\begin{table}[ht!]
\centering
\begin{tabular}{cccccc}
\toprule
& \multicolumn{5}{c}{$\max_{x,t} (\epsilon)$ (\%)} \\
Case & $\zeta$ & $u$ & $v$ & $w$ & $p$ \\
\midrule
4 & 1.85 & 0.97 & 1.26 & 0.82 & 3.46 \\ 
5 & 0.57 & 0.41 & 0.12 & 0.17 & 0.68 \\ 
\bottomrule
\end{tabular}
\caption{NASA Rotor 67: maximum spatial and temporal relative error between the off-reference ROM and FOM state variables.}
\label{t:r67_maxerr_off}
\end{table}

\section{Computational Time}

This section presents a comparison between the runtimes of the zeta-variable
ROM and the FOM solvers. 
All cases were ran on a Mac Pro with two 2.26 GHz Quad-Core Intel Xeon
processors with 24 GB of RAM. The FOM cases were ran using 6 cores.

\begin{table}[ht!]
\centering
\begin{tabular}{ccccc}
\toprule
Case & $N$ & FOM [s] & ROM [s] & Ratio \\
\midrule
2D Channel & 8514 & 15,663 & 0.30 & 52,211 \\ 
3D Nozzle & 3045 & 8778 & 0.60 & 14,630 \\ 
3D Nozzle & 22,673 & 506,635 & 5.38 & 86,901 \\ 
\bottomrule
\end{tabular}
\caption{CPU runtime comparison between the ROM and the FOM for two on-reference cases: (1) 2D channel with 7 modes, and (2) 3D nozzle with 9 modes.}
\label{t:CPU_ROMFOM}
\end{table}

Table~\ref{t:CPU_ROMFOM} shows a comparison between the CPU runtime of the zeta-variable ROM and the FOM for two cases: a two-dimensional channel flow and a three-dimensional nozzle.  The FOM solutions were generated using the UNS3D flow solver~\cite{kirketal07}.  The FOM solution of the two-dimensional channel was obtained running UNS3D as an implicit solver, while the three-dimensional nozzle solution was obtained running UNS3D as an explicit solver.  The FOM and the ROM were run such that they both predicted the flow over the same time interval.  It is apparent from Table~\ref{t:CPU_ROMFOM}  that (i) the ROM is more than four orders of magnitude faster than the FOM, and (ii) the speedup increases as the number of grid points increases.  The computational time needed to generate the POD basis functions was not included in this comparison.


%
\begin{table}[ht!]
\centering
\begin{tabular}{cccc}
\toprule
Time periods & FOM [s] & ROM [s] & Ratio \\
\midrule
3 & 15,663 & 0.30 & 52,211 \\ 
6 & 31,744 & 0.38 & 83,537 \\ 
12 & 64,616 & 0.56 & 115,386 \\ 
\bottomrule	
\end{tabular}
\caption{CPU runtime comparison between the FOM and the ROM for on-reference conditions with increasing lengths of runtime for a two-dimensional channel case with 8514 grid points and the following modes: $n^{\zeta}=1, n^{u}=n^{v}=n^{p}=2$.}
\label{t:CPU_ROMFOMtime}
\end{table}
Table~\ref{t:CPU_ROMFOMtime} shows the variation of the speedup ratio vs. the length of the time interval that is being simulated.  The results presented in this table correspond to the two-dimensional channel described in section~\ref{sec:2dc}.  The FOM solution was generated using the implicit version of the UNS3D solver.
Compared to the FOM, the zeta-variable ROM becomes more efficient as the length of the time interval increases.

\section{Conclusions}

A zeta-variable POD-based ROM has been developed and implemented for
modeling unsteady flows with or without moving boundaries.  The POD model
has been developed starting from the governing equations instead of the
discretized equations used by the FOM.  Consequently the ROM did not inherit
any of the approaches used in the FOM to generate a stable solution.  As a
result, several methods had to be pursued to obtain a stable ROM solution:
(i) the penalty method, (ii) artificial dissipation, and (iii) a variable
number of POD modes.  This approach produced a robust ROM that accurately
predicted the flow in nozzles and turbomachinery cascades.

By using the zeta variables, as opposed to the conservative or primitive variables, the coefficients of the ODE system could be precomputed and the nonlinearity of the system of ODEs was reduced.  Therefore, the computational cost of this zeta-variable POD-based ROM was more than four orders of magnitude smaller than that of the FOM.  This speedup increased as the grid size and the length of the time interval the flow was predicted increased.

\section*{Acknowledgments}
The authors would like to thank Dr. Andrew Gyekenyesi of the Ohio Aero\-spa\-ce Institute (OAI) for funding this project via a subcontract to OAIs Versatile Advanced Affordable Turbine Engines (VAATE) III, FA8650-14-D-2410, issued by USAF/AFMC, AFRL Wright Research Site. The authors gratefully acknowledges the help from James Felderhoff, who created the quasi-one-dimensional full-order model during his undergraduate research. The authors also gratefully acknowledges the feedback from Neil Matula, Ph.D. over ideas and methods within the paper.  The authors gratefully acknowledge the computational resources provided by the Texas A\&M Supercomputing Facility. This material is declared a work of the U. S. Government and is not subject to copyright protection in the United States. Approved for public release; distribution unlimited.

\section*{References}
\bibliography{jcp}

\begin{thebibliography}{10}
\expandafter\ifx\csname url\endcsname\relax
  \def\url#1{\texttt{#1}}\fi
\expandafter\ifx\csname urlprefix\endcsname\relax\def\urlprefix{URL }\fi
\expandafter\ifx\csname href\endcsname\relax
  \def\href#1#2{#2} \def\path#1{#1}\fi

\bibitem{ParkLee1998}
H.~M. Park, M.~W. Lee, An efficient method of solving the {N}avier-{S}tokes
  equations for flow control, International Journal for Numerical Methods in
  Engineering 41~(6) (1998) 1133--1151.

\bibitem{Slotnick2014}
J.~Slotnick, A.~Khodadoust, J.~Alonso, D.~Darmofal, W.~Gropp, E.~Lurie,
  D.~Mavriplis, \href{http://ntrs.nasa.gov/search.jsp?R=20140003093}{{CFD
  Vision 2030 Study: A Path to Revolutionary Computational Aerosciences}},
  {NASA} {CR}-2014-21878 (2014) 1--73\href
  {http://arxiv.org/abs/arXiv:1011.1669v3} {\path{arXiv:arXiv:1011.1669v3}},
  \href {http://dx.doi.org/10.1017/CBO9781107415324.004}
  {\path{doi:10.1017/CBO9781107415324.004}}.
\newline\urlprefix\url{http://ntrs.nasa.gov/search.jsp?R=20140003093}

\bibitem{Karhunen1946}
K.~Karhunen, {Zur Spektraltheorie Stochasticher Prozesse}, Annales Academiae
  Scientiarum Fennicae Series A 1 (1946) 34.

\bibitem{1945Loeve}
M.~Lo\`{e}ve, Functions al\'{e}atoire de second order, Comptes Rendus de
  l'Acad\'{e}mie des Sciences, Paris (1945) 220.

\bibitem{Lumley1967}
J.~Lumley, The structure of inhomogenous turbulence, in: A.~Yaglom (Ed.),
  Proceedings of the International Colloquium in the Fine Scale Structure of
  the Atmosphere and its Influence on Radio Wave Propagation, Dokl Akad Nauk
  SSSR, 1967.

\bibitem{Epureanu2000}
B.~Epureanu, E.~Dowell, K.~Hall,
  \href{http://www.sciencedirect.com/science/article/pii/S0889974600903207}{Reduced-order
  models of unsteady transonic viscous flows in turbomachinery}, Journal of
  Fluids and Structures 14~(8) (2000) 1215 -- 1234.
\newblock \href {http://dx.doi.org/https://doi.org/10.1006/jfls.2000.0320}
  {\path{doi:https://doi.org/10.1006/jfls.2000.0320}}.
\newline\urlprefix\url{http://www.sciencedirect.com/science/article/pii/S0889974600903207}

\bibitem{A.Cizmas2003}
P.~G. {A. Cizmas}, A.~Palacios, {Proper Orthogonal Decomposition of Turbine
  Rotor-Stator Interaction}, Journal of Propulsion and Power 19~(2) (2003)
  268--281.
\newblock \href {http://dx.doi.org/10.2514/2.6108} {\path{doi:10.2514/2.6108}}.

\bibitem{Nagarajan2017}
K.~Nagarajan, S.~Singha, L.~Cordier, C.~Airiau, {Open-loop control of cavity
  noise using Proper Orthogonal Decomposition Reduced-Order Model}, ASME J.
  Fluids Eng. 160 (2017) 1--13.
\newblock \href {http://dx.doi.org/10.1016/j.compfluid.2017.10.019}
  {\path{doi:10.1016/j.compfluid.2017.10.019}}.

\bibitem{Yuan2005}
T.~Yuan, P.~G. Cizmas, T.~O'Brien, {A reduced-order model for a bubbling
  fluidized bed based on proper orthogonal decomposition}, Computers and
  Chemical Engineering 30~(2) (2005) 243--259.
\newblock \href {http://dx.doi.org/10.1016/j.compchemeng.2005.09.001}
  {\path{doi:10.1016/j.compchemeng.2005.09.001}}.

\bibitem{Brenner2010}
T.~A. Brenner, R.~L. Fontenot, P.~G. Cizmas, T.~J. O'Brien, R.~W. Breault,
  \href{http://dx.doi.org/10.1016/j.powtec.2010.03.032}{{Augmented proper
  orthogonal decomposition for problems with moving discontinuities}}, Powder
  Technology 203~(1) (2010) 78--85.
\newblock \href {http://dx.doi.org/10.1016/j.powtec.2010.03.032}
  {\path{doi:10.1016/j.powtec.2010.03.032}}.
\newline\urlprefix\url{http://dx.doi.org/10.1016/j.powtec.2010.03.032}

\bibitem{Rowley2005}
C.~W. Rowley, {Model reduction for fluids using balanced proper orthogonal
  decomposition}, Int. J. Bifurcation and Chaos 15~(3) (2005) 997--1013.
\newblock \href {http://dx.doi.org/10.1142/S0218127405012429}
  {\path{doi:10.1142/S0218127405012429}}.

\bibitem{SCHMID2010}
P.~J. Schmid,
  \href{http://www.journals.cambridge.org/abstract{\_}S0022112010001217}{{Dynamic
  mode decomposition of numerical and experimental data}}, Journal of Fluid
  Mechanics 656 (2010) 5--28.
\newblock \href {http://arxiv.org/abs/arXiv:1312.0041v1}
  {\path{arXiv:arXiv:1312.0041v1}}, \href
  {http://dx.doi.org/10.1017/S0022112010001217}
  {\path{doi:10.1017/S0022112010001217}}.
\newline\urlprefix\url{http://www.journals.cambridge.org/abstract{\_}S0022112010001217}

\bibitem{Aubry1991}
N.~Aubry, {On the hidden beauty of the proper orthogonal decomposition},
  Theoretical and Computational Fluid Dynamics 2~(5-6) (1991) 339--352.
\newblock \href {http://dx.doi.org/10.1007/BF00271473}
  {\path{doi:10.1007/BF00271473}}.

\bibitem{Lucia2004}
D.~J. Lucia, P.~S. Beran, W.~A. Silva,
  \href{http://dx.doi.org/10.2514/6.2001-853}{{Reduced-order modeling: New
  approaches for computational physics}}, Progress in Aerospace Sciences
  40~(1-2) (2004) 51--117.
\newblock \href {http://dx.doi.org/10.1016/j.paerosci.2003.12.001}
  {\path{doi:10.1016/j.paerosci.2003.12.001}}.
\newline\urlprefix\url{http://dx.doi.org/10.2514/6.2001-853}

\bibitem{Barone2009}
M.~F. Barone, I.~Kalashnikova, D.~J. Segalman, H.~K. Thornquist,
  \href{http://dx.doi.org/10.1016/j.jcp.2008.11.015}{{Stable Galerkin reduced
  order models for linearized compressible flow}}, Journal of Computational
  Physics 228~(6) (2009) 1932--1946.
\newblock \href {http://dx.doi.org/10.1016/j.jcp.2008.11.015}
  {\path{doi:10.1016/j.jcp.2008.11.015}}.
\newline\urlprefix\url{http://dx.doi.org/10.1016/j.jcp.2008.11.015}

\bibitem{Freno2014a}
B.~A. Freno, N.~R. Matula, R.~L. Fontenot, P.~G.~A. Cizmas, {The Use of Dynamic
  Basis Functions in Proper Orthogonal Decomposition}, 52nd Aerospace Sciences
  Meeting (2014) 1--47.

\bibitem{Bond2008}
B.~N. Bond, L.~Daniel, {Guaranteed stable projection-based model reduction for
  indefinite and unstable linear systems}, in: IEEE/ACM International
  Conference on Computer-Aided Design, Digest of Technical Papers, ICCAD, 2008,
  pp. 728--735.
\newblock \href {http://dx.doi.org/10.1109/ICCAD.2008.4681657}
  {\path{doi:10.1109/ICCAD.2008.4681657}}.

\bibitem{Amsallem2012a}
D.~Amsallem, C.~Farhat,
  \href{https://onlinelibrary.wiley.com/doi/abs/10.1002/nme.4274}{Stabilization
  of projection-based reduced-order models}, International Journal for
  Numerical Methods in Engineering 91~(4) (2012) 358--377.
\newblock \href
  {http://arxiv.org/abs/https://onlinelibrary.wiley.com/doi/pdf/10.1002/nme.4274}
  {\path{arXiv:https://onlinelibrary.wiley.com/doi/pdf/10.1002/nme.4274}},
  \href {http://dx.doi.org/10.1002/nme.4274} {\path{doi:10.1002/nme.4274}}.
\newline\urlprefix\url{https://onlinelibrary.wiley.com/doi/abs/10.1002/nme.4274}

\bibitem{Kalashnikova2014}
I.~Kalashnikova, B.~{van Bloemen Waanders}, S.~Arunajatesan, M.~Barone,
  \href{http://dx.doi.org/10.1016/j.cma.2014.01.011}{{Stabilization of
  projection-based reduced order models for linear time-invariant systems via
  optimization-based eigenvalue reassignment}}, Computer Methods in Applied
  Mechanics and Engineering 272 (2014) 251--270.
\newblock \href {http://dx.doi.org/10.1016/j.cma.2014.01.011}
  {\path{doi:10.1016/j.cma.2014.01.011}}.
\newline\urlprefix\url{http://dx.doi.org/10.1016/j.cma.2014.01.011}

\bibitem{Tomas-Rodriguez2013}
M.~Tomas-Rodriguez, S.~P. Banks, {An iterative approach to eigenvalue
  assignment for nonlinear systems}, International Journal of Control 86~(5)
  (2013) 883--892.
\newblock \href {http://dx.doi.org/10.1080/00207179.2013.765037}
  {\path{doi:10.1080/00207179.2013.765037}}.

\bibitem{Kalashnikova2010}
I.~Kalashnikova, M.~F. Barone, {On the stability and convergence of a Galerkin
  reduced order model (ROM) of compressible flow with solid wall and far-field
  boundary treatment.}, International Journal for Numerical Methods in
  Engineering 83~(10) (2010) 1345--1375.
\newblock \href {http://dx.doi.org/10.1002/nme.2867}
  {\path{doi:10.1002/nme.2867}}.

\bibitem{Rempfer2000}
D.~Rempfer, \href{https://doi.org/10.1007/s001620050131}{{On Low-Dimensional
  Galerkin Models for Fluid Flow}}, Theoretical and Computational Fluid
  Dynamics 14~(2) (2000) 75--88.
\newblock \href {http://dx.doi.org/10.1007/s001620050131}
  {\path{doi:10.1007/s001620050131}}.
\newline\urlprefix\url{https://doi.org/10.1007/s001620050131}

\bibitem{Hesthaven1996}
J.~Hesthaven, D.~Gottlieb,
  \href{http://epubs.siam.org/doi/pdf/10.1137/S1064827594268488}{{A Stable
  Penalty Method for the Compressible Navier-Stokes Equations: I. Open Boundary
  Conditions}}, SIAM Journal of Scientific Computing 17~(3) (1996) 579--612.
\newblock \href {http://dx.doi.org/10.1137/S1064827594268488}
  {\path{doi:10.1137/S1064827594268488}}.
\newline\urlprefix\url{http://epubs.siam.org/doi/pdf/10.1137/S1064827594268488}

\bibitem{Sirisup2005}
S.~Sirisup, G.~E. Karniadakis, {Stability and accuracy of periodic flow
  solutions obtained by a POD-penalty method}, Physica D: Nonlinear Phenomena
  202~(3-4) (2005) 218--237.
\newblock \href {http://dx.doi.org/10.1016/j.physd.2005.02.006}
  {\path{doi:10.1016/j.physd.2005.02.006}}.

\bibitem{Kalashnikova2012}
I.~Kalashnikova, M.~F. Barone, {Efficient non-linear proper orthogonal
  decomposition/Galerkin reduced order models with stable penalty enforcement
  of boundary conditions}, International Journal for Numerical Methods in
  Engineering 90~(11) (2012) 1337--1362.
\newblock \href {http://arxiv.org/abs/1010.1724} {\path{arXiv:1010.1724}},
  \href {http://dx.doi.org/10.1002/nme.3366} {\path{doi:10.1002/nme.3366}}.

\bibitem{Lucia2003}
D.~J. Lucia, P.~S. Beran, {Projection methods for reduced order models of
  compressible flows}, Journal of Computational Physics 188~(1) (2003)
  252--280.
\newblock \href {http://dx.doi.org/10.1016/S0021-9991(03)00166-9}
  {\path{doi:10.1016/S0021-9991(03)00166-9}}.

\bibitem{Sirisup2004}
S.~Sirisup, G.~E. Karniadakis, {A spectral viscosity method for correcting the
  long-term behavior of POD models}, Journal of Computational Physics 194~(1)
  (2004) 92--116.
\newblock \href {http://dx.doi.org/10.1016/j.jcp.2003.08.021}
  {\path{doi:10.1016/j.jcp.2003.08.021}}.

\bibitem{Borggaard2011}
J.~Borggaard, T.~Iliescu, Z.~Wang,
  \href{http://dx.doi.org/10.1016/j.mcm.2010.08.015}{{Artificial viscosity
  proper orthogonal decomposition}}, Mathematical and Computer Modelling
  53~(1-2) (2011) 269--279.
\newblock \href {http://dx.doi.org/10.1016/j.mcm.2010.08.015}
  {\path{doi:10.1016/j.mcm.2010.08.015}}.
\newline\urlprefix\url{http://dx.doi.org/10.1016/j.mcm.2010.08.015}

\bibitem{Iollo2000}
A.~Iollo, S.~Lanteri, J.~A. D{\'{e}}sid{\'{e}}ri, {Stability properties of
  POD-Galerkin approximations for the compressible Navier-Stokes equations},
  Theoretical and Computational Fluid Dynamics 13~(6) (2000) 377--396.
\newblock \href {http://dx.doi.org/10.1007/s001620050119}
  {\path{doi:10.1007/s001620050119}}.

\bibitem{Sandia2014}
I.~Kalashnikova, S.~Arunajatesan, M.~F. Barone, B.~G. van Bloemen~Waanders,
  J.~A. Fike, Reduced order modeling for prediction and control of large-scale
  systems., Tech. rep., Sandia National Laboratories (May 2014).

\bibitem{Loeve1945}
M.~Loeve, Aleatoire de second ordre, C R Academie des Sciences, Paris, France.

\bibitem{cipaobsy03}
P.~G.~A. Cizmas, A.~Palacios, T.~O'Brien, M.~Syamlal, Proper-orthogonal
  decomposition of spatio-temporal patterns in fluidized beds, Chemical
  Engineering Science 58~(19) (2003) 4417--4427.

\bibitem{holmes96}
P.~Holmes, J.~L. Lumley, G.~Berkooz, Turbulence, Coherent Structures, Dynamical
  Systems and Symmetry, Cambridge University Press, 1996.

\bibitem{Kantorovich1964}
L.~V. Kantorovich, V.~I. Krylov, Approximate Methods of Higher Analysis,
  Interscience Publishers, Inc., New York, 1964.

\bibitem{Pinnau2008}
R.~Pinnau, {Model Reduction via Proper Orthogonal Decomposition}, Research
  Aspects and Applications\href {http://dx.doi.org/10.1007/978}
  {\path{doi:10.1007/978}}.

\bibitem{Sirovich1987}
L.~Sirovich, {Turbulence and the dynamics of coherent structures, I-III},
  Quart. Appl. Math. 45~(3) (1987) 561--590.

\bibitem{cizpal03}
P.~G.~A. Cizmas, A.~Palacios, Proper orthogonal decomposition of turbine
  rotor-stator interaction, AIAA Journal of Propulsion and Power 19~(2) (2003)
  268--281.
\newblock \href {http://dx.doi.org/http://arc.aiaa.org/doi/pdf/10.2514/2.6108}
  {\path{doi:http://arc.aiaa.org/doi/pdf/10.2514/2.6108}}.

\bibitem{Roe1986}
P.~Roe,
  \href{http://fluid.annualreviews.org/cgi/doi/10.1146/annurev.fluid.18.1.337}{{Characteristic-Based
  Schemes for the Euler Equations}}, Annual Review of Fluid Mechanics 18~(1)
  (1986) 337--365.
\newblock \href {http://dx.doi.org/10.1146/annurev.fluid.18.1.337}
  {\path{doi:10.1146/annurev.fluid.18.1.337}}.
\newline\urlprefix\url{http://fluid.annualreviews.org/cgi/doi/10.1146/annurev.fluid.18.1.337}

\bibitem{Anderson1995}
J.~D. Anderson, {Computational Fluid Dynamics: The Basics with Applications}
  (1995).
\newblock \href {http://dx.doi.org/10.1017/CBO9780511780066}
  {\path{doi:10.1017/CBO9780511780066}}.

\bibitem{Han2003}
Z.-X. Han, P.~G. Cizmas, {A CFD Method for Axial Thrust Load Prediction of
  Centrifugal Compressors}, International Journal of Turbo and Jet Engines
  20~(1) (2003) 1--16.
\newblock \href {http://dx.doi.org/10.1515/TJJ.2003.20.1.1}
  {\path{doi:10.1515/TJJ.2003.20.1.1}}.

\bibitem{ODEPACK}
A.~C. Hindmarsh, {ODEPACK, A Systematized Collection of ODE Solvers},
  Scientific Computing 1 (1983) 55--64.

\bibitem{Blazek2005}
J.~Blazek,
  \href{http://www.sciencedirect.com/science/article/pii/B9780080445069500072}{{Unstructured
  Finite Volume Schemes}}, Computational Fluid Dynamics: Principles and
  Applications (Second Edition) (2005) 131--182\href
  {http://dx.doi.org/http://dx.doi.org/10.1016/B978-008044506-9/50007-2}
  {\path{doi:http://dx.doi.org/10.1016/B978-008044506-9/50007-2}}.
\newline\urlprefix\url{http://www.sciencedirect.com/science/article/pii/B9780080445069500072}

\bibitem{yuanetal05}
T.~Yuan, P.~G. Cizmas, T.~O'Brien, A reduced-order model for a bubbling
  fluidized bed based on proper orthogonal decomposition, Computers and
  Chemical Engineering 30~(2) (2005) 243--259.

\bibitem{Brenner2012}
T.~A. Brenner, R.~L. Fontenot, P.~G. Cizmas, T.~J. O'Brien, R.~W. Breault,
  \href{http://www.sciencedirect.com/science/article/pii/S0098135412001056}{A
  reduced-order model for heat transfer in multiphase flow and practical
  aspects of the proper orthogonal decomposition}, Computers \& Chemical
  Engineering 43 (2012) 68--80.
\newblock \href {http://dx.doi.org/10.1016/j.compchemeng.2012.04.003}
  {\path{doi:10.1016/j.compchemeng.2012.04.003}}.
\newline\urlprefix\url{http://www.sciencedirect.com/science/article/pii/S0098135412001056}

\bibitem{Burkardt2006}
J.~Burkardt, M.~Gunzburger, H.~C. Lee, {POD and CVT-based reduced-order
  modeling of Navier-Stokes flows}, Computer Methods in Applied Mechanics and
  Engineering 196~(1-3) (2006) 337--355.
\newblock \href {http://dx.doi.org/10.1016/j.cma.2006.04.004}
  {\path{doi:10.1016/j.cma.2006.04.004}}.

\bibitem{Funaro1991}
D.~Funaro, D.~Gottlieb, {Convergence results for pseudospectral by a
  penalty-type boundary treatment}, Mathematics of Computation 57~(196) (1991)
  585--596.

\bibitem{Amsallem2008}
D.~Amsallem, C.~Farhat,
  \href{http://arc.aiaa.org/doi/10.2514/1.35374}{{Interpolation Method for
  Adapting Reduced-Order Models and Application to Aeroelasticity}}, AIAA
  Journal 46~(7) (2008) 1803--1813.
\newblock \href {http://dx.doi.org/10.2514/1.35374}
  {\path{doi:10.2514/1.35374}}.
\newline\urlprefix\url{http://arc.aiaa.org/doi/10.2514/1.35374}

\bibitem{Krath2018}
E.~H. Krath, J.~A. Felderhoff, N.~R. Matula, P.~G.~A. Cizmas, D.~A. Johnston, A
  reduced-order model for compressible flows based on an efficient proper
  orthogonal decomposition method, in: Proceedings of the 15th International
  Symposium on Unsteady Aerodynamics, Aeroacoustics \& Aeroelasticity of
  Turbomachines, no. ISUAAAT15-012, University of Oxford, UK, 2018.

\bibitem{Liou1987}
M.-S. Liou, \href{https://doi.org/10.2514/6.1987-355}{{A generalized procedure
  for constructing an upwind-based TVD scheme}}, in: 25th AIAA Aerospace
  Sciences Meeting, Aerospace Sciences Meetings, American Institute of
  Aeronautics and Astronautics, 1987.
\newblock \href {http://dx.doi.org/doi:10.2514/6.1987-355}
  {\path{doi:doi:10.2514/6.1987-355}}.
\newline\urlprefix\url{https://doi.org/10.2514/6.1987-355}

\bibitem{Urasek1979}
D.~C. Urasek, W.~T. Gorrell, W.~S. Cunnan, {Performance of Two-Stage Fan Having
  Low-Aspect Ratio, First-Stage Rotor Blading}, in: NASA Technical Paper, 1979.

\bibitem{Carpenter2016}
F.~L. Carpenter~IV, Practical aspects of computational fluid dynamics for
  turbomachinery, Ph.D. thesis, Texas A\&M University, College Station, Texas
  (August 2016).

\bibitem{kirketal07}
A.~Kirk, A.~Kumar, J.~I. Gargoloff, O.~Rediniotis, P.~G.~A. Cizmas, Numerical
  and experimental investigation of a serpentine inlet duct, in: 45th Aerospace
  Sciences Meeting and Exhibit, AIAA Paper 2007-842, Reno, Nevada, 2007.

\end{thebibliography}

\end{document}